\documentclass[english,a4paper,11pt]{article}

\usepackage[natbibapa]{apacite}
\usepackage{natbib}
\let\cite\citet

\usepackage{tgpagella}
\usepackage{mathpazo}
\usepackage{setspace}

\usepackage[toc,page]{appendix}
\usepackage{amssymb,amsmath,amsthm}
\usepackage{subcaption,graphicx,float,adjustbox}
\usepackage{tikz}
\usepackage{pgfplots}
\pgfplotsset{compat=1.18}
\usepgfplotslibrary{fillbetween}
\usetikzlibrary{intersections}
\usetikzlibrary{shadows,shapes,arrows.meta,positioning}

\tikzset{
    node/.style = {draw, circle, minimum size=1cm, inner sep=2pt, font=\small},
    arrow/.style = {-Stealth}
}
\usepackage{multirow}
\usepackage{stackengine}
\usepackage{booktabs,siunitx,threeparttable,dsfont}

\usepackage{microtype}
\usepackage[citecolor=blue, pdfborder={0 0 0}, unicode, pdfauthor=]{hyperref}

\theoremstyle{definition}
\newtheorem{proposition}{Proposition}
\newtheorem{corollary}{Corollary}
\newtheorem{theorem}{Theorem}

\newtheorem{prediction}{Prediction}
\newtheorem{resq}{Research Question}
\newtheorem{result}{Result}

\title{Knowledge and Freedom: Evidence on the Relationship Between Information and Paternalism}
\author{Max R. P. Grossmann\thanks{University of Cologne, Department of Economics, Albertus-Magnus-Platz, 50923 Köln, Germany. Email: \href{mailto:m@max.pm}{m@max.pm}. ORCID: 0000-0002-6152-9042.}}
\date{\today{} (\href{https://q.mg.sb/knf}{\textcolor{blue}{get latest version}})}

\newcommand{\eg}[1]{\citep[e.g.,][]{#1}}

\newcommand{\censor}[1]{#1}

\newcounter{skipextra}
\newif\ifignore

\begin{document}
\frenchspacing
\maketitle
\vspace{-2em}
\begin{center}
    \textbf{$\star$~~~Job Market Paper~~~$\star$}
\end{center}
\vspace{1em}

\begin{abstract}
    When is autonomy granted to a decision-maker based on their knowledge, and if no autonomy is granted, what form will the intervention take?
A parsimonious theoretical framework shows how policymakers can exploit decision-maker mistakes and use them as a justification for intervention.
In two experiments, policymakers (“Choice Architects”) can intervene in a choice faced by a decision-maker.
We vary the amount of knowledge decision-makers possess about the choice.
Full decision-maker knowledge causes more than a 60\% reduction in intervention rates.
Beliefs have a small, robust correlation with interventions on the intensive margin.
Choice Architects disproportionately prefer to have decision-makers make informed decisions. Interveners are less likely to provide information.
As theory predicts, the same applies to Choice Architects who believe that decision-maker mistakes align with their own preference.
When Choice Architects are informed about the decision-maker’s preference, this information is used to determine the imposed option.
However, Choice Architects employ their own preference to a similar extent.
A riskless option is causally more likely to be imposed, being correlated with but conceptually distinct from Choice Architects' own preference.
This is a qualification to what has been termed “projective paternalism.”

    \par\vspace{1em}

    \par\vspace{1em}

\end{abstract}

\newpage

\onehalfspacing

\section*{Acknowledgments}

We thank Arno Apffelstaedt, Simon Brandkamp, Marius Gramb, Vitali Gretschko, Susanna Grundmann, Sophia Hornberger, Moritz Janas, Sander Kraaij, Axel Ockenfels, Martin Ricketts, Paula Scholz, Frederik Schwerter, Itai Sher, Sven A. Simon, and Robert Slonim, audiences at Buckingham, Cologne, Maastricht, the IMEBESS in Lisbon, the ASFEE in Montpellier and the ESA World in Lyon for valuable comments. All remaining errors are our own. We thank \censor{Martin Strobel} for support in running the experiment. We acknowledge funding by the \censor{Center for Social and Economic Behavior (C-SEB) at the University of Cologne and the German Research Foundation (DFG) under Germany’s Excellence Strategy (EXC 2126/1–390838866)}.

\section*{Declarations}

Financial support was received from the Center for Social and Economic Behavior (C-SEB) at the University of Cologne and the German Research Foundation (DFG) under Germany’s Excellence Strategy, EXC 2126/1–390838866. The author has no competing interests to declare that are relevant to the content of this article.

\section*{Data availability, IRB approval}

All software and data used for this paper are freely and publicly available.
See Section \ref{availabilityknf} in the Appendix for details.

The experiments including key analyses were preregistered and IRB approval was granted. For details, see Sections \ref{exp1proc} and \ref{exp2proc} in the main text.

\newpage

\tableofcontents

\newpage

\section{Introduction}\label{sec.intro}

To successfully act on our preferences, we need to know what we are doing.
Government regulation often conditions freedom of choice on decision-maker knowledge, and decision-maker ignorance has long been understood as a justification for intervention.
Many legal rules protect decision-makers and others.
For example, driver's licenses restrict the use of high-powered vehicles on public roads to decision-makers who possess a basic understanding of rules and common road practices.
Other such regulations target only the decision-maker.
For example, many countries restrict the offering of advanced financial products to the general public.\footnote{The European Union’s MiFID II mandates that investors prove some experience or knowledge about financial markets before they can be offered certain products. Under the U.S. Securities Act of 1933, Rule 506 of Regulation D establishes rigorous requirements to become an “accredited investor,” ensuring that only those with tremendous financial resources can invest in unregistered securities.}
Parents' everyday accounts conform to this story: a recent survey found that American parents' overriding concern about tattoos is that their child may regret it later \citep{MottPoll2018}.
The possibility of decision-maker regret also appears to play an important role in the opposition to organ markets \eg{clemens2018testing,ambuehl2015more,roth2007repugnance}.
Regret is possible because we are ignorant about our future values and experiences. We lack access to these states of mind.
These examples share a common thread: individual freedom can be curtailed based on a perceived lack of knowledge.

Policymakers can exercise their authority to restrict the choices of decision-makers.
They can also choose to provide decision-makers with information central to their decision-making.
Moreover, policymakers' own knowledge can shape the rules that they impose.
Economists have tended to focus on equilibrium behavior in institutions, without much consideration for causal determinants of institutional design.
However, policy is invented through conscious and purposeful human design.
What determines elements of this design?
We use experiments and simple economic theory to investigate how and why experimental policymakers \emph{actually govern} over decision-makers in the presence of asymmetric information on either side.

Does decision-makers' knowledge cause them to obtain more autonomy from impartial policymakers?
We first formulate a parsimonious formal theory of optimal paternalism with and without decision-maker mistakes. Chooser's mistakes can arise from a lack of knowledge.
In contrast to the standard approach in welfare economics, we allow policymakers (“Choice Architects”, CAs) to be non-neutral---to prefer that the Chooser choose the CA's subjectively preferred option.
The model has only two parameters: one for the weight CAs place on their own preference, and a belief about the Chooser's preference.
CAs can elect to have the Chooser have his own choice, or to impose one option on him.\footnote{We follow the convention that the Choice Architect is female and the Chooser male.}
Our theory shows that Chooser mistakes can make freedom of choice less attractive if CAs are sufficiently neutral.
However, there can be a strategic advantage to Chooser mistakes.
The model nests John Stuart Mill's (\citeyear{mill1859on}) idea of respecting fully informed choices and a preference for information provision over imposition.

We conduct two simple experiments to test the influence of Chooser knowledge on the autonomy they are granted.
In both experiments, Choosers choose between a potentially ambiguous lottery and a safe amount.
A CA is matched to each Chooser. She can leave the Chooser free to decide or impose an option on him.
Interventions are made conditional on the Chooser's knowledge. We vary the amount of knowledge that Choosers have about the probability in the lottery.
However, CAs face no ambiguity.\footnote{Experiment 1 uses a version of partial ambiguity \eg{camerer1992recent,chipman1963stochastic,chew2017partial,gigliotti1996testing}: Choosers observe the lottery $k$ times before choosing. In experiment 2, Choosers face a fully ambiguous lottery or no ambiguity whatsoever.}
Both experiments demonstrate that CAs respect increased Chooser knowledge. Few CAs intervene without regard to knowledge.
Across both experiments, full information causes a more than 60\% reduction in intervention rates.
As in previous work, intervening CAs tend to impose their own preference between the options.

Knowledge can be enhanced or diminished by policymakers. If policymakers can send true information to decision-makers, is this opportunity used strategically?
Since Mill, information provision has long been recognized as an alternative to prohibitions.
\cite{blackwell1953equivalent} proved that true information must make a decision-maker better off according to his own subjective utility.
Thus, to most economists, decision-maker mistakes are inherently linked to the correctness of beliefs.
For example, under this criterion, it may well be the case that some individuals should smoke \emph{more} if they overestimate the risks from smoking \eg{Caplan2011ABA}. Some evidence for such overestimation has been found \eg{viscusi1990smokers}. In such cases, economists may argue for a correction of false beliefs.
On the other hand, policymakers can engage in “paternalism by omission” by deliberately keeping decision-makers uninformed if that leads to policymakers' preferred option being more likely to be chosen.
Our theory shows that mistakes are not created equal: departures from true preferences that go in the direction of policymakers' subjectively preferred option can benefit the CA, and in turn mute the provision of true information.

In experiment 2, we allow CAs to construct the information sent to a Chooser.
Overall, about 80\% of CAs do provide information to Choosers.
An additional analysis demonstrates that non-provision of information is indeed associated with a belief that Chooser mistakes go in the direction of CAs' subjectively preferred option.
The presence of this combination of beliefs and CA preferences is associated with a 13.4 p.p. reduction in information provision.
Our theoretical and empirical study of strategic information provision is related to Grossman's (\citeyear{grossman1981informational}) and Milgrom's (\citeyear{milgrom1981good}) early conceptualizations of strategic communication. In these models, a sender can withhold information but not misrepresent it because it is fundamentally verifiable \citep{kartik2009strategic}. In the present paper, verifiability arises because experimental Choosers cannot be deceived.

Policymakers can, in general, use both information provision and intervention in choices.
As outlined above, if it is certain that the Chooser will get his choice, more information must make him better off. However, if intervention can follow information provision, non-provision can be employed strategically to provide a justification for intervention.
In a treatment, we randomly enable some CAs to intervene in the resulting choice simultaneous to deciding on information provision.
In general, we find that whether or not CAs can intervene makes no difference regarding the information they provide to Choosers. However, intervening CAs are 29.7 p.p. less likely to provide information to Choosers.
A novel finding is that only a small but highly statistically significant proportion of subjects---about 3.3\% of CAs---appear to provide information strategically in this way.

So far, we have discussed instances of policymakers simply knowing more than a decision-maker, and asking how the former react to this asymmetric information under varying political economies (imposition \textit{vs.} information provision).
However, policymakers had to rely on their own beliefs or preferences when deciding what to impose onto the decision-maker.
Suppose now that the decision-maker is ignorant, but the policymaker knows about his (counterfactual) full-information preference.
That is, the policymaker actually knows what the decision-maker would have done had he been in possession of all information.
In such cases, policymakers can unambiguously resolve the problem of intervention by just implementing the decision-maker's fully informed preference for him.
Mill highlighted that this behavior complies with notions of classical liberalism, since the intervention is based on what the decision-maker actually wants.
How do policymakers react to this information?

We experimentally provide information about another Chooser's fully informed preferences to CAs.
This approach also helps us disentangle two causal models of the intensive margin: CAs' own preferences may directly cause the intensive margin \eg{ambuehl2021motivates}, or bias beliefs and in turn determine the intensive margin. Our treatment shocks CAs' beliefs: CAs know \emph{with certainty} what option the Chooser would have preferred had he been fully informed.
We find that CAs significantly “help” the Chooser by imposing the hypothetically preferred option.
But this result is to be qualified: first, the null hypothesis that the CA's own preference and the Chooser's type matter to a similar degree cannot be rejected.
Second, we establish a novel finding that CAs do not simply intervene in the direction of their own preference:

Exploratory analyses reveal that a riskless option is causally more likely to be implemented \emph{even when controlling for the CA's own preference}.
When both the CA and the Chooser prefer the riskless option, the CA imposes it in 79.6\% of cases. Yet when both prefer the lottery, the CA imposes the lottery in only 52.2\% of cases.
A similar pattern emerges by correlation for the baseline experiment, in which CAs had to rely on their beliefs about Choosers. CAs are less likely to match Choosers with the risky option than the safe option.

This novel result suggests that CAs are able to conceive a “cosmically” ideal intervention that may not always coincide with their personal preference, providing an important qualification to what has been termed “projective paternalism” \citep{ambuehl2021motivates}.
As we discuss below, our result of a cosmic ideal in intervention may be driven by interpersonal regret: in expected value terms, the lottery used in our experiment is just barely better than the safe option.\footnote{The expected value of the lottery is €16, in contrast to a safe option of €15.}
This design feature allows us to identify cosmic ideals for the first time.

There is a small, significant association between beliefs and what option is imposed. This association is independent of CAs' own preference (although that preference robustly biases beliefs).
We construct a random-utility binary choice model of our formal theory, and estimate its parameters from data. This model organizes our results by revealing that CAs consistently attach a weight of about one-third to their own preference.
Consistent with our results, this implies that CAs actually care a lot about what the Chooser would do if he were able to choose informedly.
This is true for both the baseline experiment (where we utilized incentivized beliefs) as well as for the treatment in which CAs actually knew the hypothetical choice of a Chooser (where beliefs were exogenously provided).
The binary choice model of our formal theory performs well in predicting the intensive margin despite its parsimony; indeed, it performs about as well as full-fledged linear probability models and logistic regressions.

This paper contributes to the intersection of public policy, political economy and experimental economics.
The importance of knowledge in decision-making is universally recognized in economics. Pareto taught in his \textit{Manual} that only repeated choices reveal a person’s preferences \citep[72]{pareto2014manual}. \cite{hayek1945use} highlighted that markets---unlike central planners---can efficiently aggregate the dispersed knowledge of decision-makers using prices as signals. \cite{bernheim2016good} distinguished between direct judgments (underlying preferences) and indirect judgments (e.g., choices). Both the former and, if \emph{correctly informed}, the latter should not be questioned by the analyst.
In the field of public policy, \cite{musgrave1956multiple} introduced the notion of merit goods such as education and healthcare. Policymakers are assumed to possess more knowledge about these goods, justifying their provision to an uninformed or myopic citizenry \citep{musgrave1959theory,head1966merit,kirchgassner2017soft}. The same intuition holds for the examples in the beginning of this introduction: less knowledge means more intervention.

We add to a nascent experimental literature on paternalism \eg{bartling2023free,undo,ambuehl2021motivates,ackfeld2021people,pids,kiessling2022parental,buckle2023paternalism} that investigates foundations of public policy and political economy.
We, too, focus on actions whose outcomes affect only the Chooser, not a third party. Much paternalism \citep{dworkin1972paternalism,dworkin2020paternalism} is motivated by a lack of knowledge on the side of decision-makers. Recent empirical work on paternalism has emphasized that even if the consequences of some action are precisely defined, CAs intervene. For example, a CA might remove impatient options from a choice menu to enforce a minimum of patience in intertemporal choice. CAs tend to intervene in the direction of their own preference while leaving some space for Choosers to express themselves \citep{ambuehl2021motivates,pids}.
As we take no stance on the normative assessment of governmental policy, our approach is purely descriptive or positive \citep{friedman1953methodology}.

Our finding of cosmic ideals modifies and qualifies our understanding of “projective paternalism” as introduced by \cite{ambuehl2021motivates}. Their conceptualization of paternalism as projection suggested two anchors that shape CAs' interventions: \textit{(i)} CAs impose in the direction of their own preference while \textit{(ii)} leaving some space so that Choosers can partially express their preferences.
Because our experiment features a relatively unattractive lottery in a binary choice between that lottery and a safe option, we can identify a third anchor that appears to influence CAs: \textit{(iii)} some options may simply be more “objectively correct” independent of CAs' own preferences. This implies that the projection embedded in paternalistic action is somewhat asymmetric.

A study related to our own is by \cite{bartling2023free}. They conduct a study of paternalism in the United States, and vary the feature of the choice ecology through which a Chooser makes a mistake. They show that few CAs restrict freedom of choice, but that a substantial share of CAs provides information to Choosers. However, their design only allowed information provision or intervention as substitutes, not the joint use of both tools.
We replicate their approach and their results in cases where CAs can both intervene \emph{and} provide information. We characterize the nature of non-providers theoretically and empirically, and we highlight the possibility for a strategic use of information provision if CAs are not neutral.

Another related study is by \cite{buckle2023paternalism}, where CAs impose their risk preference despite knowing the Chooser's preferences. Our results also reveal that both CA and Chooser preferences matter. CAs mix their own preference distinctly with that of the Chooser. We add a recognition that some options are more likely to be implemented beyond CA preferences. We argue that these cosmic ideals represent uncontroversially imposed options---perhaps those without the possibility of regret---that are correlated with, but conceptually distinct from, CA preferences.

We contribute to the literature by investigating central motives of experimental policymakers, with increased Chooser knowledge leading to more respect for autonomy. This paper demonstrates that freedom of choice is strikingly contingent on knowledge.
This core result has important implications for the design of norms and institutions, and sheds light on central behavioral foundations of political economy.
CA behavior is nuanced yet systematic and partially strategic. Interventions go beyond CAs' preferences, hinting at the existence of cosmic ideals that are just more likely to be imposed. Most CAs provide Choosers with crucial information, but non-providers can be empirically and theoretically characterized according to their own preferences, actions, and beliefs. CAs' knowledge also matters.
Our findings suggest that knowledge fundamentally shapes the social contract between them and policymakers. Informational asymmetries are resolved through deliberate institutional responses. Knowledge is a decisive determinant of the boundary between individual autonomy and control by policymakers and others.

The remainder of this paper proceeds as follows: in Section \ref{sec.model}, we formulate a theory of optimal self-interested paternalism. This theory is informed by other authors' normative and positive conceptions of paternalism. We use these insights to design and discuss experimental investigations of the relationship between knowledge and freedom. In Section \ref{sec.experiment}, we present the experimental designs used in our two experiments. Section \ref{results} discusses our results. Finally, we conclude.

\section{Theoretical framework}\label{sec.model}

In this paper, we have a CA govern over a Chooser: the CA decides whether to have the Chooser have his choice between two options, or to impose an option on him. This Section builds a formal model of optimal but possibly self-interested paternalism.

Our framework is most closely related to the models by \cite{NBERw31349} and \cite{bartling2023free}. In contrast to these approaches---and in line with \cite{ambuehl2021motivates}---we allow a CA to include her own preference in a measure of welfare. We investigate the implications for interventions, choices made in place of the Chooser and information provision. The Section subsequently informs model parameters with expectations from the literature in order to sketch hypotheses for experimental research.

\subsection{A formal model of paternalism with and without mistakes}
\label{patformal}

\begin{table}
    \centering

    \begin{tabular}{@{}ccc@{}}
        \toprule
        & CA prefers $x$ & CA prefers $y$ \\ \midrule
        Chooser prefers $x$ & \begin{tabular}[c]{@{}c@{}}$W(x) = 1$\\ $W(y) = 0$\end{tabular} & \begin{tabular}[c]{@{}c@{}}$W(x) = 1-\phi$\\ $W(y) = \phi$\end{tabular} \\
            \\
            Chooser prefers $y$ & \begin{tabular}[c]{@{}c@{}}$W(x) = \phi$\\ $W(y) = 1-\phi$\end{tabular} & \begin{tabular}[c]{@{}c@{}}$W(x) = 0$\\ $W(y) = 1$\end{tabular} \\ \bottomrule
    \end{tabular}
    \caption{$W$, given the Chooser's and CA's possible types}
    \label{wchooserca}
\end{table}

Consider a choice between two options: $x$ and $y$. CA and Chooser are two agents endowed with a strict preference over $x, y$. While the CA does not make an interpersonal utility comparison \citep{kolm1993impossibility,binmore2009interpersonal,hausman1996impossibility}, she does recognize the possibility of the Chooser's preferences disagreeing with her own. The CA's utility function is $U: \left\{x, y\right\} \rightarrow \left\{0, 1\right\}$, and the Chooser's utility function---as perceived by the CA---is $V: \left\{x, y\right\} \rightarrow \left\{0, 1\right\}$.

The CA aggregates these binary utilities using a welfare function, $W: \left\{ x, y, \left\{x,y\right\} \right\} \rightarrow [0, 1]$. The argument to $W$ indicates the menu of choices available to the Chooser.
She has to decide whether to impose one of these options or to let the Chooser have his choice. In this model, freedom is instrumental; i.e., it is merely a means to achieve an end.
For $z \in \left\{x, y\right\}$, $W$ is defined as follows:
\begin{equation}\label{knfcaw}
    W(z) = \phi U(z) + \left(1-\phi\right) V(z),
\end{equation}
where $\phi \in [0,1]$ is the weight placed by the CA on her own preference.
In this model, self-interest may arise from whatever connection a CA feels to the Chooser's choice.
$\phi$ may thus be a reflection of CAs' social preferences \eg{FehrCharnessForthcoming}, paternalistic projection \citep{ambuehl2021motivates}, a means of enforcing norm compliance \eg{traxler2012survey} or an attempt to change the norm by bringing others' behavior in line with the CA's own \citep{khalmetski2024why}.
While our experiments do not create any explicit link between CAs and the Chooser's decision,\footnote{Most importantly, CAs' payment is independent of the Chooser's decision.} a connection may still be present because of prominent behavioral phenomena.

The value of $W\left( \left\{x,y\right\} \right)$ depends on how Choosers actually choose; we define it in the following Sections based on conditional expectations of Equation \ref{knfcaw}.
In contrast to the usual approach in welfare economics, policymakers may be self-interested. As we demonstrate in Section \ref{millintensive}, John Stuart Mill implied that $\phi$ ought to be nil. In that case, the CA assumes the Chooser's preference. However, as shown empirically \eg{ambuehl2021motivates}, interventions are correlated with CAs' own preferences. In that case, $\phi$ may be strictly positive. Table \ref{wchooserca} gives all possible values of Equation \ref{knfcaw} if Chooser preferences are known to the CA exactly. In that case, CAs partially project their own tastes on Choosers.

\subsubsection{Beliefs}

In the following---without loss of generality---we restrict our analysis to an $x$-preferring CA.
A Chooser's type may not be known exactly, or a CA may need to evaluate $W$ for a distribution of types. Let $\widetilde{q} \in [0,1]$ denote the CA's belief about the proportion of Choosers preferring $x$. By mixing over Chooser types in Table \ref{wchooserca}, we find
\begin{align}
    W_{x}(x) &= \phi + \widetilde{q} (1-\phi), \label{camixed1}\\
    W_{x}(y) &= (1-\widetilde{q}) (1-\phi), \label{camixed2}
\end{align}
with the subscript indicating the CA's type. In Section \ref{knftheorydata}, we use Equations \ref{camixed1}--\ref{camixed2} (and those corresponding to $W_{y}(x)$ and $W_{y}(y)$) and incentivized or provided data on beliefs to estimate $\phi$.

\subsubsection{Interventions without Chooser mistakes}
\label{knfnomistakes}

Consider first the instrumental value of liberty if Choosers choose \emph{perfectly} according to their type.
Assume that Choosers choose the option implied by their true preferences \citep[for critical viewpoints of this concept, see][]{sugden2022debiasing,vspecian2019precarious}. That is, all Choosers with a preference for option $x$ choose option $x$. Similarly, all Choosers with a preference for option $y$ choose $y$.

In the absence of mistakes, $\widetilde{q}$ in Equations \ref{camixed1}--\ref{camixed2} will equal $1$ for those who choose $x$ and $0$ for those who choose $y$.\footnote{Recall how $\widetilde{q}$ is defined as a belief about preferences, not choices. However, if choices are noiseless, both concepts coincide.}
Let $\pi \in (0,1)$ be the proportion of Choosers opting for option $x$. We define freedom of choice as the Chooser being able to choose from the menu $\left\{x,y\right\}$. Then, by mixing over $W$, for all CA types $\theta \in \left\{x,y\right\}$,
\begin{equation}\label{actualw}
    W_{\theta}\left(\left\{x,y\right\}\right) = \pi W_{\theta}\left(x\,\middle\vert\,\widetilde{q} = 1\right) + \left(1-\pi\right) W_{\theta}\left(y\,\middle\vert\,\widetilde{q} = 0\right).
\end{equation}

It is straightforward to compute that $W_{x}\left(\left\{x,y\right\}\right) = 1 - \phi (1-\pi)$. The CA will compare this value of Equation \ref{actualw} against Equations \ref{camixed1}--\ref{camixed2}, with $\widetilde{q} \equiv \pi$. This ternary comparison of welfares determines the CA's governance over the Chooser. In this model, the comparison between Equations \ref{camixed1}--\ref{camixed2} (i.e., what to impose) is independent of the comparison between Equations \ref{camixed1}--\ref{camixed2} and Equation \ref{actualw} (i.e., whether to intervene, and if so, how).
This \emph{independence of irrelevant alternatives} arises because Equations \ref{camixed1}--\ref{camixed2} are independent of actual choice proportions---they refer only to beliefs about true preferences.
This finding is important for empirical analysis. It suggests that we can separately analyze the extensive and intensive margins of paternalistic intervention.

Let us reconsider the relationship between Equations \ref{camixed1}--\ref{actualw} if Choosers choose correctly. It follows from Equations \ref{camixed1}--\ref{camixed2} that an $x$-preferring CA will never impose $y$. This is because $W_{x}\left(\left\{x,y\right\}\right) - W_{x}\left(y\,\middle\vert\,\widetilde{q} = \pi\right) = \pi > 0$. (Similarly, a $y$-preferring CA will not impose $x$.)

\begin{proposition}\label{knfth1}
    Without Chooser mistakes, a CA will (weakly) impose her own preference if and only if $\phi \geq 1/2$. Otherwise, she does not intervene at all.

    \begin{proof}
        Use Equations \ref{camixed1} and \ref{actualw}, setting $\widetilde{q} = \pi$, and solve for $\phi$.
    \end{proof}
\end{proposition}

In other words, if CAs weigh their own preference heavily enough, CAs implement it. This reproduces the core empirical result of \cite{ambuehl2021motivates} in a simpler model.

\begin{figure}
\begin{tikzpicture}
\begin{axis}[
    axis lines=middle,
    xlabel=\(\phi\),
    ylabel={\(\widetilde{q} \equiv \pi\)},
    xmin=0, xmax=1,
    ymin=0, ymax=1,
    xtick={0,0.5,1},
    ytick={0,0.5,1},
    clip=false,
    domain=0:0.5,
    samples=100,
    smooth,
    area style,
    unit vector ratio=2.5 1,
    height=10cm
    ]
    \addplot[name path=F,black,thick] {((1 - 2*x)/(2*(1 - x)))};
    \path[name path=axis] (axis cs:0,0) -- (axis cs:1,0);
    \addplot [
        thick,
        color=blue,
        fill=blue,
        fill opacity=0.3
    ]
    fill between[
        of=F and axis
    ];
    \addplot[dashed] coordinates {(0.5, 0) (0.5, 1)};
    \fill[color=red, fill opacity=0.3] (axis cs:0.5,0) rectangle (axis cs:1,1);

    \node[align=center] at (axis cs:0.75, 0.5) {$x \succ \left\{x,y\right\} \succ y$};
    \node[align=center] at (axis cs:0.2, 0.7) {$\left\{x,y\right\} \succ x \succ y$};
    \node[align=center] at (axis cs:0.2, 0.15) {$\left\{x,y\right\} \succ y \succ x$};
\end{axis}
\end{tikzpicture}
    \caption{Optimal CA decisions}
    \label{optimalca}
\end{figure}
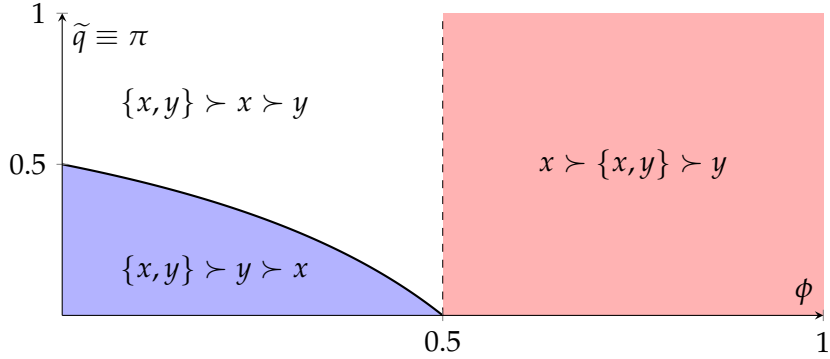

Under choice without mistakes, CAs cannot impose their non-preferred option. However, it is not true that $x$ always prevails over $y$; rather, freedom of choice prevails over $y$.
Figure \ref{optimalca} reveals why. For an $x$-preferring CA, there are two corridors at which the ranking of optimal governance changes. At very low $\widetilde{q}$, $y$ would be chosen over $x$ if the choice were only between $x$ and $y$, not $x$, $y$ and $\left\{x,y\right\}$. However, at the points where $y$ could be imposed, $\phi$ is too low to cause an intervention.

Moreover, note how the non-utilitarian nature of the model produces a majoritarian standard in the case of $\phi = 0$. If a CA were forced to intervene, theory predicts a clear trade-off between $\phi$ and $\widetilde{q}$, with the more popular option being imposed at $\phi = 0$. Section \ref{fcb} discusses this majoritarian standard further. Both $W_{x}\left(\left\{x,y\right\}\right)$ and $W_{x}\left(x\right)$ increase in $\widetilde{q}$ for all $\phi$.

\subsubsection{Mistakes and intervention}
\label{knfmistakes}

We now relax the assumption that Choosers choose correctly. In a choice between two options, mistakes can be incorporated easily.
Let $\epsilon_x, \epsilon_y \in [0,1]$. We say that a proportion of $\pi' = \pi-\epsilon_x+\epsilon_y$ Choosers chooses $x$.
Consider the following contingency table for Chooser preferences and decisions:

\begin{center}
    \begin{tabular}{@{}lccc@{}}
        \toprule
        & Choose $x$           & Choose $y$ &   \textbf{Sum}        \\ \midrule
        Prefer $x$ & $\pi-\epsilon_x$                  & $\epsilon_x$                  & $\pi$ \\
        Prefer $y$ & $\epsilon_y$                  & $1-\pi-\epsilon_y$                  & $1-\pi$ \\
        \textbf{Sum} & $\pi'$ & $1-\pi'$              & 1 \\  \bottomrule
    \end{tabular}
\end{center}

An absence of mistakes implies $\epsilon_x = \epsilon_y = 0$. We now allow arbitrary off-diagonal elements. Out of $\pi'$, only a proportion of $(\pi-\epsilon)/{\pi'}$ Choosers actually prefers $x$. Similarly, a proportion of ${\epsilon_x}/(1-\pi')$ Choosers prefers $x$ (but they choose $y$). Equations \ref{camixed1}--\ref{camixed2} are unchanged, as they only refer to $\widetilde{q}$, i.e., beliefs about true preferences. However, Equation \ref{actualw} must be adjusted to account for a change in conditional proportions.
\begin{equation}\label{actualw2}
    W'_{\theta}\left(\left\{x,y\right\}\right) = \pi' W_{\theta}\left(x\,\middle\vert\,\widetilde{q} = \frac{\pi-\epsilon_x}{\pi'}\right) + \left(1-\pi'\right) W_{\theta}\left(y\,\middle\vert\,\widetilde{q} = \frac{\epsilon_x}{1-\pi'}\right).
\end{equation}

The following Theorem establishes the welfare attained under freedom of choice.

\begin{theorem}\label{knfth2}
    \emph{Freedom with Chooser mistakes.}
    \begin{equation}
        W'_{x}\left(\left\{x,y\right\}\right) = 1 - \phi \left[ 1 - \pi - 2\epsilon_y \right] - \epsilon_x - \epsilon_y.
    \end{equation}

    \begin{proof}
        Use Equations \ref{camixed1}--\ref{camixed2} in Equation \ref{actualw2}.
    \end{proof}
\end{theorem}

Let us briefly reflect on this Theorem. Recall that the error implied by $\epsilon_x$ is worse, to an $x$-preferring CA, than that implied by $\epsilon_y$. In the case of $\epsilon_x$, Choosers who would have preferred $x$ erroneously choose $y$, the option not preferred by the CA. On the other hand, $\epsilon_y$ reflects a proportion of Choosers who now choose $x$ (although their true preferences are better reflected by $y$). This highlights that freedom of choice is purely instrumental in this model.
${\partial W'_{x}\left(\left\{x,y\right\}\right)}/{\partial \epsilon_x} = -1$, implying that an increase in the error disfavored by the CA always leads to an absolute decrease in the welfare accorded to liberty.
However, ${\partial W'_{x}\left(\left\{x,y\right\}\right)}/{\partial \epsilon_y} = -1 + 2\phi$. Not only is the negative effect on welfare attenuated by an increase in the error favored by the CA, but if $x$ is heavily weighed by the CA, more errors in $x$'s direction actually enhance the attractiveness of liberty.

Recall that the ranking of $x$ and $y$ on the intensive margin is unaffected by mistakes.
This implies that we can concern ourselves exclusively with the effect of variations to Choosers' choice ecology on the extensive margin. The following Corollary makes one statement about such an effect.

\begin{corollary}\label{knfth3}
    \emph{Chooser mistakes make freedom less attractive for sufficiently neutral CAs.}
    Suppose a CA does not intervene where Choosers choose without the possibility of mistakes. She may intervene where Choosers have the possibility of making mistakes.

    \begin{proof}
        If the CA does not intervene if Choosers cannot make mistakes, this reveals her $\phi \leq 1/2$ (Proposition \ref{knfth1}). Consider the following object:

        \begin{align*}
            \Delta &= W'_{x}\left(\left\{x,y\right\}\right) - W_{x}\left(\left\{x,y\right\}\right)\\
            &= - \epsilon_x - \epsilon_y \left[ 1 - 2\phi \right].
        \end{align*}

        Since $\phi \leq 1/2$ and $\epsilon_x, \epsilon_y \geq 0$, $\Delta \leq 0$.
    \end{proof}
\end{corollary}

This reduction in the welfare perceived by the CA is nonzero if $\pi \neq \pi'$. It may cause $\left\{x,y\right\}$ to become less desirable than the imposition of $x$ or $y$.

Chooser mistakes can cause a “peeling back” of the dashed frontier in Figure \ref{optimalca}. If the area of intervention is increased, Choosers may face the imposition of $x$ or $y$. Above, we showed that if no Chooser mistakes are possible, the intensive margin is restricted to CAs' own type. This result does not obtain here.

In some instances, CAs can choose between $W'_{x}\left(\left\{x,y\right\}\right)$ and $W_{x}\left(\left\{x,y\right\}\right)$. For example, Corollary \ref{knfth3} suggests that if information can be costlessly provided to Choosers to help them decide without mistakes, CAs can raise welfare by doing so if $\phi \leq 1/2$.
Corollary \ref{knfth3} thus also reproduces Mill's idea that providing information to Choosers is preferable over imposition (Section \ref{infoprov}).
Moreover, the Corollary implies that some CAs may consciously choose to have Choosers make mistakes by having him decide uninformedly---because those mistakes benefit the CA.

\subsection{Mill on paternalism}

We draw on the work of English economist\footnote{See \cite[ch. 6]{cowen2023goat}.} and philosopher John Stuart Mill (1806--1873) to inform economic theory and to create predictions for positive research \citep{friedman1953methodology} on paternalism. Mill's “On Liberty” (\citeyear{mill1859on}) emphasizes that \emph{(i)} people are, in general, best left free to pursue their lives unless \emph{(ii)} they have insufficient knowledge about the consequences of their actions. In this case, interventions must be based on a counterfactual assessment of true preferences.

\subsection{Extensive margin: The “if” of interventions}
\label{millextensive}

Freedom of choice allows a diverse group of individuals to satisfy every extant preference \citep{konrad2023political}. In a society with wide-ranging preferences and viewpoints, liberty will increase satisfaction compared with imposition \citep[ch. 1]{mill1859on}.

Nonetheless, Mill provides some exceptions to this general rule. Of the five identified by \cite{mabsout2022john}, arguably the most relevant for real-life policy is harm to others. We focus instead on a lack of knowledge. At least two authors \citep{arneson1980mill,scoccia2018concept} have attempted to modify the definition of paternalism so that interventions motivated by a lack of knowledge on the side of decision-makers are only paternalistic if they conflict with their full-information preference \citep[fn. 6]{mabsout2022john}. We discuss below the relevance of this counterfactual assessment in Mill.\footnote{At this point, let us offer two arguments in response to Arneson and Scoccia. First, real-life policy governs over broad swathes of the population that is differentially informed. Some of these Choosers may possess all relevant information and still decide to act in a specific way. To that extent, the policy regains a paternalistic character. Second, this argument begs the question of what relevant knowledge is. Policymakers must unavoidably make subjective judgments about choice situations and whether they feel able to improve upon them. Whether decision-makers have all objectively required knowledge is not a helpful benchmark because what information is useful is merely another subjective judgment, one that has often given rise to abuse \citep{berlin1958two}. In acting upon this subjective assessment, policymakers unavoidably accept to place their judgment upon others; they prioritize their own understanding of a situation.}
Nonetheless, as hinted at in the introduction, it is important to note that one does not need to accept Dworkin's definition of paternalism to recognize the profound role that Chooser knowledge plays in policy.

\citeauthor{mill1859on} viewed it acceptable to never intervene. However, he gives the following well-known example:

\begin{quotation}
    If either a public officer or any one else saw a person attempting to cross a bridge which had been ascertained to be unsafe, and there were no time to warn him of his danger, they might seize him and turn him back without any real infringement of his liberty; for liberty consists in doing what one desires, and he does not desire to fall into the river. \textit{\citep[172f.]{mill1859on}}
\end{quotation}

Here, a CA is neutral and reacts only to informational disadvantages: if the pedestrian knew what he was doing, it would not be justifiable to intervene. The pedestrian lacks knowledge and warning him is not possible, \emph{ipso facto} intervention is justified. In the language of our formal model, the pedestrian makes a mistake: there is a discrepancy between preferences and choices (Section \ref{knfmistakes}).

Below, we experimentally vary the amount of knowledge available to Choosers in a choice between two options. This tests the fundamental conditionality of liberty on knowledge that Mill so eloquently put forward: in a situation where knowledge is fixed and cannot be provided, do more interventions take place if the Chooser is less informed?
This aspect of our research deals with the \emph{extensive margin} of paternalism: whether or not interventions take place. As demonstrated above, theory provides a justification for the separation of the extensive and intensive margins.

\subsection{Information provision as a substitute to intervention}
\label{infoprov}

Mill argued that knowledge should be provided instead of intervening, but if that is not possible, it is legitimate to intervene even if the harms accrue only to the decision-maker. Corollary \ref{knfth3} explains why: the appeal of liberty is reduced if decision-maker mistakes are possible. (That is, if the CA is Millian and does not weigh her own preference too heavily.)

It has long been recognized in economics \citep{blackwell1953equivalent} that an expected-utility decision-maker can better his position by relying on more accurate information. Indeed, \cite{bartling2023free} found experimentally that a vast majority of CAs transmit true information to Choosers. However, information can play a strategic role for self-interested policymakers.
First, if (a lack of) information leads to Chooser mistakes in the CA's subjectively preferred direction, information may be withheld from Choosers. Corollary \ref{knfth3} illustrates this point formally. This argument hints at why policymakers do not correct the widespread pessimistic misconceptions about health risks to smoking \citep{viscusi1990smokers}: these misconceptions go in policymakers' subjectively preferred direction (fewer people smoke). By not correcting these misconceptions, policymakers engage in “paternalism by omission.”

Second, if it is certain that the Chooser obtains his choice after information is provided, the CA may help to effectuate the Chooser's preferences by providing information. In this sense, the provision may activate a utilitarian system in CAs, where they may disagree with what the Chooser does, but where they keep him informed so that he may at least achieve his ends.
However, if an opportunity to intervene presents itself to the CA, any prior provision of information may be used “behaviorally.” As an example, consider that Choosers may be grateful for intervention if they are uninformed. If a CA suspects that this may be the case, she can launder her intervention by keeping the Chooser uninformed, hiding that she controlled the information received. Far from enabling a utilitarian system in CAs, the possibility of a follow-on intervention changes the first-stage calculus of providing information. We test this idea below: in the baseline, CAs can only provide information (or not), similar to \cite{bartling2023free}. In a treatment, they may simultaneously intervene.
From a theoretical perspective, the treatment tests the independence of irrelevant alternatives: in the baseline, CAs choose between $W_{\theta}\left(\left\{x,y\right\}\right)$ and $W'_{\theta}\left(\left\{x,y\right\}\right)$. In the treatment, $W_{\theta}\left(x\right)$ and $W_{\theta}\left(y\right)$ are additionally available (both with and without information provided to Choosers).

\subsection{Intensive margin: The “how” of interventions}
\label{millintensive}

Mill makes a normative statement about the intensive margin should an intervention ever occur. It relates to what is still supposedly intended by many interventions: the implementation of true preferences. Mill makes an inference on the hypothetical preference of the “person” \textit{viz.} Chooser. The intervener is to implement the Chooser's counterfactual full-information preference, not her own personal view. If the pedestrian's goal was to get wet, intervention would not have been legitimate \citep{arneson1980mill}. This implies that $\phi = 0$ in Equations \ref{camixed1}--\ref{camixed2}, which in turn requires CAs to implement the Chooser's true preference if it is known.

If a CA is not informed about the Chooser's hypothetical full-information preference, what will they rely upon to form an intervention? Equations \ref{camixed1}--\ref{camixed2} suggest that beliefs about counterfactual choice proportions ought to be used, and perhaps a majoritarian standard. Moreover, we test how CAs use exact information on a Chooser's type. Let us now distinguish a number of positive and normative theories on how to intervene.

\subsubsection{Utilitarianism}

In the absence of hypothetical full-information choices, no information about utilities is available. Furthermore, no method for eliciting a prediction of utilities is known. Thus, this theory cannot be tested. It is not necessary at this point to delve into the manifold issues with utilitarianism and utilitarian calculation in general \eg{kolm1993impossibility}.

\subsubsection{Projective paternalism}

Previous studies on paternalism and intervention \eg{ambuehl2021motivates,pids} have emphasized a profound tendency for CAs to impose “in the direction” of their own preference as well as a similar bias in beliefs. For example, \cite{pids} show how CAs that share their data tend to believe that others also want to share their data. False consensus bias has a long history in psychology; see \cite{ross1977false} for a magisterial exposition. More recently, the idea that individuals project their tastes onto others has garnered renewed attention in experimental economics. For example, \cite{bushong2020experiment} show that workers in a real-effort task who were more (less) tired believed other workers to be less (more) willing to work.

We can test this theory by correlating the CA's own preference with the intensive margin. Ideally, we would shock the CA's own preference to provide causal estimates. However, we leave that to future work.

\subsubsection{Majoritarian counterfactuals}
\label{fcb}

In the absence of information on utilities---as in Section \ref{patformal}---CAs can rely on their beliefs in selecting the intensive margin. One simple theory is that the CA implements what she believes to be the majority choice under full information. Such an intervention could be justified by reference to the median voter theorem \citep{black1948rationale,downs1957economic}, as this voter would prevail in a simple majority election if preferences are single-peaked. In our theoretical framework, use of a majoritarian standard implies $\phi = 0$.

\begin{figure}
    \centering

    \begin{tikzpicture}[
            node distance = 1cm,
            node/.style = {draw, circle, minimum size=3cm, text width=2.5cm, align=center, inner sep=2pt, font=\footnotesize},
            arrow/.style = {-Stealth}
        ]

        \node[node] (A) {CA preference for an Option};
        \node[node, right=of A] (B) {Beliefs are biased};
        \node[node, right=of B] (C) {Impose an Option};

        \draw[arrow] (A) -- (B);
        \draw[arrow, dashed, bend left=45] (A) to (C);
        \draw[arrow] (B) -- (C);

    \end{tikzpicture}

    \caption{Two causal models of interventions based on beliefs}
    \label{dagfcb}
\end{figure}
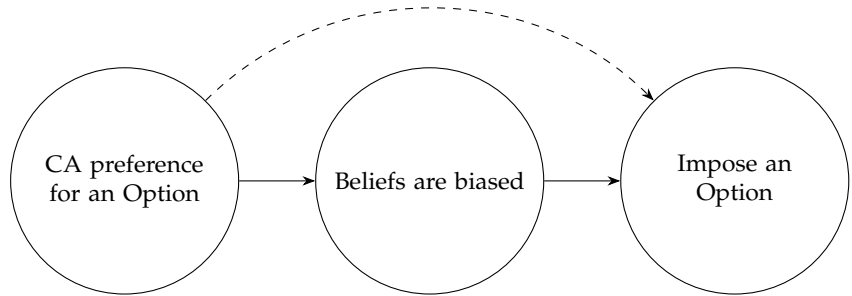

It is important to note that beliefs may be incorrect \citep{ambuehl2021motivates,ross1977false}, highlighting the subjective nature of rule-making. For this reason, it is essential to disentangle two potential causes of an intervention on the intensive margin (Figure \ref{dagfcb}). First, a CA's own preference could lead a majoritarian CA to believe that an Option is more popular at full information than it really is, leading to a systematic but unintended imposition of the subjectively preferred Option. Second, a CA's own preference could \emph{directly} cause the imposition of the preferred Option, without Mill's normative inference.

We interpret the indirect path as purely \emph{statistical or majoritarian intervention} and the direct path as \emph{taste-based intervention}. In \citeauthor{becker1957economics}’s classic discussion (\citeyear{becker1957economics}), similar language is used to distinguish two fundamental motives of discrimination. In the modern reading of this perspective, beliefs can be biased \eg{bohren2019inaccurate}, but statistical discrimination relies on perceived differences in productivity, whereas taste-based discrimination is concerned with non-objectifiable personal decider biases. The same fundamental distinction applies here. As Figure \ref{dagfcb} demonstrates, CA tastes can enter interventions on the intensive margin on two paths. Majoritarian intervention demands that the intensive margin is selected from beliefs about the most popular full-information counterfactual choice, i.e., without the dashed path. On the other hand, real CAs may intervene on taste-based grounds.
The tendency to---inadvertently---impose one's own preference must be accounted for in the analysis by including both beliefs and the CA's own preference. Otherwise, the estimates for the weight of the CA's own preference may be biased upward. Our formal model makes precise statements about the trade-off between $\phi$ and $\pi$ which we use in estimation (Section \ref{knftheorydata}).

\subsubsection{Individual counterfactuals: Testing Mill's postulate}\label{millpost}

Mill's idea of implementing what the Chooser \emph{would have done given full information} can be tested directly. CAs can be informed about the Chooser's type. \cite{ambuehl2021motivates} revealed the self-image of some Choosers to CAs, which significantly shifted the intensive margin of interventions. More recently, \cite{buckle2023paternalism} conducted an experiment in which decisions in an investment game could be overridden by financial advisors. These authors found that advisors use both their own preferences and those of the investor to decide on the intensive margin.

In the context of paternalism, focusing on a single individual also resolves the issues of utilitarian calculation mentioned above. Beliefs about true preferences are shocked to $\pi = 0$ or $\pi = 1$. If CAs are provided with information about what the Chooser would have done had they possessed full information, Mill's postulate about the intensive margin---that $\phi = 0$---can be separately tested.

\subsection{Aggregating degrees of knowledge}
\label{aggregate}

A related literature on positive welfare economics has investigated how CAs can work to aggregate the preferences of a group's members. In \cite{ambuehl2024interpreting}, group members have to work on a task. Each member possesses an individual preference rankings about the available tasks; an experimental social planner is informed about these rankings to assign tasks to each member of the group. Separately, these authors study how planners direct donations to Swiss political parties based on similar rankings.

Both preferences and degrees of knowledge can vary widely in society. Below, we study a related case: how degrees of knowledge are aggregated when a single person can either be well informed or not informed at all. A CA concerned with a Chooser who is in one of several states has to weigh the relative importance of each state. We conduct trials to explain interventions for this scenario using previous interventions in which the amount of information provided to the Chooser was known to CAs in advance.

\section{Experimental design}
\label{sec.experiment}

In our experiments we vary the amount of information obtained by a Chooser in a binary choice and let CAs govern for the Chooser. CAs face no monetary incentives whatsoever from the Chooser's actual choice.

\subsection{Operationalizing partial knowledge}\label{partialknowledge}

Consider the following binary choice faced by a Chooser:

\begin{center}
    \begin{tabular}{@{}cccc@{}}
        \toprule
        \multicolumn{2}{c}{\textbf{Option 1}}     & \multicolumn{2}{c}{\textbf{Option 2}} \\ \midrule
        Prob.              & Outcome              & Prob.            & Outcome            \\ \midrule
        \multirow{2}{*}{1} & \multirow{2}{*}{$z$} & $p$              & $y'$               \\ \cmidrule(l){3-4}
        &                      & $1-p$            & $y''$              \\ \bottomrule
    \end{tabular}
\end{center}

with $y' \leq z \leq y''$. Option 1 is a certain amount of money, while Option 2 is a simple lottery. The degree of knowledge that Choosers obtain about Option 2 can be varied. We consider only an ambiguity variant in which knowledge about $p$ is varied, but not information about $z, y', y''$.

In our experiments, we will consider three cases: \textit{(i)} the Chooser will know the value of $p$ exactly (there is no ambiguity); \textit{(ii)} the Chooser will observe exactly $k$ draws from Option 2 (knowing a prior for $p$);  or \textit{(iii)} the Chooser obtains no information whatsoever about $p$, not even the prior. In case \textit{(i)}, the Chooser is fully informed, while in \textit{(iii)} he decides in a scenario of sheer ignorance. Case \textit{(ii)} is an intermediate position. Experiment 1 will focus on the comparison between cases \textit{(i)} and \textit{(ii)}, while experiment 2 uses the contrast between cases \textit{(i)} and \textit{(iii)}. In all experiments, given this information structure, the Chooser decides between the Options under rules constructed by a CA.

\subsection{Ambiguity in the Estimation Game}
\label{partialambiguity}

Option 2 can be made partially ambiguous by having the Chooser observe $k$ draws from Option 2 before deciding. This is what we call the “Estimation Game.”

This Game lends itself well to experiments with asymmetric information. Consider the following setup: a computer draws $p$ in Option 2 from a distribution. The decision-maker (Chooser) enters the experiment. The Chooser obtains $k$ draws from Option 2 and the prior. $n \leq k$ draws show $y'$. A CA may be matched to the Chooser; this CA knows the exact value of $p$. CAs may intervene in the Choosers' choice between Options 1 and 2 (or let the Chooser have his choice).
From an economic viewpoint, Choosers estimate the utility of Option 2 to decide whether to choose it over Option 1 \citep[see][for an early example of partial ambiguity]{gigliotti1996testing}. From a statistical viewpoint, the Chooser attempts to estimate $p$.
By having CAs decide for varying values of $k$, we can disentangle the effect of Chooser knowledge on their autonomy. That is the setup of experiment 1 below (Section \ref{exp1}).

Clearly, $n \sim \text{Binomial}(k, p)$. For any $n, k$, and the uniform distribution from which $p$ is drawn, the Bayesian posterior for $p$ is Beta$(n+1, k-n+1)$. Marginalizing over $n$, we obtain a marginal posterior of $p$ with the following density function:
\begin{align}
    f_{k}(x) &= \sum_{n=0}^{k} \frac{x^n (1-x)^{k-n} }{B(n + 1, k - n + 1)} \binom{k}{n} p^n (1-p)^{k-n}\\
    &= (k + 1) (1-p)^k (1-x)^k \, { _2 F_1 \left( -k, -k, 1, \frac{px}{ (p-1) (x-1) } \right) },
\end{align}
where $_2 F_1$ is the hypergeometric function.
This is the distribution an expected-utility CA would expect an expected-utility Chooser to work with if the CA knows $p$ and the Chooser is to obtain $k$ draws from Option 2, but the precise $n$ observed by the Chooser is not yet known.

Suppose that CAs decide for the case of $p = 0.2$, $z = \text{€}15$, $y' = \text{€}0$ and $y'' = \text{€}20$ (as below). Table \ref{margpost} in the Appendix presents summary statistics of the marginal posterior for $p = 0.2$. All of these statistics converge in $k$, i.e., a higher $k$ carries superior information. These measures are important beyond \citeauthor{blackwell1953equivalent}'s (\citeyear{blackwell1953equivalent}) order, as they highlight the value of an increased $k$ even for non-expected-utility Choosers. As $k$ increases, Choosers' inference on $p$ is robustly improved because a higher $k$ is monotonically more informative. The Estimation Game is similar to the well-known “balls and urns” paradigm, but the samples from Option 2 are drawn with replacement.

\subsection{Choice Architects' information structure}

In all experiments, CAs obtain information about the Chooser's decision scenario. They know $z, y', y'', p$ and they know what Choosers know: $z, y', y'', k$ and that $n$ will be drawn once from $\text{Binomial}(k, p)$. CAs may then impose one of the Options or have the Chooser have his choice. All experiments were free of deception.

In experiment 1, CAs know that their decision for the Chooser can only be implemented if $p$ takes on the value $0.2$.\footnote{Instead of a continuous distribution for $p$, we used a discrete uniform distribution from 0\% to 100\%, inclusive, in 1\% increments, to draw from $p$.} In experiment 2, CAs know that $p$ will certainly take the value $p=0.2$ for Choosers. Their decision can only be implemented if they are randomly selected.

\subsection{Experiment 1}
\label{exp1}

Experiment 1 investigates the effect of Choosers' partial ambiguity (Section \ref{partialambiguity}) on the freedom they are granted.

\subsubsection{The Choice Architect's view}

\begin{figure}[H]
    \centering
    \begin{minipage}{0.45\textwidth}
        \begin{subfigure}[b]{\textwidth}
            \includegraphics[width=\textwidth]{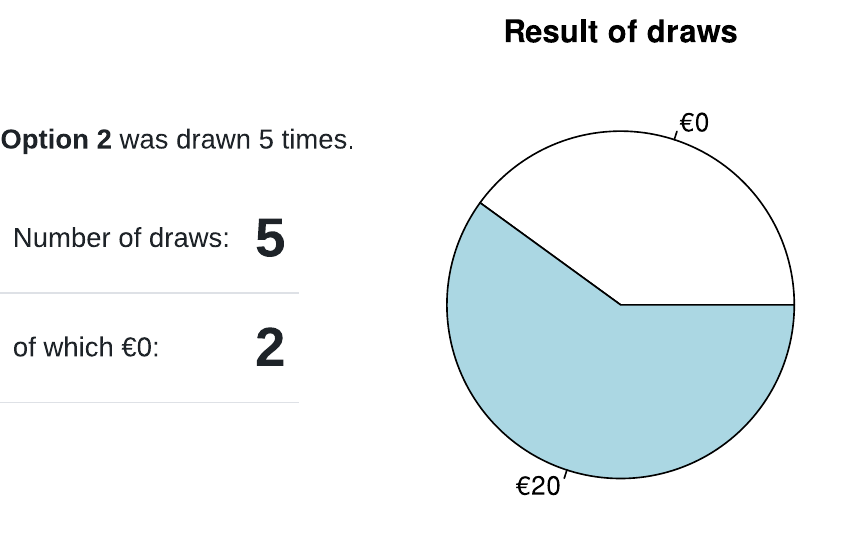}
            \caption{$k = 5$, $n = 2$}
            \label{fig:plot1}
        \end{subfigure}
    \end{minipage}
    \hfill
    \begin{minipage}{0.45\textwidth}
        \begin{subfigure}[b]{\textwidth}
            \includegraphics[width=\textwidth]{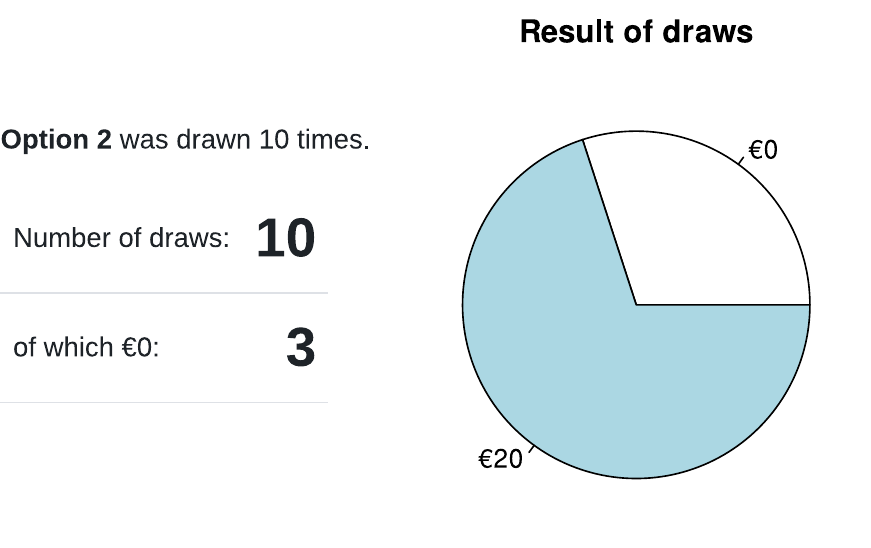}
            \caption{$k = 10$, $n = 3$}
            \label{fig:plot2}
        \end{subfigure}
    \end{minipage}
    \vskip\baselineskip
    \begin{minipage}{0.45\textwidth}
        \begin{subfigure}[b]{\textwidth}
            \includegraphics[width=\textwidth]{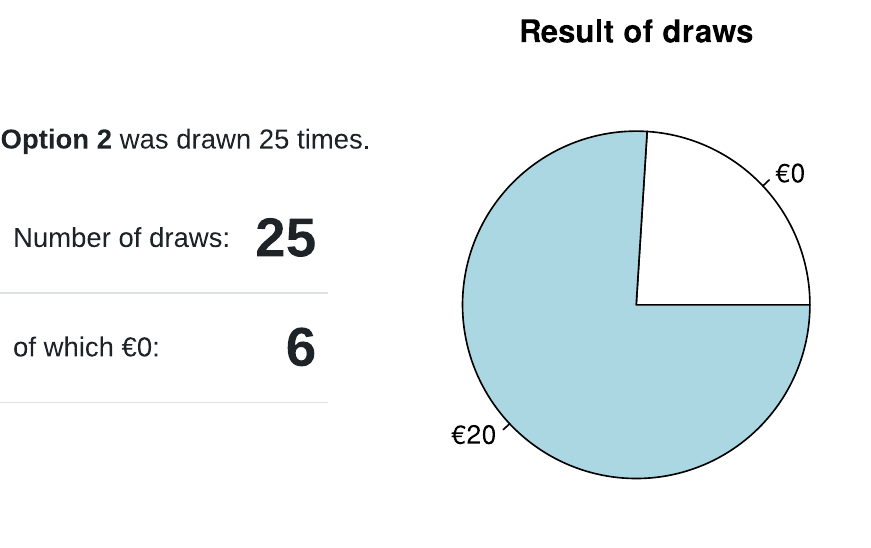}
            \caption{$k = 25$, $n = 6$}
            \label{fig:plot3}
        \end{subfigure}
    \end{minipage}
    \hfill
    \begin{minipage}{0.45\textwidth}
        \begin{subfigure}[b]{\textwidth}
            \includegraphics[width=\textwidth]{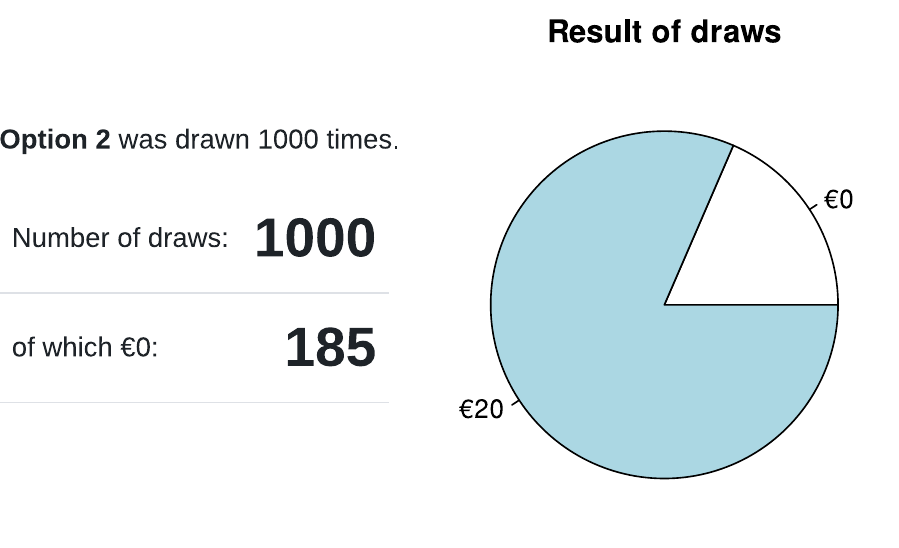}
            \caption{$k = 1000$, $n = 185$}
            \label{fig:plot4}
        \end{subfigure}
    \end{minipage}

    \caption{Examples of how the Chooser's view was presented to CAs, translated to English}
    \label{animations}
\end{figure}

CAs were able to simulate sample draws from Option 2; these draws are visualized using pie charts, similar to the approach in \cite[fig. 1]{harrison2018risk}. $n$ was drawn from the corresponding binomial distribution, given $k$ and $p$. See Figure \ref{animations} for examples. For any fixed $k$, a new draw was made about every 2 seconds. This allowed CAs to obtain a thorough understanding of the distribution of $n$ to dispel incorrect beliefs, such as in the “law of small numbers” \eg{tversky1971belief}.\footnote{As our results indicate, subjects made much use of the animations. Only 12\% of CAs never watched an animation; the median number of animations shown to a CA was $24.5$.}

Note that CAs do not know the actual value of $n$ shown to Choosers. As in the discussion on the marginal posterior in Section \ref{partialambiguity}, we decided to keep $n$ random, as we believed any other design to be difficult to implement without deceiving either CAs or Choosers. Moreover, keeping $n$ random while fixing $k$ reduces statistical noise arising from differential intervention given the many possible values of $n$. In sum, CAs only learn the distribution of $n$ for all possible $k$.

\subsubsection{Treatments}

CAs were informed that $p$ will originally be drawn from a discrete uniform distribution that includes the end points 0\% and 100\%, but that their decision can only be implemented if $p$ will randomly take the value $0.2$ (20\%). Hence, the probability of implementation was low. Each CA was presented---in random order \citep{charness2012experimental}---with the following values of $k$: $0$, $1$, $2$, $5$, $10$, $25$, $50$, $1000$ and $\infty$. $k=\infty$ represents the condition that the Chooser learns the value of $p$ exactly, as in case \textit{(i)} of Section \ref{partialknowledge}. Since all CAs participated in all of these treatments, our experiment has \textit{within} characteristics.

\subsubsection{Procedures}
\label{exp1proc}

The experiment was conducted at the \censor{Cologne Laboratory for Economic Research in Germany} in March 2023. Recruitment was done using ORSEE \citep{greiner2015subject}. All participants identified as students. Participants were not selected based on major or any demographic variable. English-language instructions are available in Appendix \ref{app.instr}. IRB approval was granted on January 30, 2023 by \censor{the WiSo Ethics Review Board at the University of Cologne (Reference 230005MG).} The experiment was preregistered at AsPredicted.\footnote{The preregistration can be viewed at \url{https://aspredicted.org/Y68_8JW}, accessed \today.}

Subjects were invited to participate in the online experiment at a date of their choosing. On that date, they were free to start the experiment at any point between 2pm and 6pm. Subjects were only allowed to intervene after passing a comprehensive set of comprehension checks. While we did not restrict the number of attempts, we offered a bonus for passing the comprehension checks on one's first attempt. A number of participants were unable to pass the comprehension checks during the time limit allocated to the page, and some participants withdrew at any point. All in all, 368 CAs and 2 Choosers participated. On average, CAs earned €5.14 and spent 22 minutes in the experiment. 301 CAs are “complete” as defined by the preregistration (they participated in all parts of the experiment).

As prescribed by the preregistration, only the data of the first 300 CAs is used in the analysis. No further exclusions are applied.

\subsection{Experiment 2}
\label{exp2}

Experiment 2 exploits cases \textit{(i)} and \textit{(iii)} of Section \ref{partialknowledge} to investigate further aspects of the relationship between paternalism and knowledge. Experiment 2 is a follow-up to experiment 1, designed after learning the results of experiment 1. As in experiment 1, CAs are allowed but not required to intervene in the decision faced by Choosers.

\subsubsection{Chooser scenarios and treatments}
\label{blocks}

\begin{table}
    \centering
    \begin{adjustbox}{max width=\textwidth}%
    \begin{tabular}{@{}ccccc@{}}
        \toprule
        Chooser & Block & Information & Type & Reference \\ \midrule
        1 & I & None & Unknown to CA & Section \ref{partialknowledge}, case \textit{(iii)} \\
        2 &  & Full & Unknown to CA & Section \ref{partialknowledge}, case \textit{(i)} \\
        3 & II & Uncertain & Unknown to CA & Section \ref{aggregate} \\
        4 &  & Decided by CA & Unknown to CA & Section \ref{infoprov} \\
        5 &  & None & Known to CA & Section \ref{millpost} \\ \bottomrule
    \end{tabular}%
    \end{adjustbox}
    \caption{Choosers in experiment 2}
    \label{choosers2}
\end{table}

In experiment 2, CAs made decisions for a total of five Choosers in two blocks (see Table \ref{choosers2}). The order of Choosers within blocks was randomized, but block II always followed block I. Block I attempted to replicate the findings of experiment 1 in a setting where Choosers did not receive information about the prior for $p$ ($p$ is not random, but indeed fixed at $0.2$).

\paragraph{Block I}

Chooser 1 obtained no information (not even a prior), as in case \textit{(iii)} of Section \ref{partialknowledge}.\footnote{Note how in Screen 9 of Section \ref{app.instr2} in the Appendix, the value of $p$ is essentially blacked out. The value of $p$ was not available through any other means, including navigating to the page's source code.} As in case \textit{(i)}, Chooser 2 received full information.

\paragraph{Block II}

CAs are told that Chooser 3 can either be uninformed or fully informed. Either of these states of nature can occur with 50\% probability, as in Section \ref{aggregate}. At the time of CA rule-making, it is yet unknown which state is the true one.

Chooser 4's degree of knowledge is determined by the CA (Section \ref{infoprov}). For this Chooser, CAs were randomly allocated to a baseline or the treatment Plus. Our treatment relates to the institutional setup of information provision for Chooser 4: In the baseline, CAs were only allowed to provide information to Choosers, as in \cite{bartling2023free}. In Plus, they were enabled to intervene in the resulting Choice in addition to providing information. Simply put, in both treatments the CA can choose between cases \textit{(i)} and \textit{(iii)} of Section \ref{partialknowledge}; in Plus they may also add an intervention in the resulting choice. On the other hand, in the baseline, it is a given that the Chooser's own choice is implemented after Chooser 4 receives the information decided upon by the CA. Both information provision and---in Plus---the intervention for the Chooser took place on the same screen.

Chooser 5 is uninformed, but there is information about his counterfactual choice (Section \ref{millpost}).
For this Chooser, CAs were randomly allocated to the Chooser's hypothetical full-information choice: what Option the Chooser preferred in the full-information counterfactual. CAs were told that Chooser 5 decided between the Options for several possible values of $p$, not knowing the true value. CAs were guaranteed that their decision could only be implemented if Chooser 5 did, in fact, prefer the given Option at $p = 0.2$. Simply put, CAs can “help” Chooser 5 get what he would obtain if he were fully informed.

\subsubsection{Procedures}
\label{exp2proc}

Experiment 2 was developed with \censor{uproot \citep{uproot}} and conducted at \censor{the Cologne Laboratory for Economic Research} with students from \censor{Cologne and Maastricht} in mid-2024. Participants were not selected based on major or any demographic variable.
Recruitment was done using ORSEE \citep{greiner2015subject}. English-language instructions are available in Appendix \ref{app.instr2}. IRB approval was granted by the Gesellschaft für experimentelle Wirtschaftsforschung e.V. on May 13, 2024 (Approval ID RNnxiot5), a German nonprofit association providing services to experimental economists.\footnote{The ethics certificate can be viewed at \url{https://gfew.de/ethik/RNnxiot5}, accessed \today.} The experiment was preregistered at AsPredicted.\footnote{The preregistration can be viewed at \url{https://aspredicted.org/625_QRC}, accessed \today.}

Subjects were invited to participate in the online experiment at a date of their choosing. On that date, they were free to start the experiment at any point between 10am and 6pm.\footnote{Students from \censor{Maastricht} were allowed to participate at any time. They were recruited by email through a lecture.} All in all, 610 CAs started the experiment. 603 CAs are “complete” as defined by the preregistration (they participated in all parts of the experiment). On average, these CAs earned €3.60 and spent 10 minutes in the experiment.

As prescribed by the preregistration, only the data of the first 600 CAs is used in the analysis. No further exclusions are applied.

\subsection{Predictions and research questions}

Given the considerations of the value of $k$ in Section \ref{partialambiguity}, we predict that, for $p = 0.2$, there will be fewer interventions as $k$ is increased.
Since---to some extent---experiment 2 is a \textit{conceptual replication} (or “many-designs replication”) of experiment 1 \citep{doi:10.1177/17456916211041116}, we predict that a similar result can be obtained for the extreme cases \textit{(i)} and \textit{(iii)} of Section \ref{partialknowledge}. The following prediction also follows directly from our formal model of Section \ref{patformal}.

\begin{prediction}\label{pred1}
    \textit{Knowledge and freedom}.
    \textit{Experiment 1}:
    There will be fewer interventions under high $k$ than under low $k$.
    \textit{Experiment 2}:
    There will be fewer interventions under full knowledge (Chooser 2) than under no knowledge (Chooser 1).
\end{prediction}

Furthermore, we predict that beliefs about behavior at $k = \infty$ (or under full knowledge) will be systematically biased (see Section \ref{fcb}):

\begin{prediction}\label{pred2}
    \textit{False consensus bias}.
    CAs will believe Option 1 to be more popular if they themselves prefer Option 1.
\end{prediction}

In research question \ref{rq1}, we assess the majoritarian and taste-based extents of interventions on the intensive margin.

\begin{resq}\label{rq1}
    \textit{Intensive margin}.
    When controlling for the CAs' own preferences, will those CAs who intervene systematically impose the more popular option full information, according to their own beliefs?
\end{resq}

Research question \ref{rq2} relates to Chooser 5.

\begin{resq}\label{rq2}
    \textit{Mill's intensive-margin postulate}.
    Is knowledge of the Chooser's hypothetical perfect-information choice able to overcome projective paternalism?
\end{resq}

Prediction \ref{pred4} is about the information provided to Chooser 4 by CAs.

\begin{prediction}\label{pred4}
    \textit{Knowledge provision}.
    CAs will provide less information to Chooser 4 when they can both inform and intervene in the Chooser's decision (as in treatment Plus), compared to when CAs can only inform without being able to intervene.
\end{prediction}

Research question \ref{rq3} relates to Chooser 3.

\begin{resq}\label{rq3}
    \textit{Aggregating degrees of knowledge}.
    Which is more predictive of the choice to intervene for Chooser 3: the choice to intervene for Chooser 1 or that for Chooser 2?
\end{resq}

\section{Results}
\label{results}

In this Section, we discuss our experiments' results. For both experiments, we follow the preregistrations exactly (except for a minor correction to the preregistration of experiment 2, see Section \ref{precorrect} in the Appendix). We reference our preregistered analyses when it comes to the evaluation of predictions and research questions. For reasons of exhibition, however, the presentation in the main text focuses on Ordinary Least Squares (OLS) regressions with HC3 standard errors used in all models (unless otherwise noted). The findings of the preregistered analyses are not contradicted.

\subsection{Chooser knowledge increases freedom}

\begin{figure}[t]
    \includegraphics[width=\textwidth]{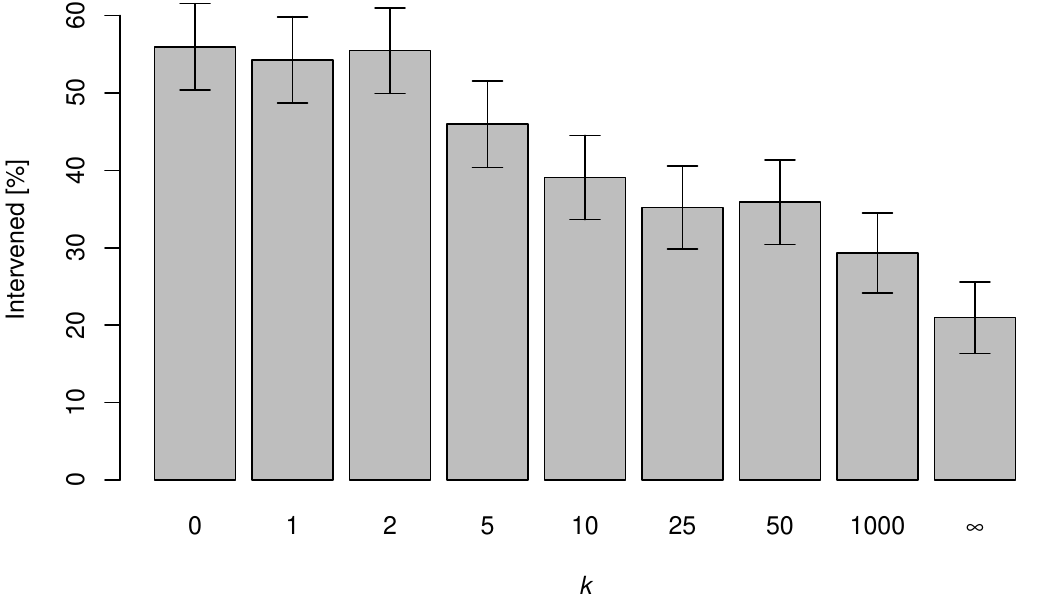}

    \caption{Percentage of interventions by $k$}
    \caption*{\footnotesize Error bars span the 95\% confidence interval (clustered by subject), calculated from the marginal effects in a logistic regression with treatment dummies, round number and demographic controls (Section \ref{intlogitapp} in the Appendix).}
    \label{intplot}
\end{figure}

How do CAs behave in experiment 1? Figure \ref{intplot} demonstrates that there is a clear trend towards fewer interventions as $k$ rises. We can also evaluate Prediction \ref{pred1} econometrically. In accordance with the preregistration, we code all values of $k$ by their rank. Table \ref{transform} in the Appendix describes the transformation. As suggested by the preregistered analysis in Section \ref{intlogitapp} in the Appendix as well as robustness checks (Table \ref{intlogit} in the Appendix), $k$ causes a highly statistically and economically significant reduction in the amount of intervention (two-tailed $z$-test with standard errors clustered on the subject level, $z = -9.95$, $p < 0.001$).

It is also possible to evaluate the null hypothesis without relying on the transformation of $k$ by using \citeauthor{pagel}'s $L$ test (\citeyear{pagel}). This non-parametric test tests for a trend given repeated measurement. The null hypothesis of no trend is again rejected ($L = 63922.5$, $\chi_1^2 = 94.8$, $p < 0.001$).

\begin{table}
    \begin{center}
        \sisetup{parse-numbers=false, table-text-alignment=center}
        \resizebox{\textwidth}{!}{%
        \begin{threeparttable}
                \begin{tabular}{l S[table-format=4.6] S[table-format=4.6] S[table-format=3.6] S[table-format=3.6] S[table-format=3.6]}
                    \toprule
                    & \multicolumn{2}{c}{Choosers~1 and 2} & \multicolumn{1}{c}{Chooser~1 or 2} & \multicolumn{2}{c}{Chooser 1} \\
                    \cmidrule(lr){2-3} \cmidrule(lr){4-4} \cmidrule(lr){5-6}
                    & {Model 1} & {Model 2} & {Model 3} & {Model 4} & {Model 5} \\
                    \midrule
                    Intercept             & 0.220^{***}  & 0.289^{***}  & 0.199^{**}   & 0.488^{***}  & 0.657^{***}  \\
                    & (0.017)      & (0.062)      & (0.069)      & (0.077)      & (0.181)      \\
                    Full Knowledge        & -0.143^{***} & -0.143^{***} & -0.103^{***} &              &              \\
                    & (0.020)      & (0.020)      & (0.024)      &              &              \\
                    $\pi$ (standardized) &              & -0.010       & -0.019       & 0.107^{*}    & 0.163^{*}    \\
                    &              & (0.017)      & (0.019)      & (0.044)      & (0.065)      \\
                    $\pi > 0.5$     &              & -0.012       & 0.014        &              & -0.198       \\
                    &              & (0.057)      & (0.066)      &              & (0.193)      \\
                    Block I order         &              & 0.034        &              &              &              \\
                    &              & (0.020)      &              &              &              \\
                    CA prefers~1          &              & -0.095^{**}  & -0.080^{*}   & 0.263^{**}   & 0.261^{**}   \\
                    &              & (0.030)      & (0.038)      & (0.096)      & (0.096)      \\
                    \midrule
                    Outcome               & {Intervened} & {Intervened} & {Intervened} & {Imposed 1}  & {Imposed 1}  \\
                    Subset                & {---}        & {---}        & {Round 1}    & {Intervened} & {Intervened} \\
                    Standard errors       & {Clustered}  & {Clustered}  & {HC3}        & {HC3}        & {HC3}        \\
                    R$^2$                 & 0.041        & 0.060        & 0.048        & 0.164        & 0.172        \\
                    Adj. R$^2$            & 0.040        & 0.056        & 0.041        & 0.151        & 0.153        \\
                    Num. obs.             &{1200}        &{1200}        &{600}         &{132}         &{132}         \\
                    \bottomrule
                \end{tabular}
                \begin{tablenotes}[flushleft]
                \scriptsize{\item $^{***}p<0.001$; $^{**}p<0.01$; $^{*}p<0.05$}
                \end{tablenotes}
        \end{threeparttable}}
        \caption{Regressions of extensive and intensive margin outcomes on Chooser knowledge, CA beliefs and preferences}
        \label{t2_1}
    \end{center}
\end{table}

Experiment 2 is able to replicate this pattern. Models 1--2 in Table \ref{t2_1} regress CA interventions on knowledge. Noteworthily, the baseline rate of intervention is much lower than in experiment 1. Recall from Section \ref{blocks} that the order of Choosers 1 and 2 was randomized between subjects. Thus, model 3 restricts the analysis to the first Chooser that was seen. Once again, the coefficient on Full Knowledge is highly significant. This implies that knowledge enhances autonomy both in within (experiment 1, experiment 2: models 1--2 and Section \ref{e2a2} in the Appendix) and between (experiment 2: model 3) analyses. From this evidence, we conclude:

\begin{result}
    The greater knowledge in experiments 1 and 2, the greater the probability of autonomy being granted. Prediction \ref{pred1} is confirmed.
\end{result}

\subsection{Testing Mill's intensive-margin postulate}

What influences the intensive margin of interventions? That is, given that an intervention is to take place, what determines the choice between Options 1 and 2? As discussed in Section \ref{millintensive}, Mill offered a solution that stands in stark contrast with the literature on projective paternalism \citep{ambuehl2021motivates}: the intervention ought to be based on an assessment of counterfactual full-information choices. We test this idea here.

\subsubsection{Beliefs and the intensive margin}
\label{knfbeliefs}

In both experiments, we asked CAs to estimate the number of people opting for one of the Options under full information.\footnote{In experiment 1, we elicited unincentivized beliefs about how many people out of 1,000 would prefer Option 2. In experiment 2, we used data from experiment 1 to elicit beliefs about the preferences of 300 subjects for Option 1. These beliefs were incentivized using the binarized scoring rule \citep{hossain2013binarized}.}
In experiment 2, we also elicited unincentivized beliefs about behavior under ambiguity (that is, case \textit{(iii)} of Section \ref{partialknowledge}).
For the models in Table \ref{tabbeliefs} in the Appendix, we put these data on a common scale (proportion of decision-makers thought to prefer Option 1). As in Section \ref{patformal}, we call this outcome $\pi$. In the models dealing with beliefs about fully informed behavior under risk, our results confirm classic findings on false consensus bias \eg{ross1977false}. CAs believe that their subjectively preferred Option is more popular.

\begin{result}
    Prediction \ref{pred2} is confirmed (although beliefs about behavior under ambiguity cannot be explained by CAs' own preferences).
\end{result}

Evidently, the reported beliefs differ strongly between experiments, with beliefs in experiment 1 implausibly small.\footnote{In experiment 1, beliefs were elicited at various values of $p$ using sliders (see Screen 4 of Section \ref{app.instr} in the Appendix).} Moreover, CAs predict ambiguity aversion (model 5).
Note how we did not elicit data on the direction of mistakes made under ambiguity (Section \ref{knfmistakes}). However, it is easy to see that if $\pi' > \pi$, $\epsilon_y > \epsilon_x$: to a CA who prefers Option 1, a belief in ambiguity aversion will augment mistakes in the subjectively preferred direction. To a CA who prefers Option 2, the same belief goes into the non-preferred direction. We exploit this fact below.

As outlined in Section \ref{fcb}, the presence of false consensus bias leads to an identification problem in motives. Under the majoritarian standard, CAs may wish to implement the Option they believe to be more popular, but if that belief is biased, beliefs \emph{and} CA preferences must be accounted for in the analyses. As can be gleaned from the standard errors in Table \ref{t1_2} in the Appendix, experiment 1 was not sufficiently powered to reliably identify the influences of ${\pi > 0.5}$ and $\pi$ on intensive-margin interventions.

In experiment 2, beliefs have a significant, but small association with this outcome. Model 4 in Table \ref{t2_1} shows that a one standard deviation increase in $\pi$ is correlated with a 10.7 p.p. increase in the probability of imposing Option 1. Model 5 demonstrates that there is no association with ${\pi > 0.5}$ (i.e., beliefs about majority behavior). As is clear from models 4--5, the CA's own preference also has a significant correlation with the intensive margin \citep{ambuehl2021motivates}. We estimate that a CA who prefers Option 1 is about 26 p.p. more likely to impose Option 1. Model 4 implies that beliefs that are $(0.263)/(0.107) \approx 2.458$ standard deviations removed from the mean are equivalent to the CA's own preference in their effect on the intensive margin. Section \ref{t2_1asec} in the Appendix assesses the relative importance of beliefs and CA preferences in imposing Option 1 formally. We conclude from these investigations that beliefs are far weaker predictors of intensive-margin interventions than CAs' preferences.

\begin{result}
    Experiment 2 shows that beliefs about majority behavior do not correlate on the intensive margin. Beliefs generally matter significantly, but much weaker as predictors than CAs' preferences.
\end{result}

\subsubsection{Providing information about Chooser type}


Even if we allow for biased beliefs---as in the previous Section---the CA's own preference appears to be an important predictor for the intensive margin.

\begin{table}
    \begin{center}
        \sisetup{parse-numbers=false, table-text-alignment=center}
        \begin{threeparttable}
            \resizebox{\textwidth}{!}{%
                \begin{tabular}{l S[table-format=3.6] S[table-format=3.6] S[table-format=3.6] S[table-format=3.5] S[table-format=3.5]}
                    \toprule
                    & \multicolumn{4}{c}{Chooser 5} & \multicolumn{1}{c}{Chooser 1} \\
                    \cmidrule(lr){2-5} \cmidrule(lr){6-6}
                    & {Model 1} & {Model 2} & {Model 3} & {Model 4} & {Model 5} \\
                    \midrule
                    Intercept             & 0.368^{***}  & 0.360        & 0.421^{***}       & 0.589^{**}        & 0.531^{**}   \\
                    & (0.084)      & (0.185)      & (0.084)           & (0.192)           & (0.189)      \\
                    $\pi$ (standardized) &              & 0.045        &                   & 0.049             & 0.001        \\
                    &              & (0.054)      &                   & (0.055)           & (0.070)      \\
                    $\pi > 0.5$     &              & 0.040        &                   & -0.185            & 0.049        \\
                    &              & (0.184)      &                   & (0.186)           & (0.198)      \\
                    CA prefers~1          & 0.127        & 0.085        & 0.141             & 0.134             & 0.203^{*}    \\
                    & (0.085)      & (0.089)      & (0.085)           & (0.090)           & (0.101)      \\
                    Chooser prefers~1     & 0.254^{***}  & 0.263^{***}  & 0.192^{**}        & 0.200^{**}        &              \\
                    & (0.070)      & (0.070)      & (0.070)           & (0.071)           &              \\
                    \midrule
                    Outcome               & {Imposed 1}  & {Imposed 1}  & {Matched Chooser} & {Matched Chooser} & {Matched CA} \\
                    Subset                & {Intervened} & {Intervened} & {Intervened}      & {Intervened}      & {Intervened} \\
                    R$^2$                 & 0.070        & 0.081        & 0.047             & 0.054             & 0.051        \\
                    Adj. R$^2$            & 0.060        & 0.060        & 0.037             & 0.033             & 0.029        \\
                    Num. obs.             &{188}         &{188}         &{188}              &{188}              &{132}         \\
                    \bottomrule
            \end{tabular}}
            \begin{tablenotes}[flushleft]
                \scriptsize{\item $^{***}p<0.001$; $^{**}p<0.01$; $^{*}p<0.05$}
            \end{tablenotes}
        \end{threeparttable}
        \caption{Intensive margin regressions when Chooser type is known}
        \label{t2_1b}
    \end{center}
\end{table}

When deciding for Chooser 5 in experiment 2, CAs were informed that the Chooser they were deciding for \emph{certainly} preferred either Option 1 or Option 2 (the Chooser's “type”). Models 1 and 2 in Table \ref{t2_1b} reveal that providing information about the Chooser's type does significantly influence the intensive margin selected by the CA. However, the preregistered Wald test in Section \ref{e2a4} in the Appendix on the relative influence of CA and Chooser preferences reveals no significant difference between the coefficients on CA and Chooser preferences in a logistic regression ($p = 0.21$). It is important to note that while CA preference has no significant effect on the intensive margin for Chooser 5, this is not evidence of absence. The standard error on CA preference is merely large enough so that the hypothesis that \emph{both} influences matter equally cannot be rejected.

\begin{result}
    The Chooser's type significantly influences the intensive margin, but this influence is not significantly different from the influence of the CA's preference (research question \ref{rq2}).
\end{result}

To investigate this issue further, we estimated models 3--4 in Table \ref{t2_1b}. Here, the outcome variable is whether the CA actually implemented the Chooser's counterfactual choice. Note that about a majority of CAs “match” the Chooser. We detect a significant influence of the \emph{Chooser's} preference. That is, if the Chooser prefers Option 1, that Option is about 20 p.p. likely to be implemented for the Chooser---ceteris paribus---than if he prefers Option 2, even when controlling for the CA's own preference.\footnote{We can verify this finding using the following highly conservative robustness check: among all CAs who intervened for Chooser 5, we can condition on their own preference and that of the Chooser. 54 of all intervening CAs preferred Option 1 and were presented with a Chooser preferring Option 1. In that case, Option 1 was implemented 43 times (79.6\%). 23 of all intervening CAs preferred Option 2 and intervened for a Chooser preferring Option 2. 12 of these CAs imposed Option 2 (52.2\%). The difference is significant (two-sided test of equal proportions, $\chi^2 = 4.69$, $p = 0.03$).} The correlational analysis in model 5 reveals that the same pattern emerges for Chooser 1.

This result stands in contrast to findings concerning decision-making for others under risk \citep{polman2020decision}, where decisions for others are more risky. Similarly, a literature has characterized the influence of social context on risk-taking \eg{schwerter2024social,BoltonOckenfelsStauf2015} or the role of social preferences \citep{BoltonOckenfels2010}.
The present finding highlights the nuances between policymakers' own preferences and the intensive margin. Some choices may be viewed as more respectable or objectively correct beyond the CA's own preference and thus more likely to be implemented. Future work can seek to disentangle CA preferences from such cosmic ideals or “bliss points.”
One possible explanation revolves around regret and guilt: any imposition in some sense makes the CA the Chooser and thus responsible for the outcome. Option 2 did yield €0 with probability 0.2, a significant possibility. Moreover, Option 2's expected value is only €1 higher than the safe amount, making it relatively unattractive. In that case, a preference for the safe Option 1 may result from an “interpersonal” kind of regret. \cite{corbett2021interpersonal} discusses pro-social risk-taking and its relation to regret and guilt further. Our own results do not allow us to make a statement about any underlying cognitive mechanisms.

\begin{result}\label{asymmetricproj}
    The riskless Option 1 is significantly more likely than Option 2 to be implemented. CAs do not simply implement the full-information counterfactual choice of the Chooser or their own preference.
\end{result}

\subsection{Connecting formal theory with data}
\label{knftheorydata}

Our theory of Section \ref{patformal} allows us to estimate $\phi$ using only three observable parameters: beliefs about fully informed preferences ($\pi$), CAs' own preferences and intensive margin interventions.
Recall that, on the intensive margin, option $x$ will be implemented if $W_{\theta}(x) > W_{\theta}(y)$ (e.g., Equations \ref{camixed1}--\ref{camixed2} for $\theta = x$). We can introduce unobserved and independent error terms on both sides of this inequality to yield a “random utility” (or Fechner-type) model. Let $\xi_x \sim N(0, \sigma_{\xi_x}^2)$, $\xi_y \sim N(0, \sigma_{\xi_y}^2)$. In that case, option $x$ will be implemented on the intensive margin if
\def\LRA{\Leftrightarrow\mkern40mu}%
\begin{alignat*}{2}
     && W_{\theta}(x) + \xi_x &> W_{\theta}(y) + \xi_y,\\
\LRA && \underbrace{\xi_y - \xi_x}_{\sim N(0, \sigma)} &< W_{\theta}(x) - W_{\theta}(y).
\end{alignat*}
Call $\Phi$ the standard-normal cumulative distribution function and let $I^1_i$ be an indicator of whether CA $i$ implemented Option 1. Using information on each CA's type, $\theta_i \in \{1, 2\}$, the following log-likelihood function can be maximized with respect to $\phi, \sigma$ to yield consistent estimates of these parameters:
\begin{align}\label{knfllik}
    &\sum_i \bigg[ I^1_i \log \Phi \left( \frac{W_{\theta_i}(1) - W_{\theta_i}(2)}{\sigma} \right) + \nonumber \\
    &\quad (1-I^1_i) \log \left( 1 - \Phi \left( \frac{W_{\theta_i}(1) - W_{\theta_i}(2)}{\sigma} \right) \right) \bigg].
\end{align}
When we maximize Equation \ref{knfllik} for intensive-margin decisions vis-á-vis Chooser 1, we find $\widehat{\phi} = 0.294$ (95\% confidence interval: $[0.081, 0.508]$) and $\widehat{\sigma} = 0.809$ (95\% confidence interval: $[0.504, 1.114]$). All confidence intervals in this Section are calculated from the Fisher information matrix.

The same can be done for Chooser 5, where CAs had certain knowledge about the Chooser's type. Here, we find $\widehat{\phi} = 0.430$ (95\% confidence interval: $[0.221, 0.640]$) and $\widehat{\sigma} = 1.767$ (95\% confidence interval: $[0.840, 2.694]$). Similar results for $\phi$ can be obtained for Chooser 3.

In sum, $\phi$ is reasonably similar between Choosers 1, 3 and 5 and, in each case, significantly different from 0. As we describe in Appendix \ref{knfmle}, these binary-choice models fit the data about as well as model 5 in Table \ref{t2_1} and model 2 in Table \ref{t2_1b}. However, they do not work well for Chooser 2 (few CAs intervened) and Chooser 4 (few CAs could intervene and few did).

\begin{result}
    The formal model of Section \ref{patformal} performs well in describing intensive-margin behavior: $\phi$ is estimated at about one-third for Choosers 1, 3 and 5. $H_0: \phi = 0$ is rejected.
\end{result}

\subsection{Knowledge uncertainty}


\newcommand{\nmci}[3]{%
    \begin{tabular}{@{}c@{}}{\small (#1)} \\ \textbf{#2} \\ #3\end{tabular}%
}
\begin{table}
    \centering
    \begin{threeparttable}
        \begin{tabular}{l c c}
            \toprule
            & {Int. for Chooser 1} & {Not Int. for Chooser 1} \\
            \midrule
            {Int. for Chooser 2} & \nmci{12}{0.833}{$[0.516, 0.979]$} & \nmci{34}{0.412}{$[0.246, 0.593]$}\\
            \\
            {Not Int. for Chooser 2} & \nmci{120}{0.400}{$[0.312, 0.493]$} & \nmci{434}{0.267}{$[0.226, 0.312]$}\\
            \bottomrule
        \end{tabular}

        \caption{Intervention rates for Chooser 3 conditional on behavior toward Choosers 1 and 2}
        \label{t2_4}

        \begin{tablenotes}
            \small
        \item Chooser 1 is uninformed, Chooser 2 is fully informed (Section \ref{blocks}). The table shows, small and in parentheses, a cross-tab of occurrences; in bold, the rate of intervention for Chooser 3 conditional on behavior for Choosers 1 and 2; the rate's $95\%$ confidence interval.
        \end{tablenotes}
    \end{threeparttable}
\end{table}

\begin{table}
    \begin{center}
        \sisetup{parse-numbers=false, table-text-alignment=center}
        \begin{threeparttable}
            \resizebox{\textwidth}{!}{%
                \begin{tabular}{l S[table-format=3.6] S[table-format=3.6] S[table-format=3.6] S[table-format=3.5]}
                    \toprule
                    & \multicolumn{4}{c}{Chooser 3} \\
                    \cmidrule(lr){2-5}
                    & {Model 1} & {Model 2} & {Model 3} & {Model 4} \\
                    \midrule
                    Intercept             & 0.313^{***}  & 0.262^{***}  & 0.221^{*}    & 0.524^{**}   \\
                    & (0.019)      & (0.021)      & (0.103)      & (0.177)      \\
                    Int. for Chooser~1    &              & 0.158^{***}  & 0.168^{***}  &              \\
                    &              & (0.047)      & (0.049)      &              \\
                    Int. for Chooser~2    &              & 0.219^{**}   & 0.210^{**}   &              \\
                    &              & (0.075)      & (0.076)      &              \\
                    $\pi$ (standardized) &              &              & 0.000        & 0.043        \\
                    &              &              & (0.030)      & (0.047)      \\
                    $\pi > 0.5$     &              &              & 0.081        & -0.049       \\
                    &              &              & (0.098)      & (0.168)      \\
                    Block I order         &              &              & -0.037       &              \\
                    &              &              & (0.038)      &              \\
                    CA prefers~1          &              &              & -0.018       & 0.288^{**}   \\
                    &              &              & (0.052)      & (0.093)      \\
                    \midrule
                    Outcome               & {Intervened} & {Intervened} & {Intervened} & {Imposed 1}  \\
                    Subset                & {---}        & {---}        & {---}        & {Intervened} \\
                    R$^2$                 & 0.000        & 0.037        & 0.041        & 0.091        \\
                    Adj. R$^2$            & 0.000        & 0.033        & 0.031        & 0.076        \\
                    Num. obs.             &{600}         &{600}         &{600}         &{188}         \\
                    \bottomrule
                \end{tabular}}
                \begin{tablenotes}[flushleft]
                \scriptsize{\item $^{***}p<0.001$; $^{**}p<0.01$; $^{*}p<0.05$}
                \end{tablenotes}
        \end{threeparttable}
        \caption{Regressions explaining behavior toward a Chooser in uncertain state of knowledge}
        \label{t2_3}
    \end{center}
\end{table}

CAs' decision-making for Chooser 3 involved a scenario in which the Chooser could be in either of two states: fully informed or not informed. Both states were equally likely. Table \ref{t2_4} gives descriptive statistics, grouped by the decisions regarding Choosers 1 and 2. The data follow a plausible pattern: those who intervened more before will intervene more for Chooser 3. However, comparing rates of intervention (model 1 in Table \ref{t2_3} gives the average) with those in models 1--3 of Table \ref{t2_1} reveals that rates of intervention for Chooser 3 vastly exceed even those for Chooser 1.\footnote{We can use the binary indicators of intervention for Choosers 1 and 3 to test whether the within-subject difference is significantly different from zero: two-sided paired $t$-test with unequal variances, $t = -3.52$, $p < 0.001$.} We can only speculate why this is the case.

\begin{result}
    Chooser 3 attains vastly higher intervention rates than Choosers 1 and 2.
\end{result}

Moreover, models 2--3 in Table \ref{t2_3} and the preregistered analysis of Section \ref{e2a3} in the Appendix reveal that the decision to intervene for Chooser 3 is about equally well explained by each of the decisions to intervene for Choosers 1 and 2. Both of these prior decisions are highly significant in predicting behavior vis-á-vis Chooser 3. Model 4 once more highlights the predictive power of CAs' own preference in shaping the intensive margin.

\begin{result}
    The decisions for Choosers 1 and 2 are about equally predictive for behavior towards Chooser 3 (research question \ref{rq3}).
\end{result}

\subsection{Providing information}

\begin{table}
    \begin{center}
        \sisetup{parse-numbers=false, table-text-alignment=center}
        \begin{threeparttable}
            \begin{tabular}{l S[table-format=3.6] S[table-format=3.6] S[table-format=3.6]}
                \toprule
                & \multicolumn{3}{c}{Chooser 4} \\
                \cmidrule(lr){2-4}
                & {Model 1} & {Model 2} & {Model 3} \\
                \midrule
                Intercept           & 0.803^{***}     & 0.974^{***}     & 0.942^{***}     \\
                & (0.023)         & (0.105)         & (0.154)         \\
                Plus                & 0.011           & 0.012           &                 \\
                & (0.032)         & (0.032)         &                 \\
                $\pi$               &                 & -0.056          & -0.056          \\
                &                 & (0.114)         & (0.169)         \\
                $\pi'$              &                 & -0.085          & -0.007          \\
                &                 & (0.121)         & (0.195)         \\
                $\pi > \pi'$        &                 & -0.087          & -0.062          \\
                &                 & (0.079)         & (0.108)         \\
                CA prefers~1        &                 & 0.048           & 0.021           \\
                &                 & (0.066)         & (0.076)         \\
                Mistakes benefit CA &                 & -0.134^{*}      & -0.094          \\
                &                 & (0.062)         & (0.071)         \\
                Intervened          &                 &                 & -0.297^{**}     \\
                &                 &                 & (0.103)         \\
                \midrule
                Outcome             & {Info provided} & {Info provided} & {Info provided} \\
                Subset              & {---}           & {---}           & {Plus}          \\
                R$^2$               & 0.000           & 0.017           & 0.064           \\
                Adj. R$^2$          & -0.001          & 0.007           & 0.044           \\
                Num. obs.           &{600}            &{600}            &{290}            \\
                \bottomrule
            \end{tabular}
            \begin{tablenotes}[flushleft]
                \scriptsize{\item $^{***}p<0.001$; $^{**}p<0.01$; $^{*}p<0.05$}
            \end{tablenotes}
        \end{threeparttable}
        \caption{Regressions explaining CAs' information provision}
        \label{t2_2}
    \end{center}
\end{table}


CAs were enabled to communicate to Chooser 4 the value of $p$ ($0.2$) in the choice between Options 1 and 2.\footnote{In our design, CAs had to deliberately choose whether to reveal $p$ or not, as in \cite{bartling2023free}. We, too, made sure that CAs' involvement in providing information is not revealed to the Chooser.} In addition, some CAs---those in the treatment Plus---were able to intervene as well.\footnote{In this Section, we use the word “intervene” only to refer to an intervention in the choice between Options 1 and 2, although some authors view information provision as an intervention \eg{bartling2023free,mabsout2022john}.} Sections \ref{knfmistakes}, \ref{infoprov} and \ref{blocks} motivated this design choice. Simply put, real-life policymakers are not restricted from using multiple policy tools simultaneously \eg{undo} and they can strategically use information provision to achieve their ends.

We can test whether information is actually strategically provided. As stated in Section \ref{infoprov}, if it is a given that the Chooser's decision will be implemented, the CA's calculus is fundamentally different then when she can combine both intervention and information provision. Model 1 in Table \ref{t2_2} and the preregistered analysis in Section \ref{e2a1} in the Appendix demonstrate that not only do CAs not exploit strategic information provision, but CAs in the Plus treatment slightly exceed the degree of information provision observed in the baseline. This difference is not significant, but standard errors are very small. We thus replicate the finding by \cite{bartling2023free} that a vast majority of CAs provide information to Chooser 4---even in the treatment.

\begin{result}
    CAs do not strategically employ information provision (prediction \ref{pred4}).
\end{result}

This result suggests that independence of irrelevant alternatives is indeed satisfied (Section \ref{knfmistakes}).

\subsubsection{Who doesn't provide information?}

This Section reports additional, exploratory analyses to further investigate the relationship between intervention and information provision.
As we noted in Section \ref{knfmistakes}, not all Chooser mistakes are created equal. It can be shown that if $\pi' > \pi$---as in the case of ambiguity aversion (Section \ref{knfbeliefs})---$\epsilon_y > \epsilon_x$. To a CA who herself prefers Option 1, that is actually a good mistake. Similarly, a CA with a preference for Option 2 would prefer $\pi' < \pi$, implying $\epsilon_y < \epsilon_x$. In other words: with self-interest, CAs benefit from Chooser mistakes that go into CAs' subjectively preferred direction.

Our experiment did not elicit data on $\epsilon_x$ and $\epsilon_y$. However, under the conditions in Corollary \ref{knfth3}, non-provision of information can be optimal for sufficiently high $\phi$ and CA-preferring Chooser mistakes. While we do not have individual-level estimates of $\phi$, we can provide evidence that this consideration is very real. We define a new variable,
\begin{align*}
    \text{Mistakes benefit CA} &= \left(\text{CA prefers Option 1} \wedge \pi' > \pi \right) \, \vee\\
    &\left(\text{CA prefers Option 2} \wedge \pi' < \pi \right),
\end{align*}
and add it to model 1 of Table \ref{t2_2}. Moreover, we add the CA's own preference, raw beliefs and a dummy about predicted ambiguity aversion. Model 2 demonstrates that if Chooser ignorance benefits CAs, the latter are about 13.4 p.p. less likely to provide information.\footnote{A two-sided test of proportions confirms this pattern, $\chi^2 = 7.31$, $p = 0.007$. However, to ensure that this difference is not driven by secular differences in the distribution of beliefs, it is important to adjust for these confounding variables using a regression model. A test on the coefficient of “Mistakes benefit CA” in the logistic regression equivalent to model 2 in Table \ref{t2_2} reveals $p = 0.011$.} However, the effect is likely too small to be detectable among CAs in the treatment group (model 3).

\begin{result}
    CAs are slightly less likely to provide information if Chooser mistakes increase the take-up of their subjectively preferred Option.
\end{result}

\begin{figure}
    \includegraphics[width=\textwidth]{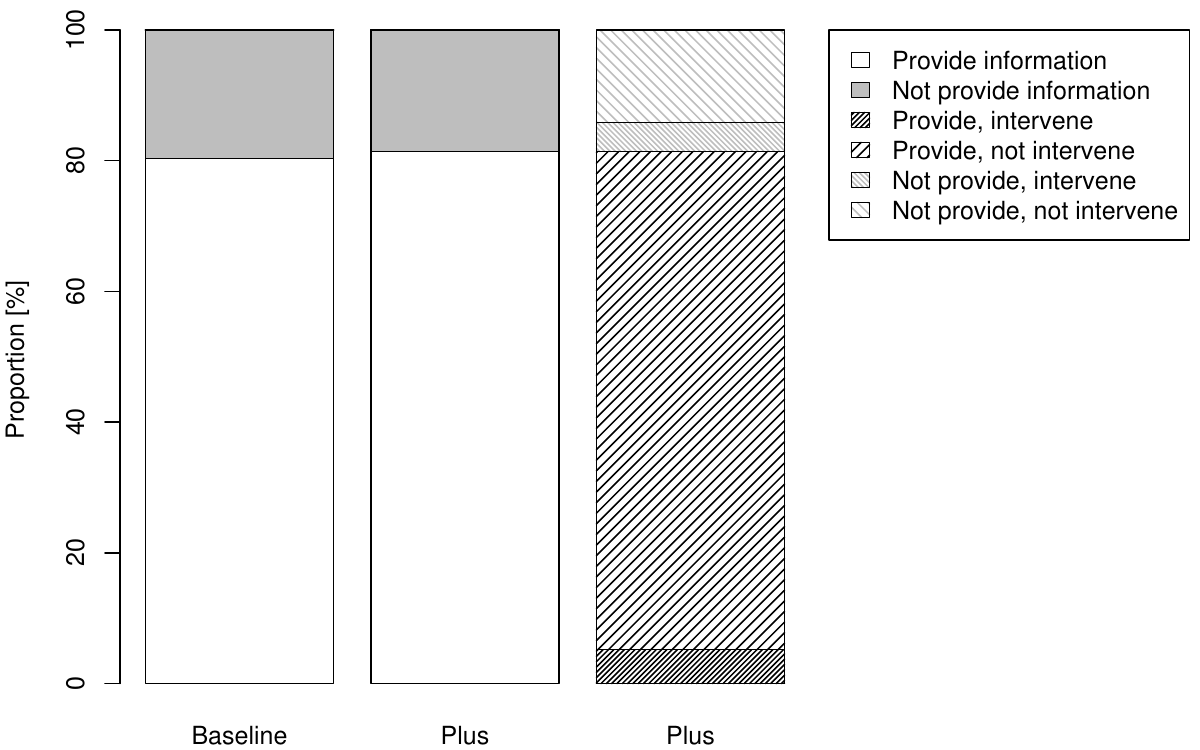}

    \caption{CA types in the population}
    \label{provplot}
\end{figure}

Model 3 in Table \ref{t2_2} restricts the analysis to CAs in the Plus treatment and regresses information provision on whether an intervention in Chooser 4's decision between Options 1 and 2 actually occurred. Although only 28 out of 290 CAs in the Plus treatment intervened, the difference in information provision between interveners and non-interveners is highly statistically and economically significant.

\begin{result}
    CAs that actually intervene are much less likely to provide information to Chooser 4.
\end{result}

Model 8 in Table \ref{t2_5} in the Appendix shows that those who intervened for Chooser 4 are especially likely to have intervened for other Choosers, too (see the discussion in Section \ref{complementarities} in the Appendix). This hints at a hidden type of CA that is characterized by not providing information and intervening if possible. Out of the 54 CAs in Plus that did not provide information, 13 (24.1\%) intervened. Out of the 236 CAs in Plus that did provide information, 15 (6.4\%) intervened.\footnote{This difference is statistically different: two-sided test of equal proportions, $\chi^2 = 13.8$, $p < 0.001$.} Recalling Model 1 in Table \ref{t2_1}, these values are almost identical to rates of intervention for Choosers 1 and 2.

The treatment thus helps us to uncover a minority of CAs that creates a lack of knowledge on the side of Choosers to intervene (Figure \ref{provplot}). This group of CAs makes up $(1-\left[0.803+0.011\right])\cdot(0.241-0.064) \approx 3.3\%$ of the sample. While small, this proportion is highly statistically significant.\footnote{Bootstrapping with 250,000 replicates, we find a 99\% confidence interval of $[0.005, 0.0666]$, $p < 0.002$.} This type of CA is not revealed by the baseline.

\section{Conclusion}\label{sec.conclusion}

This work explored the role of knowledge in paternalism. We found across two experiments that more knowledge on the side of Choosers causes a vast increase in the autonomy they are granted by impartial Choice Architects (CAs). Information helps Choosers make the right decision and CAs overwhelmingly respect that. On the other hand, a \emph{lack} of knowledge is taken by CAs as a right to intervene and prevent incorrect inference. Most CAs do not wish to override the Chooser's choice. They prefer to provide information, even when they would be able to obscure their intervention through the non-provision of knowledge. However, there is a minority of CAs that strategically abstains from providing information.

Paternalistic action is highly nuanced and context-dependent. If CAs do not know the Chooser's type, they rely on a proxy---their own preference---to select the intensive margin. If provided with the type, they use this information in conjunction with their own preference to arrive at the intensive margin. In both of these cases, a riskless Option was much more likely to be implemented. This hints at the existence of ideals for intervention that may be correlated with, but conceptually distinct from, CAs' own preferences. This is a qualification to our model and recent findings on “projective paternalism” \citep{ambuehl2021motivates}.

Policymakers' and decision-makers' beliefs, knowledge and preferences matter profoundly for regulation. Mill's arguments have stood the test of time. The idea that an intervention's intensive margin must reflect a full-information counterfactual underscores a central tenet of many classically liberal views of governance: interventions can be justified based on a lack of knowledge on the extensive margin, but their precise embodiment on the intensive margin ought to be value-free. Mill's exposition does not leave much room for taste-based intervention. Yet, as we demonstrate experimentally, policymakers distinctly mix in their personal vision of what's right even when informed about what the Chooser would have done if he had full information. As polities consider the implementation of new paternalistic policies, understanding the subjective nature of regulation is an essential condition for creating legitimate institutions and laws.

\bibliography{bib}

\begin{thebibliography}{}

\bibitem [\protect \citeauthoryear {%
Ackfeld%
\ \BBA {} Ockenfels%
}{%
Ackfeld%
\ \BBA {} Ockenfels%
}{%
{\protect \APACyear {2021}}%
}]{%
ackfeld2021people}
\APACinsertmetastar {%
ackfeld2021people}%
\begin{APACrefauthors}%
Ackfeld, V.%
\BCBT {}\ \BBA {} Ockenfels, A.%
\end{APACrefauthors}%
\unskip\
\newblock
\APACrefYearMonthDay{2021}{}{}.
\newblock
{\BBOQ}\APACrefatitle {{Do people intervene to make others behave
  prosocially?}} {{Do people intervene to make others behave
  prosocially?}}{\BBCQ}
\newblock
\APACjournalVolNumPages{Games and Economic Behavior}{128}{}{58--72}.
\PrintBackRefs{\CurrentBib}

\bibitem [\protect \citeauthoryear {%
Ambuehl%
\ \BBA {} Bernheim%
}{%
Ambuehl%
\ \BBA {} Bernheim%
}{%
{\protect \APACyear {2024}}%
}]{%
ambuehl2024interpreting}
\APACinsertmetastar {%
ambuehl2024interpreting}%
\begin{APACrefauthors}%
Ambuehl, S.%
\BCBT {}\ \BBA {} Bernheim, B\BPBI D.%
\end{APACrefauthors}%
\unskip\
\newblock
\APACrefYearMonthDay{2024}{}{}.
\newblock
\APACrefbtitle {{Interpreting the will of the people: a positive analysis of
  ordinal preference aggregation}.} {{Interpreting the will of the people: a
  positive analysis of ordinal preference aggregation}.}
\PrintBackRefs{\CurrentBib}

\bibitem [\protect \citeauthoryear {%
Ambuehl%
, Bernheim%
\BCBL {}\ \BBA {} Ockenfels%
}{%
Ambuehl%
\ \protect \BOthers {.}}{%
{\protect \APACyear {2021}}%
}]{%
ambuehl2021motivates}
\APACinsertmetastar {%
ambuehl2021motivates}%
\begin{APACrefauthors}%
Ambuehl, S.%
, Bernheim, B\BPBI D.%
\BCBL {}\ \BBA {} Ockenfels, A.%
\end{APACrefauthors}%
\unskip\
\newblock
\APACrefYearMonthDay{2021}{}{}.
\newblock
{\BBOQ}\APACrefatitle {{What motivates paternalism? An experimental study}}
  {{What motivates paternalism? An experimental study}}.{\BBCQ}
\newblock
\APACjournalVolNumPages{{A}merican {E}conomic {R}eview}{111}{3}{787--830}.
\PrintBackRefs{\CurrentBib}

\bibitem [\protect \citeauthoryear {%
Ambuehl%
, Niederle%
\BCBL {}\ \BBA {} Roth%
}{%
Ambuehl%
\ \protect \BOthers {.}}{%
{\protect \APACyear {2015}}%
}]{%
ambuehl2015more}
\APACinsertmetastar {%
ambuehl2015more}%
\begin{APACrefauthors}%
Ambuehl, S.%
, Niederle, M.%
\BCBL {}\ \BBA {} Roth, A\BPBI E.%
\end{APACrefauthors}%
\unskip\
\newblock
\APACrefYearMonthDay{2015}{}{}.
\newblock
{\BBOQ}\APACrefatitle {{More money, more problems? Can high pay be coercive and
  repugnant?}} {{More money, more problems? Can high pay be coercive and
  repugnant?}}{\BBCQ}
\newblock
\APACjournalVolNumPages{American Economic Review}{105}{5}{357--360}.
\PrintBackRefs{\CurrentBib}

\bibitem [\protect \citeauthoryear {%
Arneson%
}{%
Arneson%
}{%
{\protect \APACyear {1980}}%
}]{%
arneson1980mill}
\APACinsertmetastar {%
arneson1980mill}%
\begin{APACrefauthors}%
Arneson, R\BPBI J.%
\end{APACrefauthors}%
\unskip\
\newblock
\APACrefYearMonthDay{1980}{}{}.
\newblock
{\BBOQ}\APACrefatitle {{Mill versus paternalism}} {{Mill versus
  paternalism}}.{\BBCQ}
\newblock
\APACjournalVolNumPages{Ethics}{90}{4}{470--489}.
\PrintBackRefs{\CurrentBib}

\bibitem [\protect \citeauthoryear {%
Bartling%
, Cappelen%
, Hermes%
, Skivenes%
\BCBL {}\ \BBA {} Tungodden%
}{%
Bartling%
\ \protect \BOthers {.}}{%
{\protect \APACyear {2023}}%
}]{%
bartling2023free}
\APACinsertmetastar {%
bartling2023free}%
\begin{APACrefauthors}%
Bartling, B.%
, Cappelen, A\BPBI W.%
, Hermes, H.%
, Skivenes, M.%
\BCBL {}\ \BBA {} Tungodden, B.%
\end{APACrefauthors}%
\unskip\
\newblock
\APACrefYearMonthDay{2023}{}{}.
\newblock
\APACrefbtitle {{Free to Fail? Paternalistic Preferences in the United
  States}.} {{Free to Fail? Paternalistic Preferences in the United States}.}
\PrintBackRefs{\CurrentBib}

\bibitem [\protect \citeauthoryear {%
Becker%
}{%
Becker%
}{%
{\protect \APACyear {1957}}%
}]{%
becker1957economics}
\APACinsertmetastar {%
becker1957economics}%
\begin{APACrefauthors}%
Becker, G\BPBI S.%
\end{APACrefauthors}%
\unskip\
\newblock
\APACrefYear{1957}.
\newblock
\APACrefbtitle {{The Economics of Discrimination}} {{The Economics of
  Discrimination}}.
\newblock
\APACaddressPublisher{}{University of Chicago Press}.
\PrintBackRefs{\CurrentBib}

\bibitem [\protect \citeauthoryear {%
Berlin%
}{%
Berlin%
}{%
{\protect \APACyear {1958}}%
}]{%
berlin1958two}
\APACinsertmetastar {%
berlin1958two}%
\begin{APACrefauthors}%
Berlin, I.%
\end{APACrefauthors}%
\unskip\
\newblock
\APACrefYear{1958}.
\newblock
\APACrefbtitle {{Two Concepts of Liberty}} {{Two Concepts of Liberty}}.
\newblock
\APACaddressPublisher{}{Clarendon Press}.
\PrintBackRefs{\CurrentBib}

\bibitem [\protect \citeauthoryear {%
Bernheim%
}{%
Bernheim%
}{%
{\protect \APACyear {2016}}%
}]{%
bernheim2016good}
\APACinsertmetastar {%
bernheim2016good}%
\begin{APACrefauthors}%
Bernheim, B\BPBI D.%
\end{APACrefauthors}%
\unskip\
\newblock
\APACrefYearMonthDay{2016}{}{}.
\newblock
{\BBOQ}\APACrefatitle {{The good, the bad, and the ugly: A unified approach to
  behavioral welfare economics}} {{The good, the bad, and the ugly: A unified
  approach to behavioral welfare economics}}.{\BBCQ}
\newblock
\APACjournalVolNumPages{Journal of Benefit-Cost Analysis}{7}{1}{12--68}.
\PrintBackRefs{\CurrentBib}

\bibitem [\protect \citeauthoryear {%
Binmore%
}{%
Binmore%
}{%
{\protect \APACyear {2009}}%
}]{%
binmore2009interpersonal}
\APACinsertmetastar {%
binmore2009interpersonal}%
\begin{APACrefauthors}%
Binmore, K.%
\end{APACrefauthors}%
\unskip\
\newblock
\APACrefYearMonthDay{2009}{03}{}.
\newblock
{\BBOQ}\APACrefatitle {{Interpersonal Comparison of Utility}} {{Interpersonal
  Comparison of Utility}}.{\BBCQ}
\newblock
\BIn{} \APACrefbtitle {{The Oxford Handbook of Philosophy of Economics}.} {{The
  Oxford Handbook of Philosophy of Economics}.}
\newblock
\APACaddressPublisher{}{Oxford University Press}.
\newblock
\begin{APACrefDOI} \doi{10.1093/oxfordhb/9780195189254.003.0020}
  \end{APACrefDOI}
\PrintBackRefs{\CurrentBib}

\bibitem [\protect \citeauthoryear {%
Black%
}{%
Black%
}{%
{\protect \APACyear {1948}}%
}]{%
black1948rationale}
\APACinsertmetastar {%
black1948rationale}%
\begin{APACrefauthors}%
Black, D.%
\end{APACrefauthors}%
\unskip\
\newblock
\APACrefYearMonthDay{1948}{}{}.
\newblock
{\BBOQ}\APACrefatitle {{On the rationale of group decision-making}} {{On the
  rationale of group decision-making}}.{\BBCQ}
\newblock
\APACjournalVolNumPages{Journal of Political Economy}{56}{1}{23--34}.
\PrintBackRefs{\CurrentBib}

\bibitem [\protect \citeauthoryear {%
Blackwell%
}{%
Blackwell%
}{%
{\protect \APACyear {1953}}%
}]{%
blackwell1953equivalent}
\APACinsertmetastar {%
blackwell1953equivalent}%
\begin{APACrefauthors}%
Blackwell, D.%
\end{APACrefauthors}%
\unskip\
\newblock
\APACrefYearMonthDay{1953}{}{}.
\newblock
{\BBOQ}\APACrefatitle {{Equivalent comparisons of experiments}} {{Equivalent
  comparisons of experiments}}.{\BBCQ}
\newblock
\APACjournalVolNumPages{The Annals of Mathematical Statistics}{}{}{265--272}.
\PrintBackRefs{\CurrentBib}

\bibitem [\protect \citeauthoryear {%
Bohren%
, Haggag%
, Imas%
\BCBL {}\ \BBA {} Pope%
}{%
Bohren%
\ \protect \BOthers {.}}{%
{\protect \APACyear {2023}}%
}]{%
bohren2019inaccurate}
\APACinsertmetastar {%
bohren2019inaccurate}%
\begin{APACrefauthors}%
Bohren, J\BPBI A.%
, Haggag, K.%
, Imas, A.%
\BCBL {}\ \BBA {} Pope, D\BPBI G.%
\end{APACrefauthors}%
\unskip\
\newblock
\APACrefYearMonthDay{2023}{09}{}.
\newblock
{\BBOQ}\APACrefatitle {{Inaccurate Statistical Discrimination: An
  Identification Problem}} {{Inaccurate Statistical Discrimination: An
  Identification Problem}}.{\BBCQ}
\newblock
\APACjournalVolNumPages{The Review of Economics and Statistics}{}{}{1-45}.
\newblock
\begin{APACrefDOI} \doi{10.1162/rest_a_01367} \end{APACrefDOI}
\PrintBackRefs{\CurrentBib}

\bibitem [\protect \citeauthoryear {%
Bolton%
\ \BBA {} Ockenfels%
}{%
Bolton%
\ \BBA {} Ockenfels%
}{%
{\protect \APACyear {2010}}%
}]{%
BoltonOckenfels2010}
\APACinsertmetastar {%
BoltonOckenfels2010}%
\begin{APACrefauthors}%
Bolton, G\BPBI E.%
\BCBT {}\ \BBA {} Ockenfels, A.%
\end{APACrefauthors}%
\unskip\
\newblock
\APACrefYearMonthDay{2010}{mar}{}.
\newblock
{\BBOQ}\APACrefatitle {{Betrayal Aversion: Evidence from Brazil, China, Oman,
  Switzerland, Turkey, and the United States: Comment}} {{Betrayal Aversion:
  Evidence from Brazil, China, Oman, Switzerland, Turkey, and the United
  States: Comment}}.{\BBCQ}
\newblock
\APACjournalVolNumPages{American Economic Review}{100}{1}{628--633}.
\newblock
\begin{APACrefDOI} \doi{10.1257/aer.100.1.628} \end{APACrefDOI}
\PrintBackRefs{\CurrentBib}

\bibitem [\protect \citeauthoryear {%
Bolton%
, Ockenfels%
\BCBL {}\ \BBA {} Stauf%
}{%
Bolton%
\ \protect \BOthers {.}}{%
{\protect \APACyear {2015}}%
}]{%
BoltonOckenfelsStauf2015}
\APACinsertmetastar {%
BoltonOckenfelsStauf2015}%
\begin{APACrefauthors}%
Bolton, G\BPBI E.%
, Ockenfels, A.%
\BCBL {}\ \BBA {} Stauf, J.%
\end{APACrefauthors}%
\unskip\
\newblock
\APACrefYearMonthDay{2015}{oct}{}.
\newblock
{\BBOQ}\APACrefatitle {{Social Responsibility Promotes Conservative Risk
  Behavior}} {{Social Responsibility Promotes Conservative Risk
  Behavior}}.{\BBCQ}
\newblock
\APACjournalVolNumPages{European Economic Review}{74}{}{109--127}.
\newblock
\begin{APACrefDOI} \doi{10.1016/j.euroecorev.2014.10.002} \end{APACrefDOI}
\PrintBackRefs{\CurrentBib}

\bibitem [\protect \citeauthoryear {%
Buckle%
\ \BBA {} Luhan%
}{%
Buckle%
\ \BBA {} Luhan%
}{%
{\protect \APACyear {2023}}%
}]{%
buckle2023paternalism}
\APACinsertmetastar {%
buckle2023paternalism}%
\begin{APACrefauthors}%
Buckle, G\BPBI E.%
\BCBT {}\ \BBA {} Luhan, W\BPBI J.%
\end{APACrefauthors}%
\unskip\
\newblock
\APACrefYearMonthDay{2023}{}{}.
\newblock
\APACrefbtitle {{Do as I Do: Paternalism and Preference Differences in
  Decision-Making for Others}.} {{Do as I Do: Paternalism and Preference
  Differences in Decision-Making for Others}.}
\newblock
\APACrefnote{Univ. of Portsmouth}
\PrintBackRefs{\CurrentBib}

\bibitem [\protect \citeauthoryear {%
Bushong%
\ \BBA {} Gagnon-Bartsch%
}{%
Bushong%
\ \BBA {} Gagnon-Bartsch%
}{%
{\protect \APACyear {2024}}%
}]{%
bushong2020experiment}
\APACinsertmetastar {%
bushong2020experiment}%
\begin{APACrefauthors}%
Bushong, B.%
\BCBT {}\ \BBA {} Gagnon-Bartsch, T.%
\end{APACrefauthors}%
\unskip\
\newblock
\APACrefYearMonthDay{2024}{}{}.
\newblock
{\BBOQ}\APACrefatitle {{Failures in Forecasting: An Experiment on Interpersonal
  Projection Bias}} {{Failures in Forecasting: An Experiment on Interpersonal
  Projection Bias}}.{\BBCQ}
\newblock
\APACjournalVolNumPages{Management Science}{}{}{}.
\PrintBackRefs{\CurrentBib}

\bibitem [\protect \citeauthoryear {%
{C. S. Mott Children's Hospital}%
}{%
{C. S. Mott Children's Hospital}%
}{%
{\protect \APACyear {2018}}%
}]{%
MottPoll2018}
\APACinsertmetastar {%
MottPoll2018}%
\begin{APACrefauthors}%
{C. S. Mott Children's Hospital}.%
\end{APACrefauthors}%
\unskip\
\newblock
\APACrefYearMonthDay{2018}{8}{20}.
\newblock
\APACrefbtitle {{Tensions over Teen Tattoos}} {{Tensions over Teen Tattoos}}\
  \APACbVolEdTR {\BVOL~32}{Mott Poll Report}.
\newblock
\APACaddressInstitution{}{National Poll on Children's Health, University of
  Michigan}.
\newblock
\begin{APACrefURL}
  \url{https://mottpoll.org/sites/default/files/documents/082018_Tattoos_0.pdf}
  \end{APACrefURL}
\PrintBackRefs{\CurrentBib}

\bibitem [\protect \citeauthoryear {%
Camerer%
\ \BBA {} Weber%
}{%
Camerer%
\ \BBA {} Weber%
}{%
{\protect \APACyear {1992}}%
}]{%
camerer1992recent}
\APACinsertmetastar {%
camerer1992recent}%
\begin{APACrefauthors}%
Camerer, C.%
\BCBT {}\ \BBA {} Weber, M.%
\end{APACrefauthors}%
\unskip\
\newblock
\APACrefYearMonthDay{1992}{}{}.
\newblock
{\BBOQ}\APACrefatitle {{Recent developments in modeling preferences:
  Uncertainty and ambiguity}} {{Recent developments in modeling preferences:
  Uncertainty and ambiguity}}.{\BBCQ}
\newblock
\APACjournalVolNumPages{Journal of Risk and Uncertainty}{5}{}{325--370}.
\PrintBackRefs{\CurrentBib}

\bibitem [\protect \citeauthoryear {%
Caplan%
}{%
Caplan%
}{%
{\protect \APACyear {2011}}%
}]{%
Caplan2011ABA}
\APACinsertmetastar {%
Caplan2011ABA}%
\begin{APACrefauthors}%
Caplan, B.%
\end{APACrefauthors}%
\unskip\
\newblock
\APACrefYearMonthDay{2011}{}{}.
\newblock
\APACrefbtitle {{A.B.A.: Always Be Advising}.} {{A.B.A.: Always Be Advising}.}
\newblock
\begin{APACrefURL}
  \url{https://www.econlib.org/archives/2011/07/aba_always_be_a.html}
  \end{APACrefURL}
\PrintBackRefs{\CurrentBib}

\bibitem [\protect \citeauthoryear {%
Charness%
, Gneezy%
\BCBL {}\ \BBA {} Kuhn%
}{%
Charness%
\ \protect \BOthers {.}}{%
{\protect \APACyear {2012}}%
}]{%
charness2012experimental}
\APACinsertmetastar {%
charness2012experimental}%
\begin{APACrefauthors}%
Charness, G.%
, Gneezy, U.%
\BCBL {}\ \BBA {} Kuhn, M\BPBI A.%
\end{APACrefauthors}%
\unskip\
\newblock
\APACrefYearMonthDay{2012}{}{}.
\newblock
{\BBOQ}\APACrefatitle {{Experimental methods: Between-subject and
  within-subject design}} {{Experimental methods: Between-subject and
  within-subject design}}.{\BBCQ}
\newblock
\APACjournalVolNumPages{Journal of Economic Behavior \&
  Organization}{81}{1}{1--8}.
\PrintBackRefs{\CurrentBib}

\bibitem [\protect \citeauthoryear {%
Chew%
, Miao%
\BCBL {}\ \BBA {} Zhong%
}{%
Chew%
\ \protect \BOthers {.}}{%
{\protect \APACyear {2017}}%
}]{%
chew2017partial}
\APACinsertmetastar {%
chew2017partial}%
\begin{APACrefauthors}%
Chew, S\BPBI H.%
, Miao, B.%
\BCBL {}\ \BBA {} Zhong, S.%
\end{APACrefauthors}%
\unskip\
\newblock
\APACrefYearMonthDay{2017}{}{}.
\newblock
{\BBOQ}\APACrefatitle {{Partial ambiguity}} {{Partial ambiguity}}.{\BBCQ}
\newblock
\APACjournalVolNumPages{Econometrica}{85}{4}{1239--1260}.
\PrintBackRefs{\CurrentBib}

\bibitem [\protect \citeauthoryear {%
Chipman%
}{%
Chipman%
}{%
{\protect \APACyear {1963}}%
}]{%
chipman1963stochastic}
\APACinsertmetastar {%
chipman1963stochastic}%
\begin{APACrefauthors}%
Chipman, J\BPBI S.%
\end{APACrefauthors}%
\unskip\
\newblock
\APACrefYearMonthDay{1963}{}{}.
\newblock
{\BBOQ}\APACrefatitle {{Stochastic choice and subjective probability}}
  {{Stochastic choice and subjective probability}}.{\BBCQ}
\newblock
\APACjournalVolNumPages{Decisions, Values and Groups}{1}{}{70--95}.
\PrintBackRefs{\CurrentBib}

\bibitem [\protect \citeauthoryear {%
Clemens%
}{%
Clemens%
}{%
{\protect \APACyear {2018}}%
}]{%
clemens2018testing}
\APACinsertmetastar {%
clemens2018testing}%
\begin{APACrefauthors}%
Clemens, M\BPBI A.%
\end{APACrefauthors}%
\unskip\
\newblock
\APACrefYearMonthDay{2018}{}{}.
\newblock
{\BBOQ}\APACrefatitle {{Testing for repugnance in economic transactions:
  Evidence from guest work in the Gulf}} {{Testing for repugnance in economic
  transactions: Evidence from guest work in the Gulf}}.{\BBCQ}
\newblock
\APACjournalVolNumPages{The Journal of Legal Studies}{47}{S1}{S5--S44}.
\PrintBackRefs{\CurrentBib}

\bibitem [\protect \citeauthoryear {%
Corbett%
, Feeney%
\BCBL {}\ \BBA {} McCormack%
}{%
Corbett%
\ \protect \BOthers {.}}{%
{\protect \APACyear {2021}}%
}]{%
corbett2021interpersonal}
\APACinsertmetastar {%
corbett2021interpersonal}%
\begin{APACrefauthors}%
Corbett, B.%
, Feeney, A.%
\BCBL {}\ \BBA {} McCormack, T.%
\end{APACrefauthors}%
\unskip\
\newblock
\APACrefYearMonthDay{2021}{}{}.
\newblock
{\BBOQ}\APACrefatitle {{Interpersonal regret and prosocial risk taking in
  children}} {{Interpersonal regret and prosocial risk taking in
  children}}.{\BBCQ}
\newblock
\APACjournalVolNumPages{Cognitive Development}{58}{}{101036}.
\PrintBackRefs{\CurrentBib}

\bibitem [\protect \citeauthoryear {%
Cowen%
}{%
Cowen%
}{%
{\protect \APACyear {2023}}%
}]{%
cowen2023goat}
\APACinsertmetastar {%
cowen2023goat}%
\begin{APACrefauthors}%
Cowen, T.%
\end{APACrefauthors}%
\unskip\
\newblock
\APACrefYear{2023}.
\newblock
\APACrefbtitle {{GOAT: Who is the Greatest Economist of all Time, and Why Does
  it Matter?}} {{GOAT: Who is the Greatest Economist of all Time, and Why Does
  it Matter?}}
\newblock
\begin{APACrefURL} \url{https://econgoat.ai} \end{APACrefURL}
\PrintBackRefs{\CurrentBib}

\bibitem [\protect \citeauthoryear {%
Derksen%
\ \BBA {} Morawski%
}{%
Derksen%
\ \BBA {} Morawski%
}{%
{\protect \APACyear {2022}}%
}]{%
doi:10.1177/17456916211041116}
\APACinsertmetastar {%
doi:10.1177/17456916211041116}%
\begin{APACrefauthors}%
Derksen, M.%
\BCBT {}\ \BBA {} Morawski, J.%
\end{APACrefauthors}%
\unskip\
\newblock
\APACrefYearMonthDay{2022}{}{}.
\newblock
{\BBOQ}\APACrefatitle {Kinds of Replication: Examining the Meanings of
  "Conceptual Replication" and "Direct Replication"} {Kinds of replication:
  Examining the meanings of "conceptual replication" and "direct
  replication"}.{\BBCQ}
\newblock
\APACjournalVolNumPages{Perspectives on Psychological
  Science}{17}{5}{1490-1505}.
\newblock
\begin{APACrefDOI} \doi{10.1177/17456916211041116} \end{APACrefDOI}
\PrintBackRefs{\CurrentBib}

\bibitem [\protect \citeauthoryear {%
Downs%
}{%
Downs%
}{%
{\protect \APACyear {1957}}%
}]{%
downs1957economic}
\APACinsertmetastar {%
downs1957economic}%
\begin{APACrefauthors}%
Downs, A.%
\end{APACrefauthors}%
\unskip\
\newblock
\APACrefYearMonthDay{1957}{}{}.
\newblock
{\BBOQ}\APACrefatitle {{An economic theory of democracy}} {{An economic theory
  of democracy}}.{\BBCQ}
\newblock
\APACjournalVolNumPages{Harper and Row}{28}{}{}.
\PrintBackRefs{\CurrentBib}

\bibitem [\protect \citeauthoryear {%
Dworkin%
}{%
Dworkin%
}{%
{\protect \APACyear {1972}}%
}]{%
dworkin1972paternalism}
\APACinsertmetastar {%
dworkin1972paternalism}%
\begin{APACrefauthors}%
Dworkin, G.%
\end{APACrefauthors}%
\unskip\
\newblock
\APACrefYearMonthDay{1972}{}{}.
\newblock
{\BBOQ}\APACrefatitle {{Paternalism}} {{Paternalism}}.{\BBCQ}
\newblock
\APACjournalVolNumPages{the Monist}{}{}{64--84}.
\PrintBackRefs{\CurrentBib}

\bibitem [\protect \citeauthoryear {%
Dworkin%
}{%
Dworkin%
}{%
{\protect \APACyear {2020}}%
}]{%
dworkin2020paternalism}
\APACinsertmetastar {%
dworkin2020paternalism}%
\begin{APACrefauthors}%
Dworkin, G.%
\end{APACrefauthors}%
\unskip\
\newblock
\APACrefYearMonthDay{2020}{}{}.
\newblock
{\BBOQ}\APACrefatitle {{{Paternalism}}} {{{Paternalism}}}.{\BBCQ}
\newblock
\BIn{} E\BPBI N.~Zalta\ (\BED), \APACrefbtitle {{The {Stanford}} Encyclopedia
  of Philosophy} {{The {Stanford}} encyclopedia of philosophy}\
  (\PrintOrdinal{{F}all 2020}\ \BEd).
\newblock
\APACaddressPublisher{}{Metaphysics Research Lab, Stanford University}.
\newblock
\APAChowpublished
  {\url{https://plato.stanford.edu/archives/fall2020/entries/paternalism/}}.
\PrintBackRefs{\CurrentBib}

\bibitem [\protect \citeauthoryear {%
Fehr%
\ \BBA {} Charness%
}{%
Fehr%
\ \BBA {} Charness%
}{%
{\protect \APACyear {2024}}%
}]{%
FehrCharnessForthcoming}
\APACinsertmetastar {%
FehrCharnessForthcoming}%
\begin{APACrefauthors}%
Fehr, E.%
\BCBT {}\ \BBA {} Charness, G.%
\end{APACrefauthors}%
\unskip\
\newblock
\APACrefYearMonthDay{2024}{}{}.
\newblock
{\BBOQ}\APACrefatitle {{Social Preferences: Fundamental Characteristics and
  Economic Consequences}} {{Social Preferences: Fundamental Characteristics and
  Economic Consequences}}.{\BBCQ}
\newblock
\APACjournalVolNumPages{Journal of Economic Literature}{}{}{}.
\PrintBackRefs{\CurrentBib}

\bibitem [\protect \citeauthoryear {%
Friedman%
}{%
Friedman%
}{%
{\protect \APACyear {1953}}%
}]{%
friedman1953methodology}
\APACinsertmetastar {%
friedman1953methodology}%
\begin{APACrefauthors}%
Friedman, M.%
\end{APACrefauthors}%
\unskip\
\newblock
\APACrefYearMonthDay{1953}{}{}.
\newblock
{\BBOQ}\APACrefatitle {{The Methodology of Positive Economics}} {{The
  Methodology of Positive Economics}}.{\BBCQ}
\newblock
\BIn{} \APACrefbtitle {{Essays in Positive Economics}} {{Essays in Positive
  Economics}}\ (\BCHAP~1).
\newblock
\APACaddressPublisher{}{University of Chicago Press}.
\PrintBackRefs{\CurrentBib}

\bibitem [\protect \citeauthoryear {%
Gigliotti%
}{%
Gigliotti%
}{%
{\protect \APACyear {1996}}%
}]{%
gigliotti1996testing}
\APACinsertmetastar {%
gigliotti1996testing}%
\begin{APACrefauthors}%
Gigliotti, G.%
\end{APACrefauthors}%
\unskip\
\newblock
\APACrefYearMonthDay{1996}{}{}.
\newblock
{\BBOQ}\APACrefatitle {{The testing principle: Inductive reasoning and the
  Ellsberg paradox}} {{The testing principle: Inductive reasoning and the
  Ellsberg paradox}}.{\BBCQ}
\newblock
\APACjournalVolNumPages{Thinking \& Reasoning}{2}{1}{33--49}.
\PrintBackRefs{\CurrentBib}

\bibitem [\protect \citeauthoryear {%
Greiner%
}{%
Greiner%
}{%
{\protect \APACyear {2015}}%
}]{%
greiner2015subject}
\APACinsertmetastar {%
greiner2015subject}%
\begin{APACrefauthors}%
Greiner, B.%
\end{APACrefauthors}%
\unskip\
\newblock
\APACrefYearMonthDay{2015}{}{}.
\newblock
{\BBOQ}\APACrefatitle {{Subject pool recruitment procedures: organizing
  experiments with ORSEE}} {{Subject pool recruitment procedures: organizing
  experiments with ORSEE}}.{\BBCQ}
\newblock
\APACjournalVolNumPages{Journal of the Economic Science
  Association}{1}{1}{114--125}.
\PrintBackRefs{\CurrentBib}

\bibitem [\protect \citeauthoryear {%
Grossman%
}{%
Grossman%
}{%
{\protect \APACyear {1981}}%
}]{%
grossman1981informational}
\APACinsertmetastar {%
grossman1981informational}%
\begin{APACrefauthors}%
Grossman, S\BPBI J.%
\end{APACrefauthors}%
\unskip\
\newblock
\APACrefYearMonthDay{1981}{}{}.
\newblock
{\BBOQ}\APACrefatitle {The informational role of warranties and private
  disclosure about product quality} {The informational role of warranties and
  private disclosure about product quality}.{\BBCQ}
\newblock
\APACjournalVolNumPages{{The Journal of Law and Economics}}{24}{3}{461--483}.
\PrintBackRefs{\CurrentBib}

\bibitem [\protect \citeauthoryear {%
Grossmann%
}{%
Grossmann%
}{%
{\protect \APACyear {2024}}%
}]{%
undo}
\APACinsertmetastar {%
undo}%
\begin{APACrefauthors}%
Grossmann, M\BPBI R\BPBI P.%
\end{APACrefauthors}%
\unskip\
\newblock
\APACrefYearMonthDay{2024}{}{}.
\newblock
\APACrefbtitle {{Paternalism and Deliberation: An Experiment on Making Formal
  Rules}.} {{Paternalism and Deliberation: An Experiment on Making Formal
  Rules}.}
\newblock
\APACrefnote{Mimeo}
\PrintBackRefs{\CurrentBib}

\bibitem [\protect \citeauthoryear {%
Grossmann%
\ \BBA {} Gerhardt%
}{%
Grossmann%
\ \BBA {} Gerhardt%
}{%
{\protect \APACyear {{\protect \bibnodate {}}}}%
}]{%
uproot}
\APACinsertmetastar {%
uproot}%
\begin{APACrefauthors}%
Grossmann, M\BPBI R\BPBI P.%
\BCBT {}\ \BBA {} Gerhardt, H.%
\end{APACrefauthors}%
\unskip\
\newblock
\APACrefYearMonthDay{{\protect \bibnodate {}}}{}{}.
\newblock
\APACrefbtitle {{uproot}.} {{uproot}.}
\newblock
\APACrefnote{Mimeo}
\PrintBackRefs{\CurrentBib}

\bibitem [\protect \citeauthoryear {%
Grossmann%
\ \BBA {} Ockenfels%
}{%
Grossmann%
\ \BBA {} Ockenfels%
}{%
{\protect \APACyear {2024}}%
}]{%
pids}
\APACinsertmetastar {%
pids}%
\begin{APACrefauthors}%
Grossmann, M\BPBI R\BPBI P.%
\BCBT {}\ \BBA {} Ockenfels, A.%
\end{APACrefauthors}%
\unskip\
\newblock
\APACrefYearMonthDay{2024}{}{}.
\newblock
\APACrefbtitle {{Paternalism in Data Sharing}.} {{Paternalism in Data
  Sharing}.}
\newblock
\APACrefnote{Mimeo}
\PrintBackRefs{\CurrentBib}

\bibitem [\protect \citeauthoryear {%
Harrison%
, Hofmeyr%
, Ross%
\BCBL {}\ \BBA {} Swarthout%
}{%
Harrison%
\ \protect \BOthers {.}}{%
{\protect \APACyear {2018}}%
}]{%
harrison2018risk}
\APACinsertmetastar {%
harrison2018risk}%
\begin{APACrefauthors}%
Harrison, G\BPBI W.%
, Hofmeyr, A.%
, Ross, D.%
\BCBL {}\ \BBA {} Swarthout, J\BPBI T.%
\end{APACrefauthors}%
\unskip\
\newblock
\APACrefYearMonthDay{2018}{}{}.
\newblock
{\BBOQ}\APACrefatitle {{Risk preferences, time preferences, and smoking
  behavior}} {{Risk preferences, time preferences, and smoking
  behavior}}.{\BBCQ}
\newblock
\APACjournalVolNumPages{Southern Economic Journal}{85}{2}{313--348}.
\PrintBackRefs{\CurrentBib}

\bibitem [\protect \citeauthoryear {%
Hausman%
}{%
Hausman%
}{%
{\protect \APACyear {1995}}%
}]{%
hausman1996impossibility}
\APACinsertmetastar {%
hausman1996impossibility}%
\begin{APACrefauthors}%
Hausman, D\BPBI M.%
\end{APACrefauthors}%
\unskip\
\newblock
\APACrefYearMonthDay{1995}{07}{}.
\newblock
{\BBOQ}\APACrefatitle {{The Impossibility of Interpersonal Utility
  Comparisons}} {{The Impossibility of Interpersonal Utility
  Comparisons}}.{\BBCQ}
\newblock
\APACjournalVolNumPages{Mind}{104}{415}{473-490}.
\newblock
\begin{APACrefDOI} \doi{10.1093/mind/104.415.473} \end{APACrefDOI}
\PrintBackRefs{\CurrentBib}

\bibitem [\protect \citeauthoryear {%
Head%
}{%
Head%
}{%
{\protect \APACyear {1966}}%
}]{%
head1966merit}
\APACinsertmetastar {%
head1966merit}%
\begin{APACrefauthors}%
Head, J\BPBI G.%
\end{APACrefauthors}%
\unskip\
\newblock
\APACrefYearMonthDay{1966}{}{}.
\newblock
{\BBOQ}\APACrefatitle {{On Merit Goods}} {{On Merit Goods}}.{\BBCQ}
\newblock
\APACjournalVolNumPages{FinanzArchiv/Public Finance Analysis}{25}{}{1--29}.
\PrintBackRefs{\CurrentBib}

\bibitem [\protect \citeauthoryear {%
Hossain%
\ \BBA {} Okui%
}{%
Hossain%
\ \BBA {} Okui%
}{%
{\protect \APACyear {2013}}%
}]{%
hossain2013binarized}
\APACinsertmetastar {%
hossain2013binarized}%
\begin{APACrefauthors}%
Hossain, T.%
\BCBT {}\ \BBA {} Okui, R.%
\end{APACrefauthors}%
\unskip\
\newblock
\APACrefYearMonthDay{2013}{}{}.
\newblock
{\BBOQ}\APACrefatitle {{The binarized scoring rule}} {{The binarized scoring
  rule}}.{\BBCQ}
\newblock
\APACjournalVolNumPages{Review of Economic Studies}{80}{3}{984--1001}.
\PrintBackRefs{\CurrentBib}

\bibitem [\protect \citeauthoryear {%
Kartik%
}{%
Kartik%
}{%
{\protect \APACyear {2009}}%
}]{%
kartik2009strategic}
\APACinsertmetastar {%
kartik2009strategic}%
\begin{APACrefauthors}%
Kartik, N.%
\end{APACrefauthors}%
\unskip\
\newblock
\APACrefYearMonthDay{2009}{}{}.
\newblock
{\BBOQ}\APACrefatitle {Strategic communication with lying costs} {Strategic
  communication with lying costs}.{\BBCQ}
\newblock
\APACjournalVolNumPages{{The Review of Economic Studies}}{76}{4}{1359--1395}.
\PrintBackRefs{\CurrentBib}

\bibitem [\protect \citeauthoryear {%
Khalmetski%
\ \BBA {} Ockenfels%
}{%
Khalmetski%
\ \BBA {} Ockenfels%
}{%
{\protect \APACyear {2024}}%
}]{%
khalmetski2024why}
\APACinsertmetastar {%
khalmetski2024why}%
\begin{APACrefauthors}%
Khalmetski, K.%
\BCBT {}\ \BBA {} Ockenfels, A.%
\end{APACrefauthors}%
\unskip\
\newblock
\APACrefYearMonthDay{2024}{}{}.
\newblock
\APACrefbtitle {{Why people ban others' actions: A norm conformism
  hypothesis}.} {{Why people ban others' actions: A norm conformism
  hypothesis}.}
\newblock
\APACrefnote{Mimeo}
\PrintBackRefs{\CurrentBib}

\bibitem [\protect \citeauthoryear {%
Kiessling%
, Chowdhury%
, Schildberg-H\"{o}risch%
\BCBL {}\ \BBA {} Sutter%
}{%
Kiessling%
\ \protect \BOthers {.}}{%
{\protect \APACyear {2021}}%
}]{%
kiessling2022parental}
\APACinsertmetastar {%
kiessling2022parental}%
\begin{APACrefauthors}%
Kiessling, L.%
, Chowdhury, S\BPBI K.%
, Schildberg-H\"{o}risch, H.%
\BCBL {}\ \BBA {} Sutter, M.%
\end{APACrefauthors}%
\unskip\
\newblock
\APACrefYearMonthDay{2021}{}{}.
\newblock
\APACrefbtitle {{Parental paternalism and patience}} {{Parental paternalism and
  patience}}\ (\BNUM~358).
\newblock
\begin{APACrefURL} \url{https://hdl.handle.net/10419/228523} \end{APACrefURL}
\newblock
\APACrefnote{DICE Discussion Paper № 358}
\PrintBackRefs{\CurrentBib}

\bibitem [\protect \citeauthoryear {%
Kirchg{\"a}ssner%
}{%
Kirchg{\"a}ssner%
}{%
{\protect \APACyear {2017}}%
}]{%
kirchgassner2017soft}
\APACinsertmetastar {%
kirchgassner2017soft}%
\begin{APACrefauthors}%
Kirchg{\"a}ssner, G.%
\end{APACrefauthors}%
\unskip\
\newblock
\APACrefYearMonthDay{2017}{}{}.
\newblock
{\BBOQ}\APACrefatitle {{Soft paternalism, merit goods, and normative
  individualism}} {{Soft paternalism, merit goods, and normative
  individualism}}.{\BBCQ}
\newblock
\APACjournalVolNumPages{European Journal of Law and Economics}{43}{}{125--152}.
\PrintBackRefs{\CurrentBib}

\bibitem [\protect \citeauthoryear {%
Kolm%
}{%
Kolm%
}{%
{\protect \APACyear {1993}}%
}]{%
kolm1993impossibility}
\APACinsertmetastar {%
kolm1993impossibility}%
\begin{APACrefauthors}%
Kolm, S\BHBI C.%
\end{APACrefauthors}%
\unskip\
\newblock
\APACrefYearMonthDay{1993}{}{}.
\newblock
{\BBOQ}\APACrefatitle {{The Impossibility of Utilitarianism}} {{The
  Impossibility of Utilitarianism}}.{\BBCQ}
\newblock
\BIn{} \APACrefbtitle {{The Good and the Economical: Ethical Choices in
  Economics and Management}} {{The Good and the Economical: Ethical Choices in
  Economics and Management}}\ (\BPGS\ 30--69).
\newblock
\APACaddressPublisher{}{Springer}.
\PrintBackRefs{\CurrentBib}

\bibitem [\protect \citeauthoryear {%
Konrad%
}{%
Konrad%
}{%
{\protect \APACyear {2024}}%
}]{%
konrad2023political}
\APACinsertmetastar {%
konrad2023political}%
\begin{APACrefauthors}%
Konrad, K\BPBI A.%
\end{APACrefauthors}%
\unskip\
\newblock
\APACrefYearMonthDay{2024}{}{}.
\newblock
{\BBOQ}\APACrefatitle {{The political economy of paternalism}} {{The political
  economy of paternalism}}.{\BBCQ}
\newblock
\APACjournalVolNumPages{Public Choice}{201}{}{61--81}.
\PrintBackRefs{\CurrentBib}

\bibitem [\protect \citeauthoryear {%
Leifeld%
}{%
Leifeld%
}{%
{\protect \APACyear {2013}}%
}]{%
texreg}
\APACinsertmetastar {%
texreg}%
\begin{APACrefauthors}%
Leifeld, P.%
\end{APACrefauthors}%
\unskip\
\newblock
\APACrefYearMonthDay{2013}{}{}.
\newblock
{\BBOQ}\APACrefatitle {{{texreg}}: Conversion of Statistical Model Output in
  {R} to {\LaTeX} and {HTML} Tables} {{{texreg}}: Conversion of statistical
  model output in {R} to {\LaTeX} and {HTML} tables}.{\BBCQ}
\newblock
\APACjournalVolNumPages{Journal of Statistical Software}{55}{8}{1--24}.
\PrintBackRefs{\CurrentBib}

\bibitem [\protect \citeauthoryear {%
Mabsout%
}{%
Mabsout%
}{%
{\protect \APACyear {2022}}%
}]{%
mabsout2022john}
\APACinsertmetastar {%
mabsout2022john}%
\begin{APACrefauthors}%
Mabsout, R.%
\end{APACrefauthors}%
\unskip\
\newblock
\APACrefYearMonthDay{2022}{}{}.
\newblock
{\BBOQ}\APACrefatitle {{John Stuart Mill, soft paternalist}} {{John Stuart
  Mill, soft paternalist}}.{\BBCQ}
\newblock
\APACjournalVolNumPages{Social Choice and Welfare}{58}{1}{161--186}.
\PrintBackRefs{\CurrentBib}

\bibitem [\protect \citeauthoryear {%
Manski%
\ \BBA {} Sheshinski%
}{%
Manski%
\ \BBA {} Sheshinski%
}{%
{\protect \APACyear {2023}}%
}]{%
NBERw31349}
\APACinsertmetastar {%
NBERw31349}%
\begin{APACrefauthors}%
Manski, C\BPBI F.%
\BCBT {}\ \BBA {} Sheshinski, E.%
\end{APACrefauthors}%
\unskip\
\newblock
\APACrefYearMonthDay{2023}{}{}.
\newblock
\APACrefbtitle {Optimal Paternalism in a Population with Bounded Rationality}
  {Optimal paternalism in a population with bounded rationality}\ \APACbVolEdTR
  {}{Working Paper\ \BNUM\ 31349}.
\newblock
\APACaddressInstitution{}{National Bureau of Economic Research}.
\newblock
\begin{APACrefURL} \url{https://www.nber.org/papers/w31349} \end{APACrefURL}
\PrintBackRefs{\CurrentBib}

\bibitem [\protect \citeauthoryear {%
Milgrom%
}{%
Milgrom%
}{%
{\protect \APACyear {1981}}%
}]{%
milgrom1981good}
\APACinsertmetastar {%
milgrom1981good}%
\begin{APACrefauthors}%
Milgrom, P\BPBI R.%
\end{APACrefauthors}%
\unskip\
\newblock
\APACrefYearMonthDay{1981}{}{}.
\newblock
{\BBOQ}\APACrefatitle {Good news and bad news: Representation theorems and
  applications} {Good news and bad news: Representation theorems and
  applications}.{\BBCQ}
\newblock
\APACjournalVolNumPages{The Bell Journal of Economics}{}{}{380--391}.
\PrintBackRefs{\CurrentBib}

\bibitem [\protect \citeauthoryear {%
Mill%
}{%
Mill%
}{%
{\protect \APACyear {1869}}%
}]{%
mill1859on}
\APACinsertmetastar {%
mill1859on}%
\begin{APACrefauthors}%
Mill, J\BPBI S.%
\end{APACrefauthors}%
\unskip\
\newblock
\APACrefYear{1869}.
\newblock
\APACrefbtitle {{On Liberty}} {{On Liberty}}\ (\PrintOrdinal{4}\ \BEd).
\newblock
\APACaddressPublisher{London}{Longmans, Green, Reader and Dyer}.
\newblock
\begin{APACrefURL} \url{https://en.wikisource.org/wiki/On_Liberty}
  \end{APACrefURL}
\PrintBackRefs{\CurrentBib}

\bibitem [\protect \citeauthoryear {%
Musgrave%
}{%
Musgrave%
}{%
{\protect \APACyear {1956}}%
}]{%
musgrave1956multiple}
\APACinsertmetastar {%
musgrave1956multiple}%
\begin{APACrefauthors}%
Musgrave, R\BPBI A.%
\end{APACrefauthors}%
\unskip\
\newblock
\APACrefYearMonthDay{1956}{}{}.
\newblock
{\BBOQ}\APACrefatitle {{A Multiple Theory of Budget Determination}} {{A
  Multiple Theory of Budget Determination}}.{\BBCQ}
\newblock
\APACjournalVolNumPages{Fi\-nanz\-Ar\-chiv/Pub\-lic Finance
  Analysis}{17}{}{333--343}.
\PrintBackRefs{\CurrentBib}

\bibitem [\protect \citeauthoryear {%
Musgrave%
}{%
Musgrave%
}{%
{\protect \APACyear {1959}}%
}]{%
musgrave1959theory}
\APACinsertmetastar {%
musgrave1959theory}%
\begin{APACrefauthors}%
Musgrave, R\BPBI A.%
\end{APACrefauthors}%
\unskip\
\newblock
\APACrefYear{1959}.
\newblock
\APACrefbtitle {{The Theory of Public Finance---A Study in Public Economy}}
  {{The Theory of Public Finance---A Study in Public Economy}}.
\newblock
\APACaddressPublisher{}{McGraw-Hill}.
\PrintBackRefs{\CurrentBib}

\bibitem [\protect \citeauthoryear {%
Page%
}{%
Page%
}{%
{\protect \APACyear {1963}}%
}]{%
pagel}
\APACinsertmetastar {%
pagel}%
\begin{APACrefauthors}%
Page, E\BPBI B.%
\end{APACrefauthors}%
\unskip\
\newblock
\APACrefYearMonthDay{1963}{}{}.
\newblock
{\BBOQ}\APACrefatitle {{Ordered Hypotheses for Multiple Treatments: A
  Significance Test for Linear Ranks}} {{Ordered Hypotheses for Multiple
  Treatments: A Significance Test for Linear Ranks}}.{\BBCQ}
\newblock
\APACjournalVolNumPages{Journal of the American Statistical
  Association}{58}{301}{216-230}.
\newblock
\begin{APACrefDOI} \doi{10.1080/01621459.1963.10500843} \end{APACrefDOI}
\PrintBackRefs{\CurrentBib}

\bibitem [\protect \citeauthoryear {%
Pareto%
}{%
Pareto%
}{%
{\protect \APACyear {2014}}%
}]{%
pareto2014manual}
\APACinsertmetastar {%
pareto2014manual}%
\begin{APACrefauthors}%
Pareto, V.%
\end{APACrefauthors}%
\unskip\
\newblock
\APACrefYear{2014}.
\newblock
\APACrefbtitle {{Manual of Political Economy: A Critical and Variorum Edition}}
  {{Manual of Political Economy: A Critical and Variorum Edition}}.
\newblock
\APACaddressPublisher{}{Oxford University Press}.
\PrintBackRefs{\CurrentBib}

\bibitem [\protect \citeauthoryear {%
Polman%
\ \BBA {} Wu%
}{%
Polman%
\ \BBA {} Wu%
}{%
{\protect \APACyear {2020}}%
}]{%
polman2020decision}
\APACinsertmetastar {%
polman2020decision}%
\begin{APACrefauthors}%
Polman, E.%
\BCBT {}\ \BBA {} Wu, K.%
\end{APACrefauthors}%
\unskip\
\newblock
\APACrefYearMonthDay{2020}{}{}.
\newblock
{\BBOQ}\APACrefatitle {{Decision making for others involving risk: A review and
  meta-analysis}} {{Decision making for others involving risk: A review and
  meta-analysis}}.{\BBCQ}
\newblock
\APACjournalVolNumPages{Journal of Economic Psychology}{77}{}{102184}.
\PrintBackRefs{\CurrentBib}

\bibitem [\protect \citeauthoryear {%
Ross%
, Greene%
\BCBL {}\ \BBA {} House%
}{%
Ross%
\ \protect \BOthers {.}}{%
{\protect \APACyear {1977}}%
}]{%
ross1977false}
\APACinsertmetastar {%
ross1977false}%
\begin{APACrefauthors}%
Ross, L.%
, Greene, D.%
\BCBL {}\ \BBA {} House, P.%
\end{APACrefauthors}%
\unskip\
\newblock
\APACrefYearMonthDay{1977}{}{}.
\newblock
{\BBOQ}\APACrefatitle {{The "false consensus effect": An egocentric bias in
  social perception and attribution processes}} {{The "false consensus effect":
  An egocentric bias in social perception and attribution processes}}.{\BBCQ}
\newblock
\APACjournalVolNumPages{Journal of Experimental Social
  Psychology}{13}{3}{279--301}.
\PrintBackRefs{\CurrentBib}

\bibitem [\protect \citeauthoryear {%
Roth%
}{%
Roth%
}{%
{\protect \APACyear {2007}}%
}]{%
roth2007repugnance}
\APACinsertmetastar {%
roth2007repugnance}%
\begin{APACrefauthors}%
Roth, A\BPBI E.%
\end{APACrefauthors}%
\unskip\
\newblock
\APACrefYearMonthDay{2007}{}{}.
\newblock
{\BBOQ}\APACrefatitle {{Repugnance as a Constraint on Markets}} {{Repugnance as
  a Constraint on Markets}}.{\BBCQ}
\newblock
\APACjournalVolNumPages{Journal of Economic Perspectives}{21}{3}{37--58}.
\PrintBackRefs{\CurrentBib}

\bibitem [\protect \citeauthoryear {%
Schwerter%
}{%
Schwerter%
}{%
{\protect \APACyear {2024}}%
}]{%
schwerter2024social}
\APACinsertmetastar {%
schwerter2024social}%
\begin{APACrefauthors}%
Schwerter, F.%
\end{APACrefauthors}%
\unskip\
\newblock
\APACrefYearMonthDay{2024}{}{}.
\newblock
{\BBOQ}\APACrefatitle {{Social Reference Points and Risk Taking}} {{Social
  Reference Points and Risk Taking}}.{\BBCQ}
\newblock
\APACjournalVolNumPages{Management Science}{70}{1}{616--632}.
\PrintBackRefs{\CurrentBib}

\bibitem [\protect \citeauthoryear {%
Scoccia%
}{%
Scoccia%
}{%
{\protect \APACyear {2018}}%
}]{%
scoccia2018concept}
\APACinsertmetastar {%
scoccia2018concept}%
\begin{APACrefauthors}%
Scoccia, D.%
\end{APACrefauthors}%
\unskip\
\newblock
\APACrefYearMonthDay{2018}{}{}.
\newblock
{\BBOQ}\APACrefatitle {{The concept of paternalism}} {{The concept of
  paternalism}}.{\BBCQ}
\newblock
\BIn{} \APACrefbtitle {{The Routledge Handbook of the Philosophy of
  Paternalism}} {{The Routledge Handbook of the Philosophy of Paternalism}}\
  (\BPGS\ 11--23).
\newblock
\APACaddressPublisher{}{Routledge}.
\PrintBackRefs{\CurrentBib}

\bibitem [\protect \citeauthoryear {%
{\v{S}}peci{\'a}n%
}{%
{\v{S}}peci{\'a}n%
}{%
{\protect \APACyear {2019}}%
}]{%
vspecian2019precarious}
\APACinsertmetastar {%
vspecian2019precarious}%
\begin{APACrefauthors}%
{\v{S}}peci{\'a}n, P.%
\end{APACrefauthors}%
\unskip\
\newblock
\APACrefYearMonthDay{2019}{}{}.
\newblock
{\BBOQ}\APACrefatitle {{The Precarious Case of the True Preferences}} {{The
  Precarious Case of the True Preferences}}.{\BBCQ}
\newblock
\APACjournalVolNumPages{Society}{56}{}{267--272}.
\PrintBackRefs{\CurrentBib}

\bibitem [\protect \citeauthoryear {%
Sugden%
}{%
Sugden%
}{%
{\protect \APACyear {2022}}%
}]{%
sugden2022debiasing}
\APACinsertmetastar {%
sugden2022debiasing}%
\begin{APACrefauthors}%
Sugden, R.%
\end{APACrefauthors}%
\unskip\
\newblock
\APACrefYearMonthDay{2022}{}{}.
\newblock
{\BBOQ}\APACrefatitle {{Debiasing or regularisation? Two interpretations of the
  concept of 'true preference' in behavioural economics}} {{Debiasing or
  regularisation? Two interpretations of the concept of 'true preference' in
  behavioural economics}}.{\BBCQ}
\newblock
\APACjournalVolNumPages{Theory and Decision}{92}{3-4}{765--784}.
\PrintBackRefs{\CurrentBib}

\bibitem [\protect \citeauthoryear {%
Traxler%
\ \BBA {} Winter%
}{%
Traxler%
\ \BBA {} Winter%
}{%
{\protect \APACyear {2012}}%
}]{%
traxler2012survey}
\APACinsertmetastar {%
traxler2012survey}%
\begin{APACrefauthors}%
Traxler, C.%
\BCBT {}\ \BBA {} Winter, J.%
\end{APACrefauthors}%
\unskip\
\newblock
\APACrefYearMonthDay{2012}{}{}.
\newblock
{\BBOQ}\APACrefatitle {{Survey evidence on conditional norm enforcement}}
  {{Survey evidence on conditional norm enforcement}}.{\BBCQ}
\newblock
\APACjournalVolNumPages{European Journal of Political
  Economy}{28}{3}{390--398}.
\PrintBackRefs{\CurrentBib}

\bibitem [\protect \citeauthoryear {%
Tversky%
\ \BBA {} Kahneman%
}{%
Tversky%
\ \BBA {} Kahneman%
}{%
{\protect \APACyear {1971}}%
}]{%
tversky1971belief}
\APACinsertmetastar {%
tversky1971belief}%
\begin{APACrefauthors}%
Tversky, A.%
\BCBT {}\ \BBA {} Kahneman, D.%
\end{APACrefauthors}%
\unskip\
\newblock
\APACrefYearMonthDay{1971}{}{}.
\newblock
{\BBOQ}\APACrefatitle {{Belief in the law of small numbers}} {{Belief in the
  law of small numbers}}.{\BBCQ}
\newblock
\APACjournalVolNumPages{Psychological Bulletin}{76}{2}{105}.
\PrintBackRefs{\CurrentBib}

\bibitem [\protect \citeauthoryear {%
Viscusi%
}{%
Viscusi%
}{%
{\protect \APACyear {1990}}%
}]{%
viscusi1990smokers}
\APACinsertmetastar {%
viscusi1990smokers}%
\begin{APACrefauthors}%
Viscusi, W\BPBI K.%
\end{APACrefauthors}%
\unskip\
\newblock
\APACrefYearMonthDay{1990}{}{}.
\newblock
{\BBOQ}\APACrefatitle {{Do Smokers Underestimate Risks?}} {{Do Smokers
  Underestimate Risks?}}{\BBCQ}
\newblock
\APACjournalVolNumPages{Journal of Political Economy}{98}{6}{1253--1269}.
\PrintBackRefs{\CurrentBib}

\bibitem [\protect \citeauthoryear {%
von Hayek%
}{%
von Hayek%
}{%
{\protect \APACyear {1945}}%
}]{%
hayek1945use}
\APACinsertmetastar {%
hayek1945use}%
\begin{APACrefauthors}%
von Hayek, F\BPBI A.%
\end{APACrefauthors}%
\unskip\
\newblock
\APACrefYearMonthDay{1945}{}{}.
\newblock
{\BBOQ}\APACrefatitle {{The Use of Knowledge in Society}} {{The Use of
  Knowledge in Society}}.{\BBCQ}
\newblock
\APACjournalVolNumPages{The American Economic Review}{35}{4}{519--530}.
\PrintBackRefs{\CurrentBib}

\end{thebibliography}

\begin{center}
    \textbf{Note}: All URLs were last accessed on \today.
\end{center}

\appendix



\section{Data availability}
\label{availabilityknf}

Software, materials, data and analysis code used in this article are freely available at \censor{\url{https://gitlab.com/gr0ssmann/knf}}. We welcome attempts at replication and reproduction.

\section{Additional tables}

\begin{table}[H]
    \centering
    \begin{threeparttable}
        \begin{tabular}{ccccccccc}
            \toprule
            $k$ & Mean & Median & Mode & Var. & MAE & RMSE & $D_{\text{KL}}$ & $W_1$ \\
            \midrule
            0 & $0.500$ & $0.500$ & --- & $0.083$ & $0.340$ & $0.416$ & $0.000$ & $0.000$ \\
            1 & $0.400$ & $0.362$ & $0.000$ & $0.073$ & $0.261$ & $0.337$ & $0.062$ & $0.100$ \\
            2 & $0.350$ & $0.307$ & $0.000$ & $0.062$ & $0.222$ & $0.290$ & $0.148$ & $0.150$ \\
            5 & $0.286$ & $0.252$ & $0.083$ & $0.040$ & $0.167$ & $0.217$ & $0.338$ & $0.214$ \\
            10 & $0.250$ & $0.228$ & $0.163$ & $0.025$ & $0.128$ & $0.165$ & $0.526$ & $0.250$ \\
            25 & $0.222$ & $0.212$ & $0.187$ & $0.011$ & $0.086$ & $0.109$ & $0.852$ & $0.280$ \\
            50 & $0.212$ & $0.206$ & $0.194$ & $0.006$ & $0.062$ & $0.079$ & $1.152$ & $0.297$ \\
            1000 & $0.201$ & $0.200$ & $0.200$ & $0.000$ & $0.014$ & $0.018$ & $2.607$ & $0.330$ \\
            $\infty$ & $0.200$ & $0.200$ & $0.200$ & $0.000$ & $0.000$ & $0.000$ & $\infty$ & $0.340$ \\
            \bottomrule
        \end{tabular}
        \caption{Properties of the marginal posterior distribution of $p$}
        \label{margpost}
        \begin{tablenotes}
        \item MAE and RMSE are calculated as departures from $p = 0.2$.
        \item The Kullback–Leibler divergence and the Wasserstein 1-distance are calculated as differences between the marginal posterior and U$(0,1)$.
        \item U$(0,1)$ is the distribution used for $k = 0$.
        \end{tablenotes}
    \end{threeparttable}
\end{table}

\begin{table}
    \begin{center}
        \sisetup{parse-numbers=false, table-text-alignment=center}
        \resizebox{\textwidth}{!}{%
        \begin{threeparttable}
                \begin{tabular}{l S[table-format=3.6] S[table-format=3.6] S[table-format=3.6] S[table-format=3.6] S[table-format=3.6]}
                    \toprule
                    & {Model 1} & {Model 2} & {Model 3} & {Model 4} & {Model 5} \\
                    \midrule
                    Intercept                 & 0.210^{***} & 0.042^{*}         & 0.599^{***} & 0.732^{***}       & 0.829^{***} \\
                    & (0.017)     & (0.019)           & (0.018)     & (0.040)           & (0.018)     \\
                    CA prefers~1              & 0.094^{***} & 0.113^{***}       & 0.157^{***} & 0.220^{***}       & -0.012      \\
                    & (0.023)     & (0.033)           & (0.019)     & (0.041)           & (0.020)     \\
                    \midrule
                    Experiment                & {1}         & {1}               & {2}         & {2}               & {2}         \\
                    Outcome                   & {$\pi$}       & {${\pi > 0.5}$} & {$\pi$}       & {${\pi > 0.5}$} & {$\pi'$}       \\
                    Belief about choice under & {Risk}      & {Risk}            & {Risk}      & {Risk}            & {Ambiguity} \\
                    Standard errors           & {HC3}       & {HC3}             & {HC3}       & {HC3}             & {HC3}       \\
                    R$^2$                     & 0.048       & 0.031             & 0.142       & 0.093             & 0.001       \\
                    Adj. R$^2$                & 0.045       & 0.028             & 0.141       & 0.092             & -0.001      \\
                    Num. obs.                 &{300}        &{300}              &{600}        &{600}              &{600}        \\
                    \bottomrule
                \end{tabular}
                \begin{tablenotes}[flushleft]
                \scriptsize{\item $^{***}p<0.001$; $^{**}p<0.01$; $^{*}p<0.05$}
                \end{tablenotes}
        \end{threeparttable}}
        \caption{False consensus bias across experiments}
        \caption*{The binary outcome ${\pi > 0.5}$ indicates whether Option 1 is viewed to be the majority choice.}
        \label{tabbeliefs}
    \end{center}
\end{table}

\section{Experiment 1: Analyses}

\subsection{Transformation of independent variable}
\label{transform}

\begin{center}
    \begin{tabular}{@{}cccccccccc@{}}
        \toprule
        $k$             & 0 & 1 & 2 & 5 & 10 & 25 & 50 & 1000 & $\infty$ \\ \midrule
        $k_\text{rank}$ & 0 & 1 & 2 & 3 & 4  & 5  & 6  & 7    & 8        \\ \bottomrule
    \end{tabular}
\end{center}

\subsection{Main analysis}
\label{intlogitapp}

\begin{table}[H]
    \begin{center}
        \sisetup{parse-numbers=false, table-text-alignment=center}
        \resizebox{\textwidth}{!}{%
        \begin{tabular}{l S[table-format=5.6] S[table-format=5.6] S[table-format=4.6]}
            \toprule
            & {Model 1$^{\dagger}$} & {Model 2} & {Model 3} \\
            \midrule
            $k_{\text{rank}}$ & -0.188^{***} & -0.192^{***} & -0.139^{***}       \\
            & (0.019)      & (0.019)      & (0.040)            \\
            Round             &              & 0.043^{***}  & 0.057^{*}          \\
            &              & (0.014)      & (0.032)            \\
            \midrule
            Subset            & {---}        & {---}        & {Inconsistent CAs$^{\ddagger}$} \\
            Controls          & {No}         & {Yes}        & {Yes}              \\
            AIC               & 3518.186     & 3478.862     & 756.761            \\
            BIC               & 3529.988     & 3520.169     & 787.032            \\
            Log Likelihood    & -1757.093    & -1732.431    & -371.381           \\
            Deviance          & 3514.186     & 3464.862     & 742.761            \\
            Num. obs.         &{2700}        &{2700}        &{558}               \\
            \bottomrule
            \multicolumn{4}{l}{\scriptsize{$^{***}p<0.01$; $^{**}p<0.05$; $^{*}p<0.1$}}
        \end{tabular}}

        \caption{Logistic regressions on the extensive margin}
        \caption*{All standard errors are clustered by subject. $^{\dagger}$Original preregistered specification. $^{\ddagger}$Consistency is defined in the preregistration: “a [CA] is consistent if and only if they never intervened or if they imposed Option 1 OR Option 2, but never switched between the Options.”}
        \label{intlogit}
    \end{center}
\end{table}

Define by $\text{intervene}_{ij}$ a binary variable that equals ‘1’ if subject $i$ intervened in round $j$, and ‘0’ otherwise. We can run a logistic regression of $\text{intervene}_{ij}$ on $k_\text{rank}$, with standard errors clustered by subject. As Table \ref{intlogit} indicates,\footnote{Regression outputs courtesy of \texttt{texreg}, see \cite{texreg}.} there is a statistically significant reduction in $\text{intervene}_{ij}$ along with $k_\text{rank}$, and thus $k$. This holds both for the “naked”, preregistered, regression (Model 1) as well as when including the round and control variables.\footnote{Control variables are the university entrance exam grade or a dummy variable if it was not provided to us, a dummy indicating whether the subject ever took a class on introductory microeconomics and a dummy indicating whether the subject identified as male.}

\subsection{Beliefs and the intensive margin}

\begin{table}[H]
    \begin{center}
        \sisetup{parse-numbers=false, table-text-alignment=center}
        \begin{threeparttable}
            \resizebox{\textwidth}{!}{%
                \begin{tabular}{l S[table-format=3.6] S[table-format=3.6] S[table-format=3.6] S[table-format=3.6] S[table-format=3.6] S[table-format=3.6] S[table-format=3.6] S[table-format=3.6] S[table-format=3.6]}
                    \toprule
                    & {Model 1} & {Model 2} & {Model 3} & {Model 4} & {Model 5} & {Model 6} & {Model 7} & {Model 8} & {Model 9} \\
                    \midrule
                    Intercept             & 0.512^{***} & 0.484^{***} & 0.503^{***} & 0.234^{***} & 0.223^{***} & 0.222^{***} & 0.223       & 0.219       & 0.213       \\
                    & (0.038)     & (0.041)     & (0.045)     & (0.050)     & (0.050)     & (0.057)     & (0.145)     & (0.145)     & (0.146)     \\
                    $\pi$ (standardized) & 0.087^{*}   &             & 0.067       & 0.025       &             & -0.001      & 0.018       &             & -0.017      \\
                    & (0.037)     &             & (0.067)     & (0.031)     &             & (0.063)     & (0.031)     &             & (0.067)     \\
                    $\pi > 0.5$     &             & 0.253^{*}   & 0.079       &             & 0.099       & 0.101       &             & 0.089       & 0.131       \\
                    &             & (0.114)     & (0.210)     &             & (0.092)     & (0.192)     &             & (0.092)     & (0.200)     \\
                    CA prefers~1          &             &             &             & 0.492^{***} & 0.493^{***} & 0.493^{***} & 0.486^{***} & 0.484^{***} & 0.488^{***} \\
                    &             &             &             & (0.070)     & (0.069)     & (0.071)     & (0.073)     & (0.072)     & (0.073)     \\
                    \midrule
                    Outcome               & {Imposed 1} & {Imposed 1} & {Imposed 1} & {Imposed 1} & {Imposed 1} & {Imposed 1} & {Imposed 1} & {Imposed 1} & {Imposed 1} \\
                    Subset                & {c.n.l.}    & {c.n.l.}    & {c.n.l.}    & {c.n.l.}    & {c.n.l.}    & {c.n.l.}    & {c.n.l.}    & {c.n.l.}    & {c.n.l.}    \\
                    Controls              & {No}        & {No}        & {No}        & {No}        & {No}        & {No}        & {Yes}       & {Yes}       & {Yes}       \\
                    R$^2$                 & 0.030       & 0.025       & 0.031       & 0.254       & 0.255       & 0.255       & 0.272       & 0.273       & 0.274       \\
                    Adj. R$^2$            & 0.025       & 0.019       & 0.020       & 0.245       & 0.246       & 0.242       & 0.245       & 0.247       & 0.243       \\
                    Num. obs.             &{172}        &{172}        &{172}        &{172}        &{172}        &{172}        &{172}        &{172}        &{172}        \\
                    \bottomrule
            \end{tabular}}
            \begin{tablenotes}[flushleft]
                \scriptsize{\item $^{***}p<0.001$; $^{**}p<0.01$; $^{*}p<0.05$%
                \item c.n.l. refers to “consistent non-libertarians.”}
            \end{tablenotes}
        \end{threeparttable}
        \caption{Regressions of intensive margin on CA beliefs and preferences}
        \label{t1_2}
    \end{center}
\end{table}

\section{Experiment 2}

\subsection{Relative importance of beliefs and preferences}\label{t2_1asec}

\begin{table}[h]
    \begin{center}
        \sisetup{parse-numbers=false, table-text-alignment=center}
        \begin{tabular}{l S[table-format=3.6]}
            \toprule
            & {Model} \\
            \midrule
            Intercept       & 0.115        \\
            & (0.150)      \\
            $\pi$       & 0.544^{**}   \\
            & (0.224)      \\
            CA prefers~1    & 0.263^{***}  \\
            & (0.096)      \\
            \midrule
            Outcome         & {Imposed 1}  \\
            Subset          & {Intervened} \\
            Standard errors & {HC3}        \\
            R$^2$           & 0.164        \\
            Adj. R$^2$      & 0.151        \\
            Num. obs.       &{132}         \\
            \bottomrule
            \multicolumn{2}{l}{\scriptsize{$^{***}p<0.01$; $^{**}p<0.05$; $^{*}p<0.1$}}
        \end{tabular}
        \caption{Regression explaining the intensive margin with CA preference and raw beliefs}
        \label{t2_1a}
    \end{center}
\end{table}

Let $\pi_{1}$ be the belief of an Option 1-preferring CA, and $\pi_{2}$ the belief of an Option 2-preferring CA.

We can use the coefficients of Table \ref{t2_1a} to find the $\pi_{2}$ that solves $0.115 + 0.544 \pi_{2} = 0.115 + 0.544 \pi_{1} + 0.263$. This would be the belief of the Option 2-preferring CA required to attain the same probability of imposing Option 1 as when the CA prefers Option 1. We find $\pi_{2} \approx 0.484 + \pi_{1}$. The average $\pi_{1}$ in our sample (Table \ref{tabbeliefs}) is $0.756$. Obviously, such a large $\pi_{2}$ would not be achievable (since it exceeds 1). The counterfactual Option 1-preferring CA would have to have much lower beliefs. Thus, it is difficult to achieve parity through beliefs alone. Beliefs are weak predictors of interventions when compared to CAs' preferences.

\subsection{Performance of binary-choice model}
\label{knfmle}

Predictions (i.e., fitted values) for the probability of the outcome “Imposed (Option) 1” can be obtained from model 5 in Table \ref{t2_1} and model 2 in Table \ref{t2_1b}. Call these predictions $\widehat{v}_{1,i}$, with $i$ indicating the CA's index. Similarly, we can calculate
\begin{equation}
    \widehat{v}_{2,i} = \Phi \left( \frac{W_{\theta_i}(1) - W_{\theta_i}(2)}{\sigma} \right)
\end{equation}
for each CA after maximization of Equation \ref{knfllik}.

The threshold $1/2$ is then used to form categorical predictions. If $\widehat{v}_{m,i} \geq 1/2$ for any $m, i$, we predict “Imposed 1.” Otherwise, we predict “Imposed 2.” We can classify whether the predictions are correct by considering the actual outcome. A comparison of these classifiers reveals the following result for Chooser 1 (model 1 is model 5 of Table \ref{t2_1}):

\begin{table}[H]
\begin{tabular}{@{}lccc@{}}
    \toprule
    & Model 2 wrong & Model 2 correct & \textbf{Sum} \\ \midrule
    Model 1 wrong & 32 & 4 & 36 \\
    Model 1 correct & 2 & 94 & 96 \\
    \textbf{Sum} & 34 & 98 & 132 \\ \bottomrule
\end{tabular}
\end{table}

For Chooser 5 (model 1 is model 2 of Table \ref{t2_1b}):

\begin{table}[H]
\begin{tabular}{@{}lccc@{}}
    \toprule
    & Model 2 wrong & Model 2 correct & \textbf{Sum} \\ \midrule
    Model 1 wrong & 50 & 23 & 73 \\
    Model 1 correct & 24 & 91 & 115 \\
    \textbf{Sum} & 74 & 114 & 188 \\ \bottomrule
\end{tabular}
\end{table}

Clearly, both our binary-choice models and the linear probability models perform about equally well. In both cases, correlations between the raw values of $\widehat{v}_{m,i}$ exceed 0.9. Similar results can be obtained by running logistic regressions in place of the linear probability models. Our data package provides code and data for these analyses.

\subsection{Complementarities}
\label{complementarities}

\begin{table}[H]
    \begin{center}
        \sisetup{parse-numbers=false, table-text-alignment=center}
        \resizebox{\textwidth}{!}{%
            \begin{tabular}{l S[table-format=3.6] S[table-format=3.6] S[table-format=3.6] S[table-format=3.6] S[table-format=3.6] S[table-format=3.6] S[table-format=3.6] S[table-format=3.6] S[table-format=3.6]}
                \toprule
                & {Model 1} & {Model 2} & {Model 3} & {Model 4} & {Model 5} & {Model 6} & {Model 7} & {Model 8} & {Model 9} \\
                \midrule
                Intercept          & 0.635^{***} & 0.819^{***} & 0.522^{***} & 0.544^{***} & 0.699^{***} & 0.914^{***} & 0.598^{***} & 0.882^{***} & 0.653^{***} \\
                & (0.034)     & (0.037)     & (0.033)     & (0.036)     & (0.057)     & (0.059)     & (0.055)     & (0.057)     & (0.064)     \\
                Int. for Chooser~1 & 1.312^{***} &             &             &             & 1.442^{***} &             &             &             &             \\
                & (0.078)     &             &             &             & (0.128)     &             &             &             &             \\
                Int. for Chooser~2 &             & 1.354^{***} &             &             &             & 1.670^{***} &             &             &             \\
                &             & (0.124)     &             &             &             & (0.212)     &             &             &             \\
                Int. for Chooser~3 &             &             & 1.281^{***} &             &             &             & 1.446^{***} &             &             \\
                &             &             & (0.066)     &             &             &             & (0.113)     &             &             \\
                Int. for Chooser~4 &             &             &             &             &             &             &             & 1.761^{***} &             \\
                &             &             &             &             &             &             &             & (0.212)     &             \\
                Int. for Chooser~5 &             &             &             & 1.212^{***} &             &             &             &             & 1.270^{***} \\
                &             &             &             & (0.066)     &             &             &             &             & (0.111)     \\
                \midrule
                Outcome            & {IC$_{-4}$} & {IC$_{-4}$} & {IC$_{-4}$} & {IC$_{-4}$} & {IC$_{+4}$} & {IC$_{+4}$} & {IC$_{+4}$} & {IC$_{+4}$} & {IC$_{+4}$} \\
                Subset             & {---}       & {---}       & {---}       & {---}       & {Plus}      & {Plus}      & {Plus}      & {Plus}      & {Plus}      \\
                Standard errors    & {HC3}       & {HC3}       & {HC3}       & {HC3}       & {HC3}       & {HC3}       & {HC3}       & {HC3}       & {HC3}       \\
                R$^2$              & 0.345       & 0.151       & 0.412       & 0.368       & 0.340       & 0.187       & 0.398       & 0.239       & 0.307       \\
                Adj. R$^2$         & 0.344       & 0.150       & 0.411       & 0.367       & 0.337       & 0.184       & 0.396       & 0.236       & 0.304       \\
                Num. obs.          &{600}        &{600}        &{600}        &{600}        &{290}        &{290}        &{290}        &{290}        &{290}        \\
                \bottomrule
                \multicolumn{10}{l}{\scriptsize{$^{***}p<0.01$; $^{**}p<0.05$; $^{*}p<0.1$}}
        \end{tabular}}
        \caption{Regression of intervention counts on intervention behavior}
        \label{t2_5}
    \end{center}
\end{table}

We can define the following variable:

\begin{equation}\label{eqic}
    IC_{\mathbb{S}} \equiv \sum_{i \in \mathbb{S}} \left[ \text{Intervened for Chooser $i$} \right].
\end{equation}

This sum of indicators is essentially a counter. Table \ref{t2_5} presents regressions of Equation \ref{eqic} on the individual indicators, with $\mathbb{S} = +4 = \{1, 2, 3, 4, 5\}$, $\mathbb{S} = -4 = \{1, 2, 3, 5\}$. Obviously, each coefficient must be at least $1$ if the Chooser on the right-hand side of the regression equation is included in $\mathbb{S}$. To the extent that the coefficient exceeds $1$, this hints at a general pattern where those who intervene for a Chooser are more likely to intervene for \emph{other} Choosers, too.

\subsection{Preregistered analyses}

\subsubsection{Correction to preregistration}
\label{precorrect}

In the preregistration, analysis A4 was specified as follows:

\begin{quotation}
    \texttt{glm(I(intervene\_known == 1) $\sim$ I(own\_preference == 1) + I(other == 1), family = binomial, data = data)}
\end{quotation}

However, as is clear from the context (“This is to test for the relative importance of knowing the other's counterfactual perfect-knowledge choice.”) this analysis relates to the intensive margin. Hence, the data used for analysis must be restricted to interveners. The corrected code used below is

\begin{quotation}
    \texttt{... data = data[data\$intervene\_known > 0, ])}
\end{quotation}

\subsubsection{A1}
\label{e2a1}

Out of 600 CAs (Section \ref{exp2proc}), 290 are in treatment “Plus.” 310 are in the baseline.

Out of the 290 CAs in “Plus,” 236 (81.3\%) provided information.

Out of the 310 CAs in the baseline, 249 (80.3\%) provided information.

Using a one-sided test of equal proportions (the alternative being that fewer CAs provide information in “Plus”), we find $\chi^2 = 0.0506$, $p = 0.59$ (95\% confidence interval for the difference: $[-1, 0.067]$).

\subsubsection{A2}
\label{e2a2}

\begin{center}
    \sisetup{parse-numbers=false, table-text-alignment=center}
    \begin{tabular}{l S[table-format=4.6]}
        \toprule
        & {Model 1} \\
        \midrule
        (Intercept)    & -1.266^{***} \\
        & (0.099)      \\
        Full Knowledge & -1.223^{***} \\
        & (0.180)      \\
        \midrule
        AIC            & 960.952      \\
        BIC            & 971.132      \\
        Log Likelihood & -478.476     \\
        Deviance       & 956.952      \\
        Num. obs.      &{1200}        \\
        \bottomrule
        \multicolumn{2}{l}{\scriptsize{$^{***}p<0.01$; $^{**}p<0.05$; $^{*}p<0.1$}}
    \end{tabular}
\end{center}

Here, $z = -1.223/0.180 = -6.79$, $p < 0.001$.

\subsubsection{A3}
\label{e2a3}

\begin{center}
    \sisetup{parse-numbers=false, table-text-alignment=center}
    \begin{tabular}{l S[table-format=4.6]}
        \toprule
        & {Model 1} \\
        \midrule
        (Intercept)              & -1.035^{***} \\
        & (0.108)      \\
        Int. for Chooser 1   & 0.706^{***}  \\
        & (0.205)      \\
        Int. for Chooser 2 & 0.942^{***}  \\
        & (0.313)      \\
        \midrule
        AIC                      & 731.148      \\
        BIC                      & 744.339      \\
        Log Likelihood           & -362.574     \\
        Deviance                 & 725.148      \\
        Num. obs.                &{600}         \\
        \bottomrule
        \multicolumn{2}{l}{\scriptsize{$^{***}p<0.01$; $^{**}p<0.05$; $^{*}p<0.1$}}
    \end{tabular}
\end{center}

A Wald test on equality of the two non-Intercept independent variables reveals $\chi^2 = 0.4$. With 1 degree of freedom, $p = 0.53$.

\subsubsection{A4}
\label{e2a4}

\emph{Note the correction in Section \ref{precorrect}.}

\begin{center}
    \sisetup{parse-numbers=false, table-text-alignment=center}
    \begin{tabular}{l S[table-format=4.6]}
        \toprule
        & {Model 1} \\
        \midrule
        (Intercept)              & -0.571^{*}  \\
        & (0.343)     \\
        CA prefers 1 & 0.557       \\
        & (0.354)     \\
        Chooser prefers 1           & 1.094^{***} \\
        & (0.321)     \\
        \midrule
        AIC                      & 249.466     \\
        BIC                      & 259.175     \\
        Log Likelihood           & -121.733    \\
        Deviance                 & 243.466     \\
        Num. obs.                &{188}        \\
        \bottomrule
        \multicolumn{2}{l}{\scriptsize{$^{***}p<0.01$; $^{**}p<0.05$; $^{*}p<0.1$}}
    \end{tabular}
\end{center}

A Wald test on equality of the two non-Intercept independent variables reveals $\chi^2 = 1.5$. With 1 degree of freedom, $p = 0.21$.

\section{Instructions}

\emph{Note}: The following screenshots represent English translations of experiments that were wholly (experiment 1) or partially (experiment 2) conducted in German.

\subsection{Experiment 1}
\label{app.instr}

\subsubsection*{Screen 1}

(Consent form.)

\subsubsection*{Screen 2}

\begin{center}
    \includegraphics[width=\textwidth]{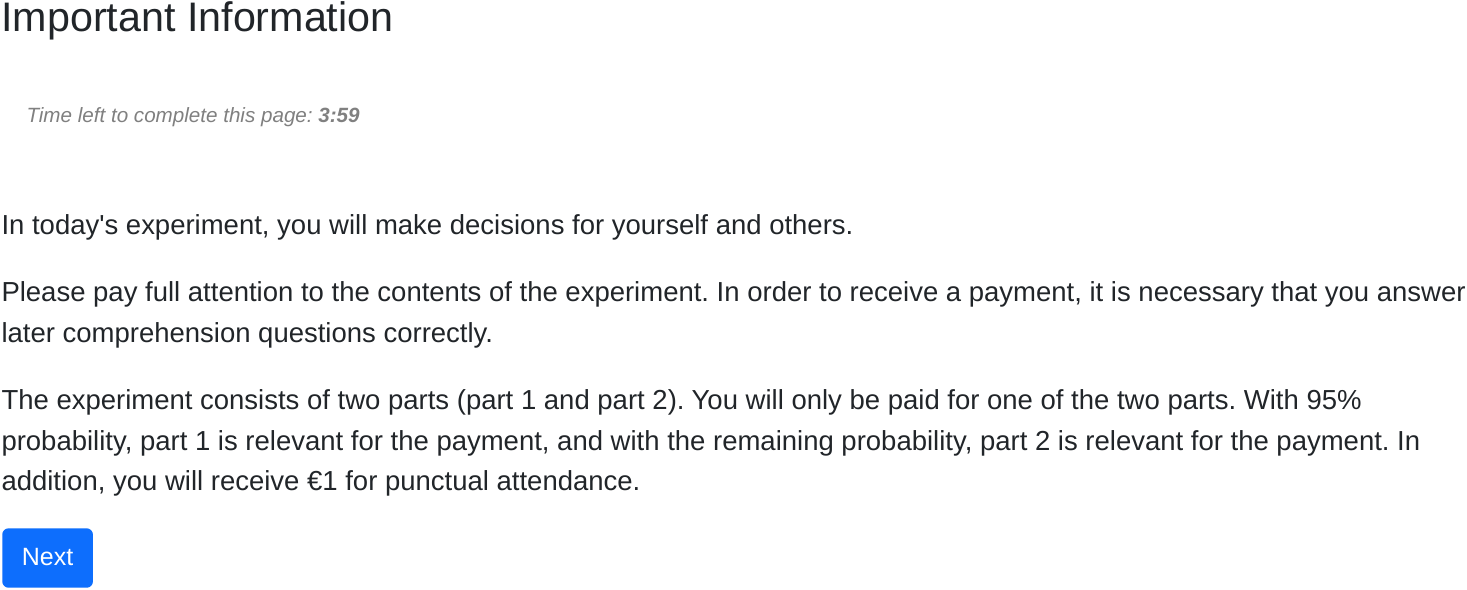}
\end{center}

\subsubsection*{Screen 3}

\begin{center}
    \includegraphics[width=\textwidth]{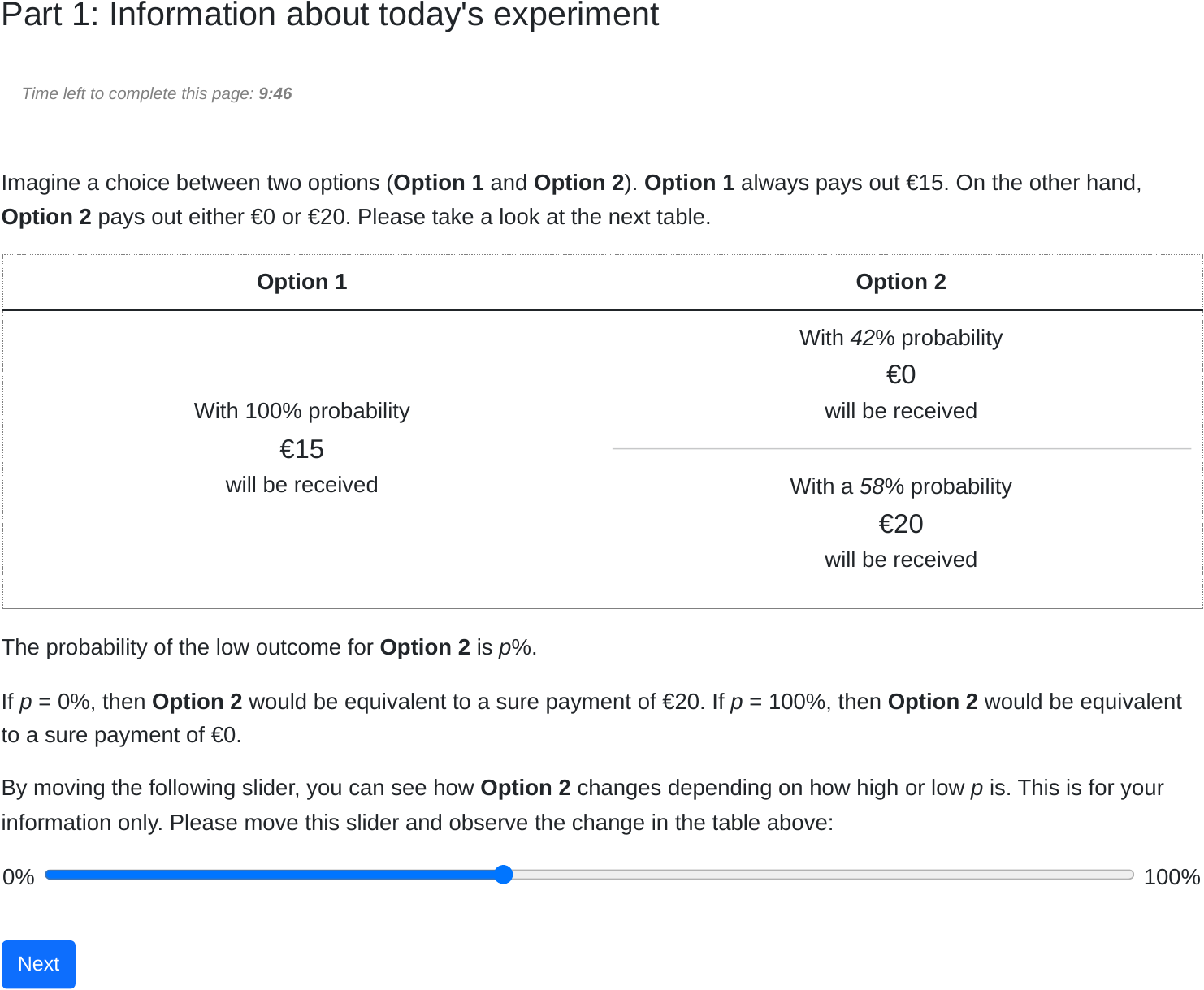}
\end{center}

\subsubsection*{Screen 4}

\begin{center}
    \includegraphics[width=0.5\textwidth]{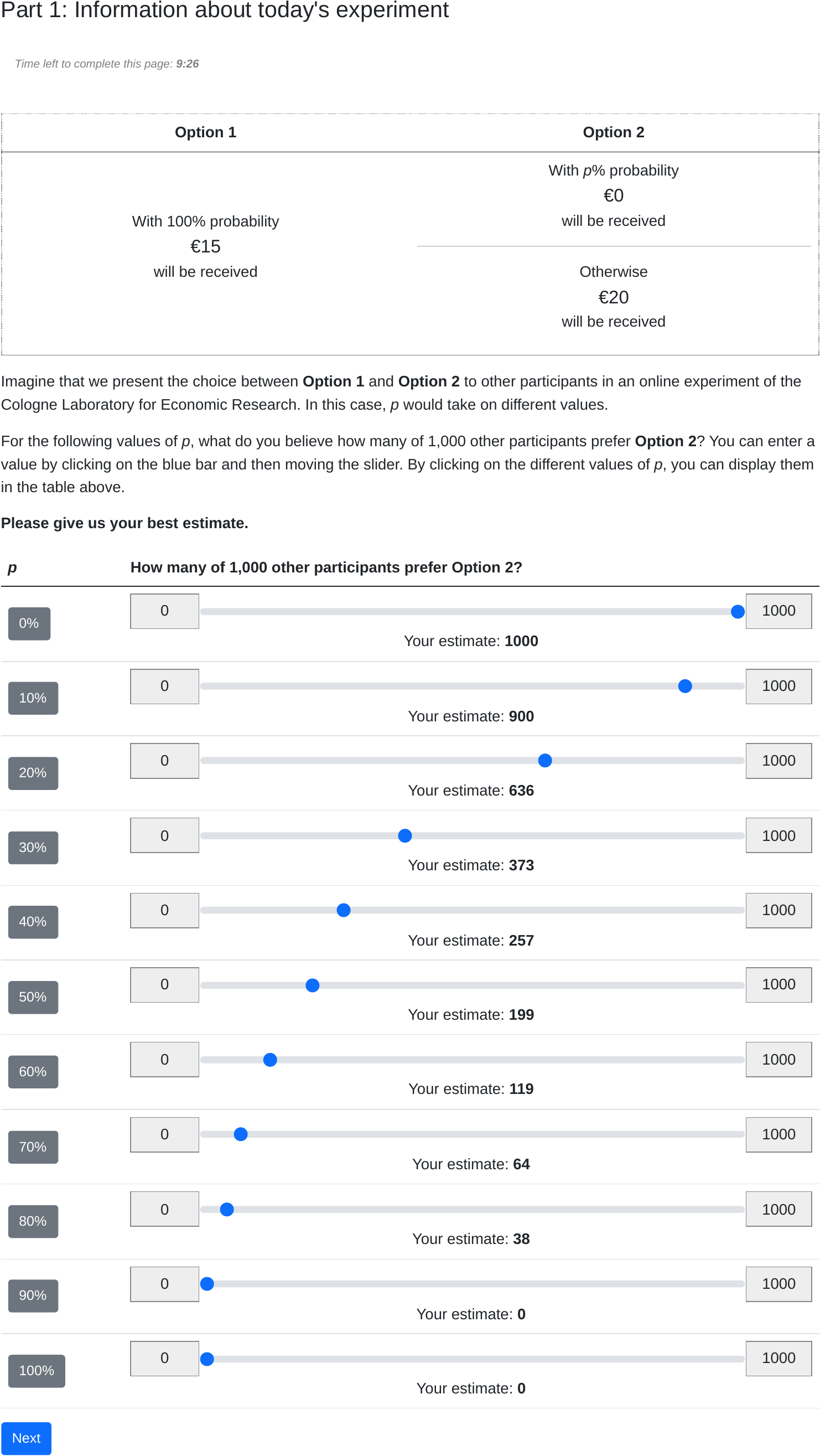}
\end{center}

\subsubsection*{Screen 5}

\begin{center}
    \includegraphics[width=\textwidth]{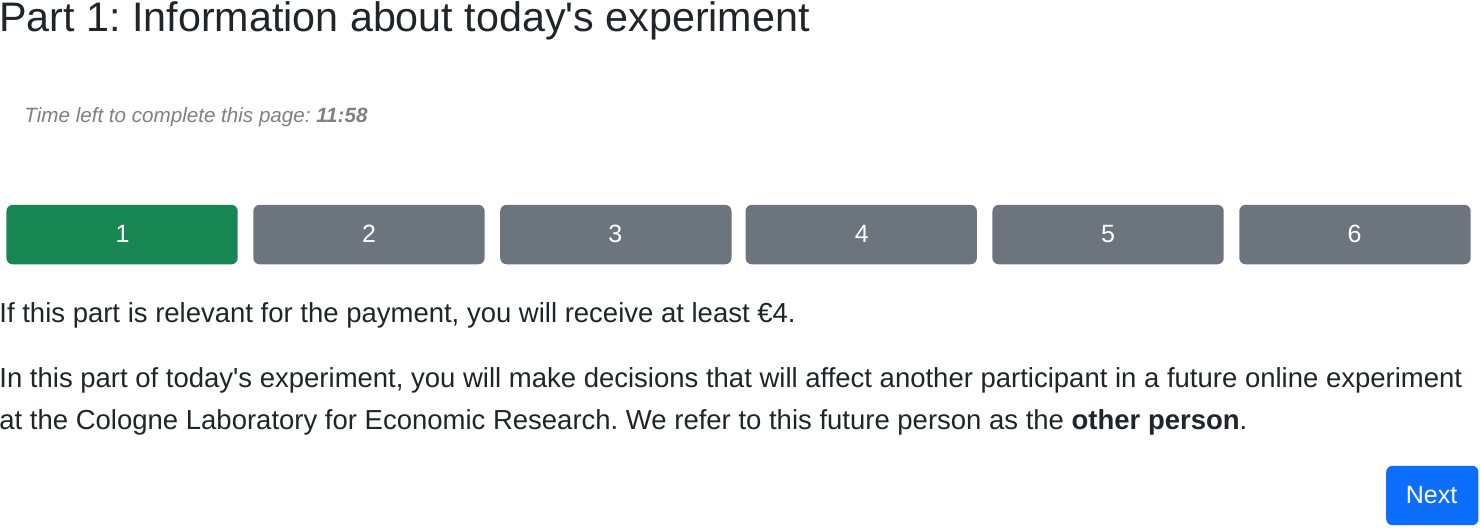}
\end{center}

\subsubsection*{Screen 6}

\begin{center}
    \includegraphics[width=\textwidth]{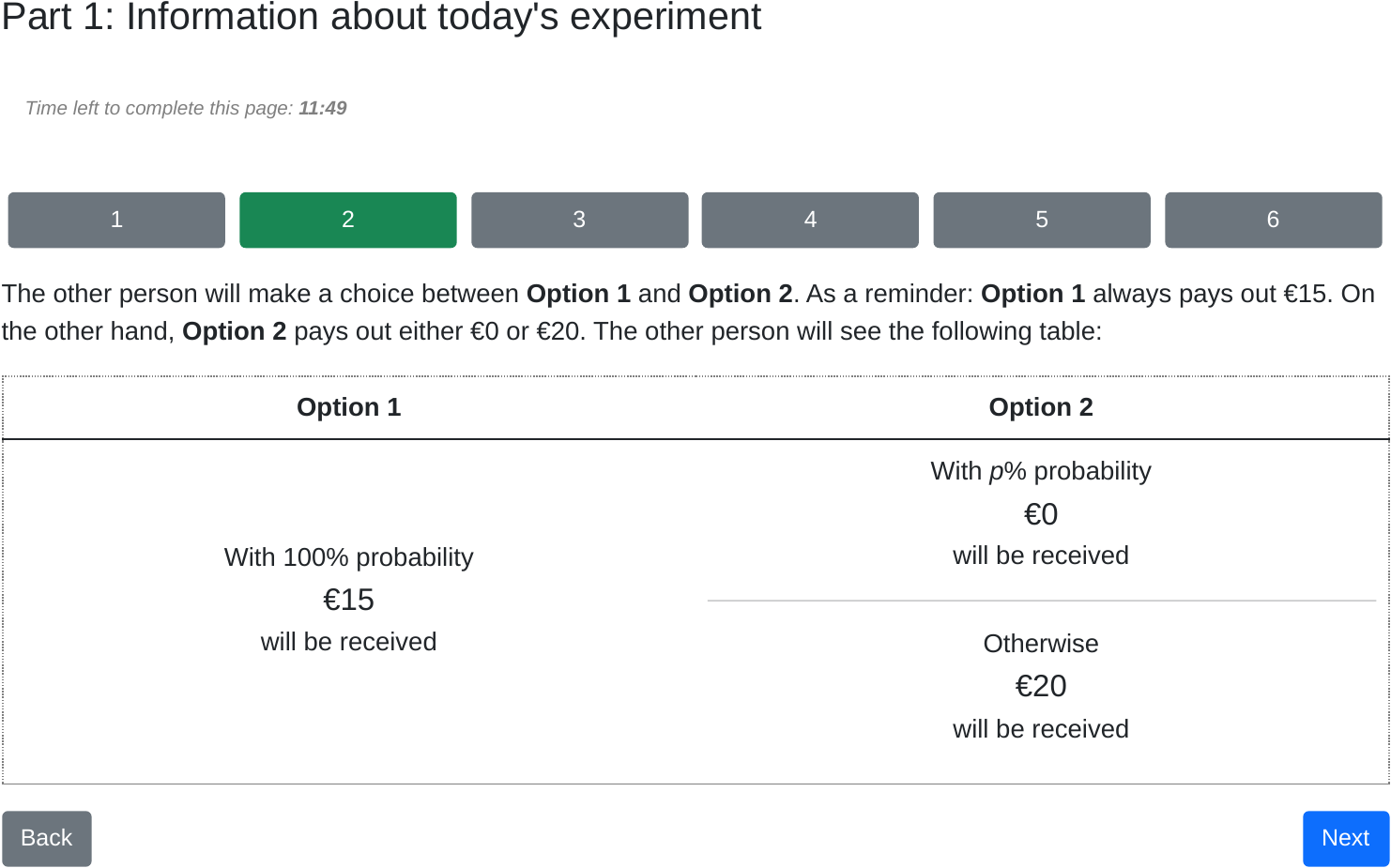}
\end{center}

\subsubsection*{Screen 7}

\begin{center}
    \includegraphics[width=\textwidth]{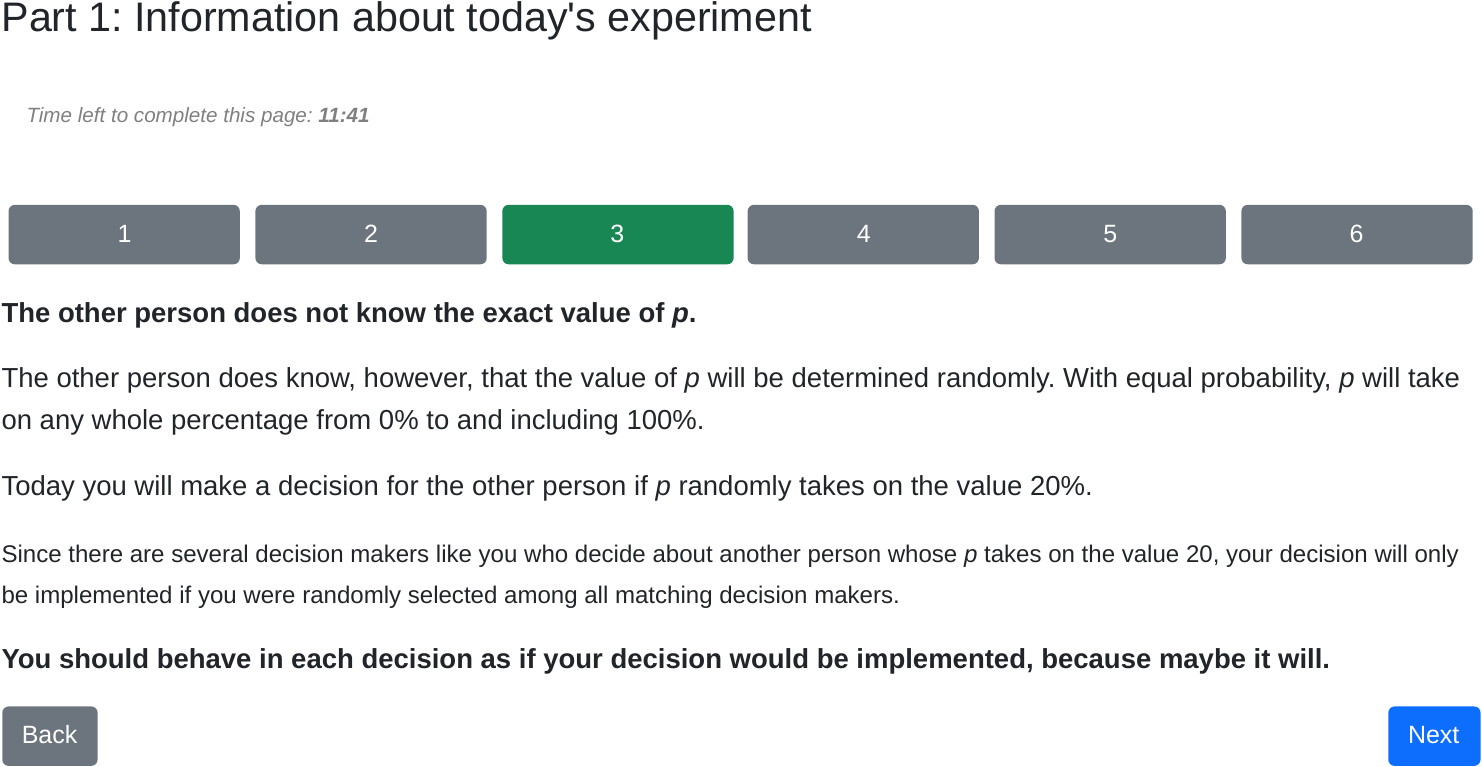}
\end{center}

\subsubsection*{Screen 8}

\begin{center}
    \includegraphics[width=\textwidth]{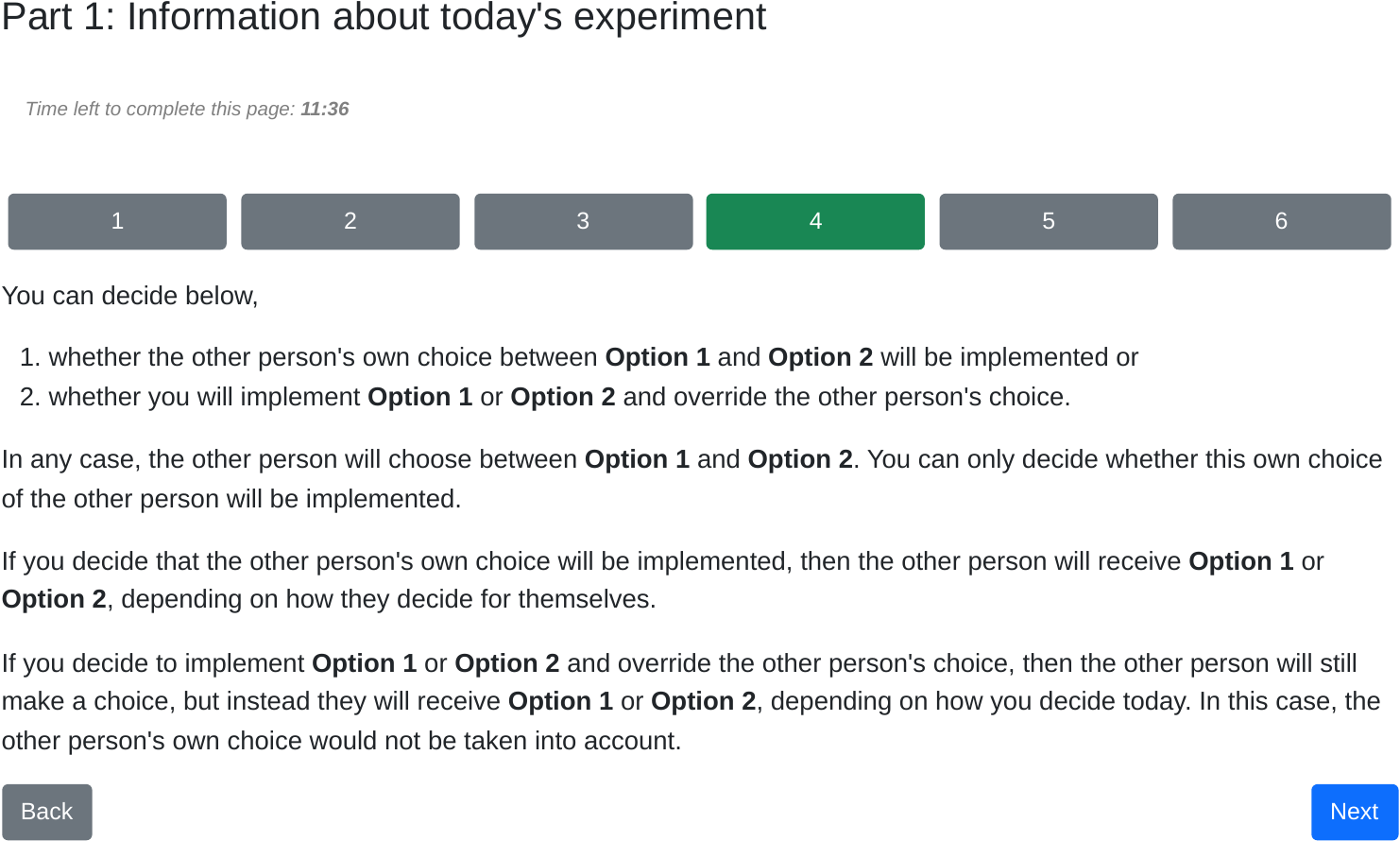}
\end{center}

\subsubsection*{Screen 9}

\begin{center}
    \includegraphics[width=\textwidth]{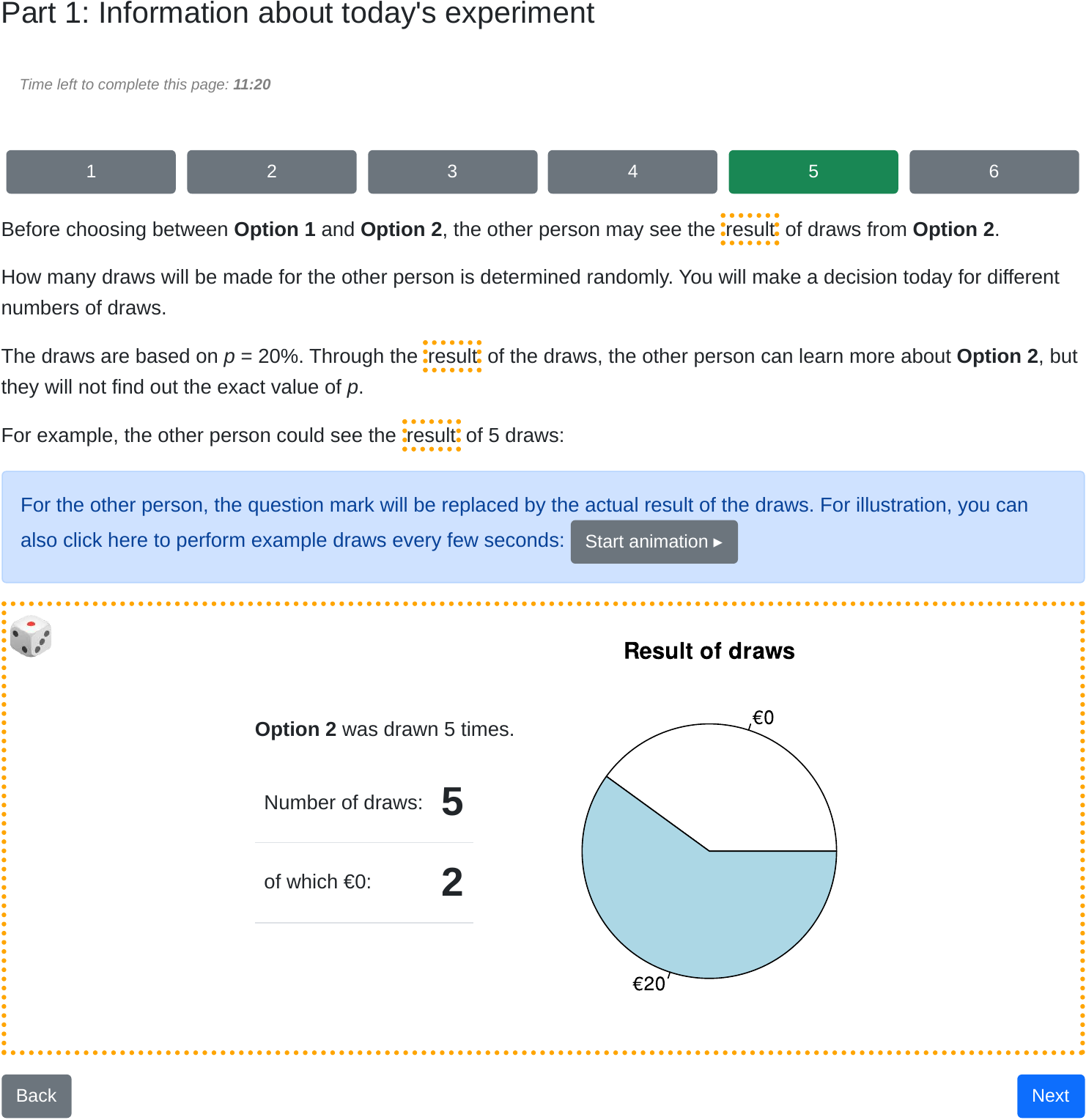}
\end{center}

\subsubsection*{Screen 10}

\begin{center}
    \includegraphics[width=\textwidth]{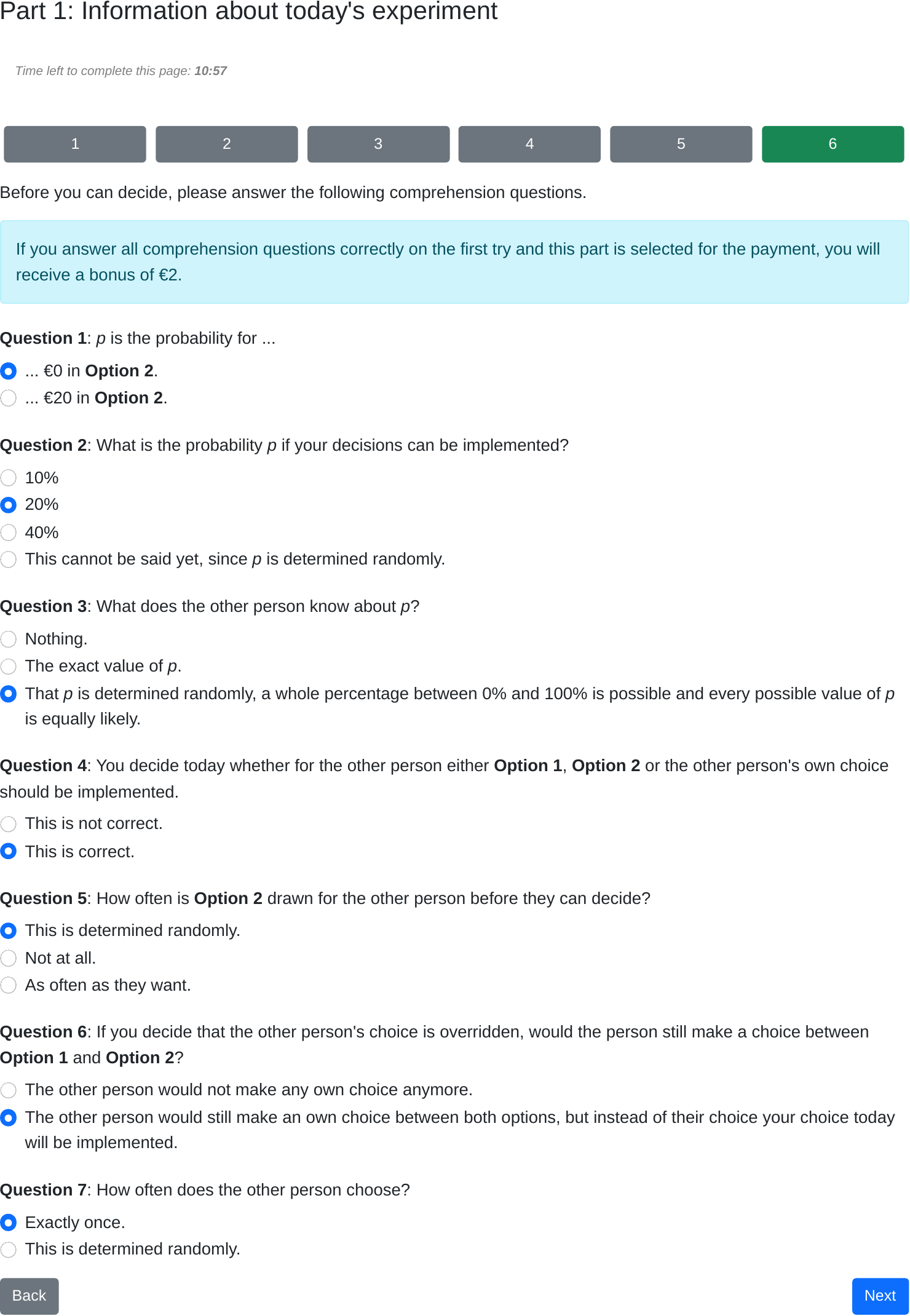}
\end{center}

\subsubsection*{Screen 11}

\begin{center}
    \includegraphics[width=\textwidth]{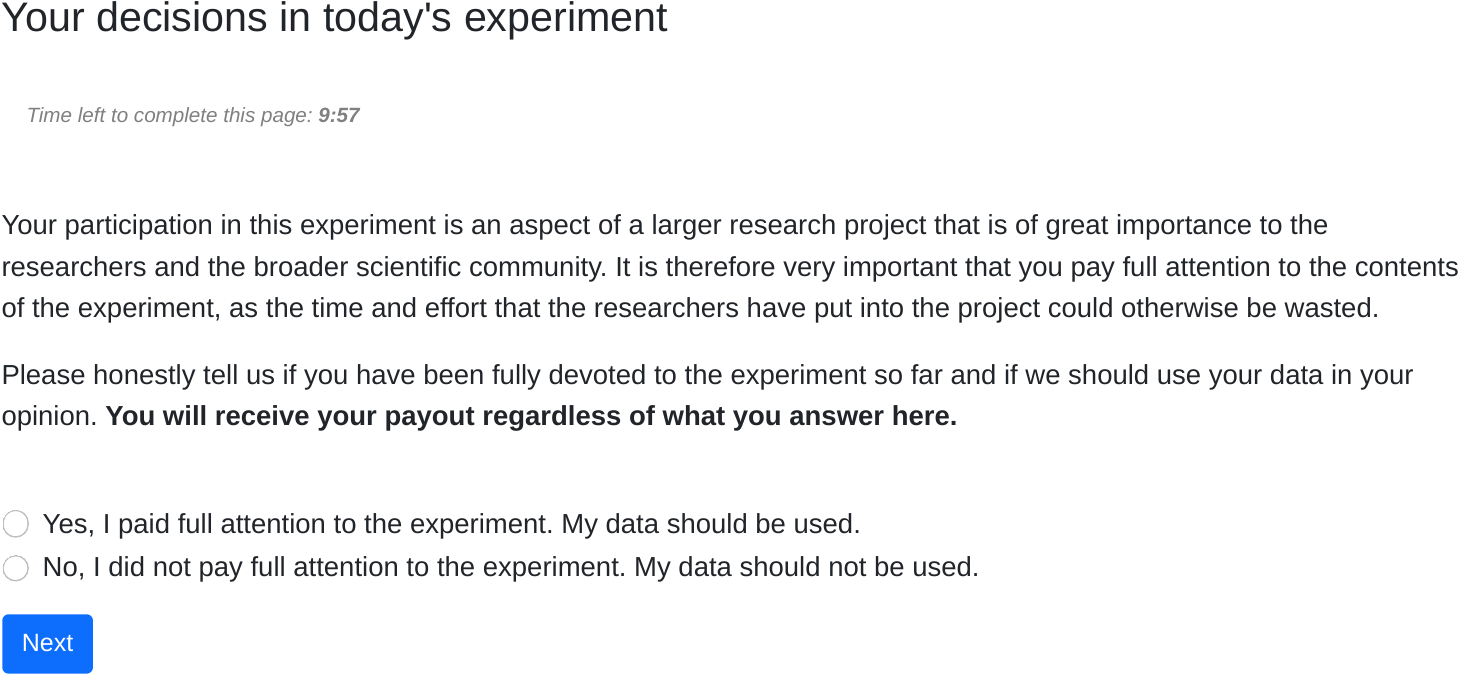}
\end{center}

\subsubsection*{Screen 12}

\begin{center}
    \includegraphics[width=\textwidth]{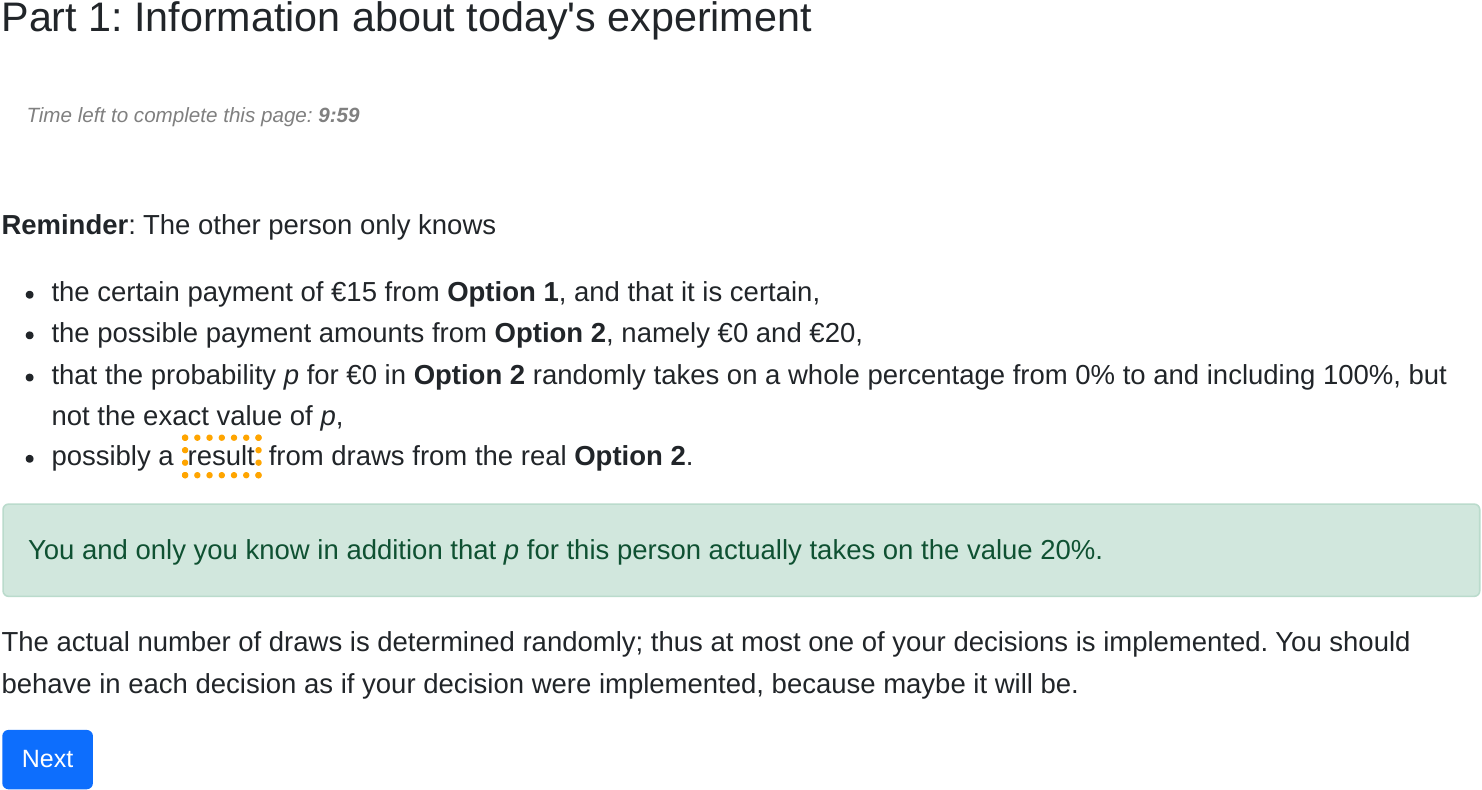}
\end{center}

\subsubsection*{Screen 13}

\begin{center}
    \includegraphics[width=\textwidth]{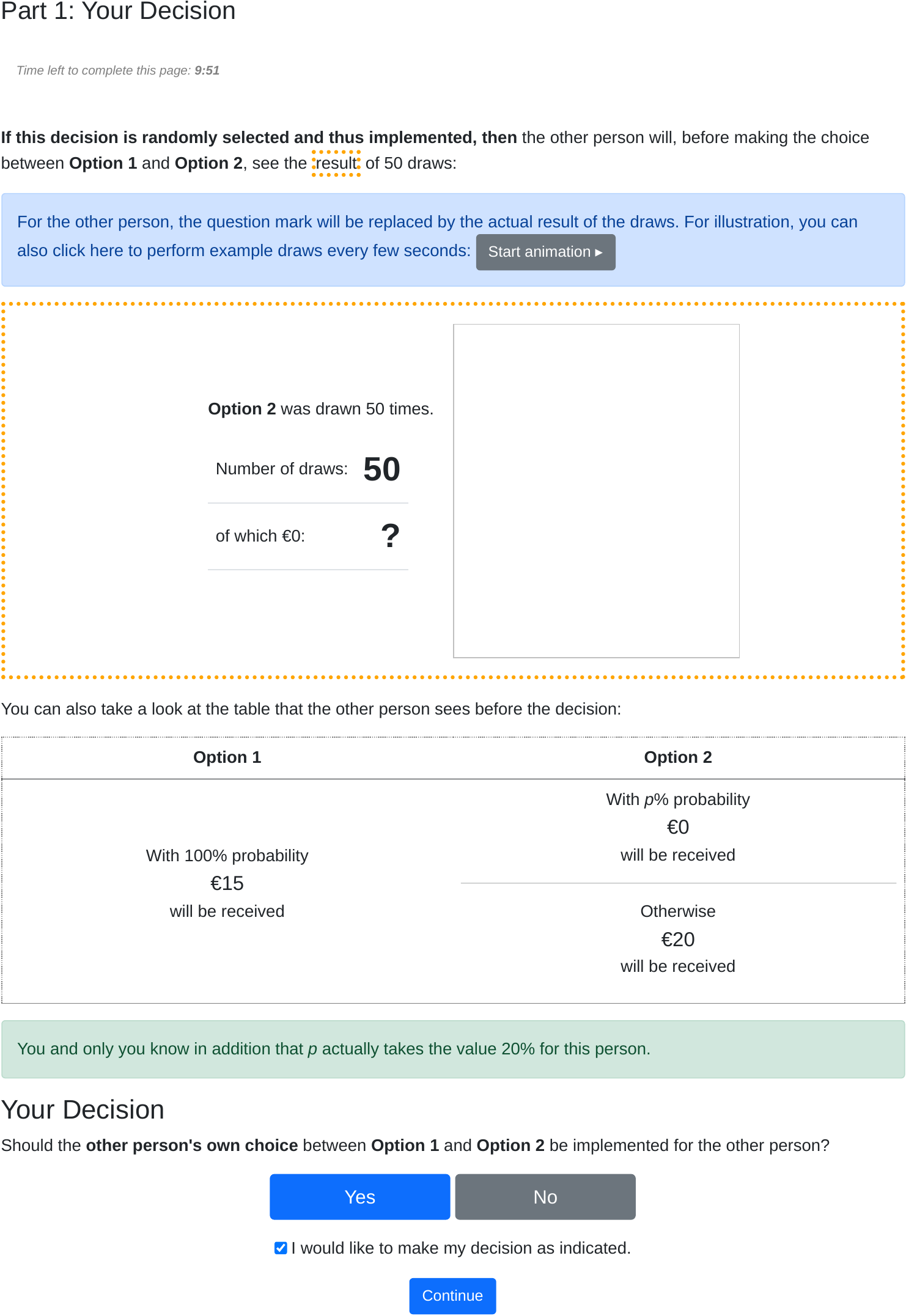}
\end{center}

\subsubsection*{Screen 14}

\begin{center}
    \includegraphics[width=\textwidth]{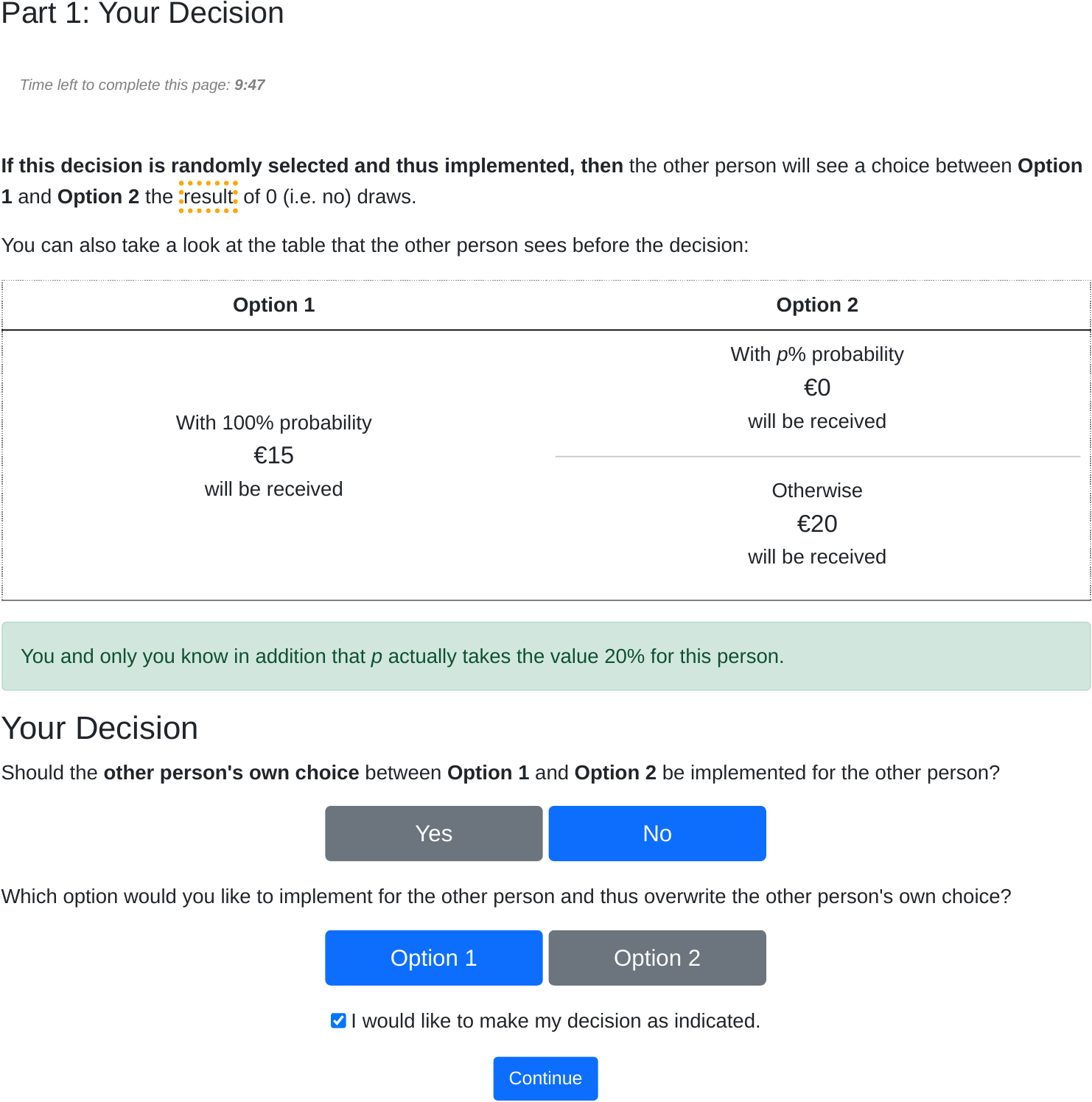}
\end{center}

\subsubsection*{Screen 15}

\begin{center}
    \includegraphics[width=\textwidth]{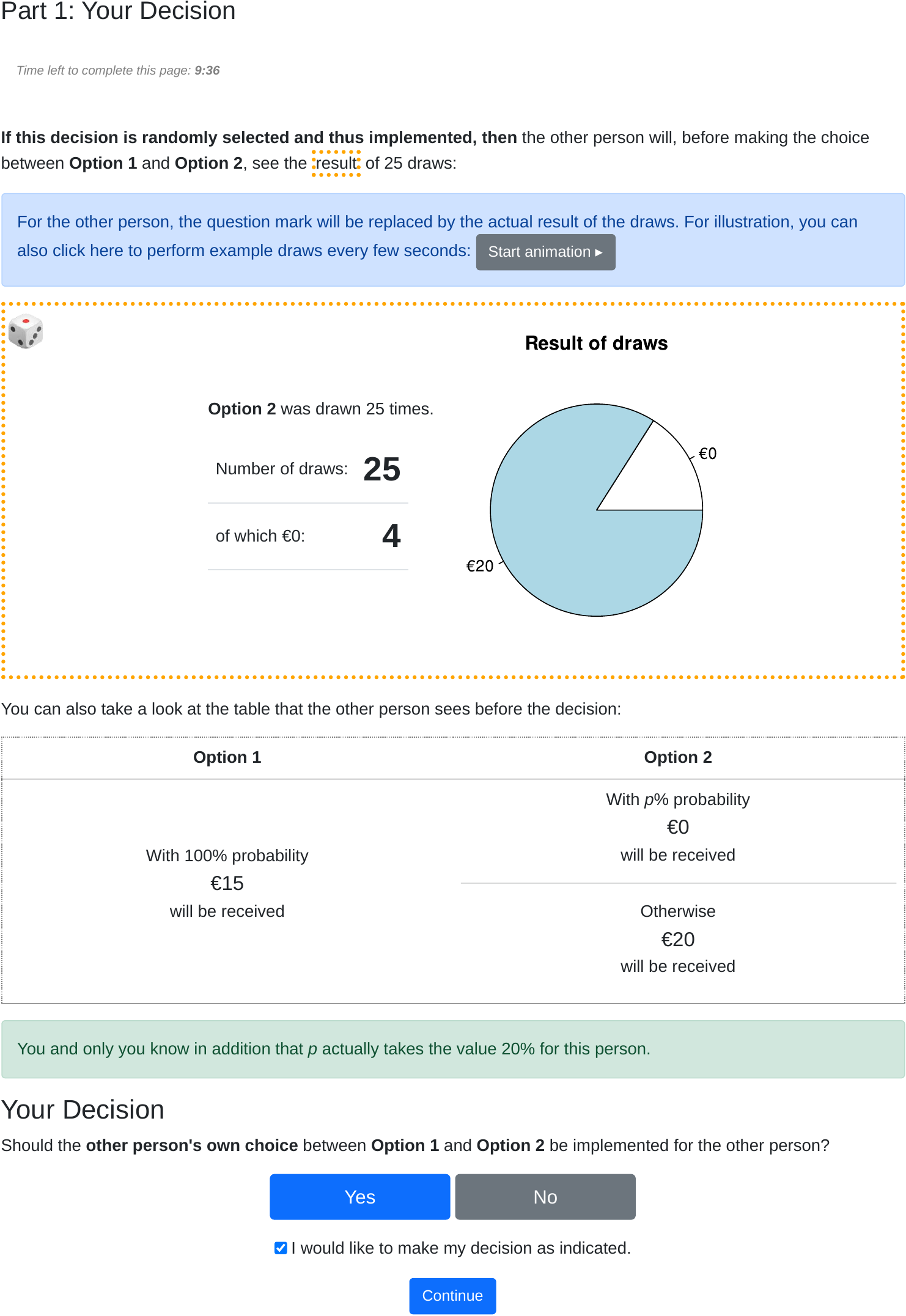}
\end{center}

\subsubsection*{Screen 16}

\begin{center}
    \includegraphics[width=\textwidth]{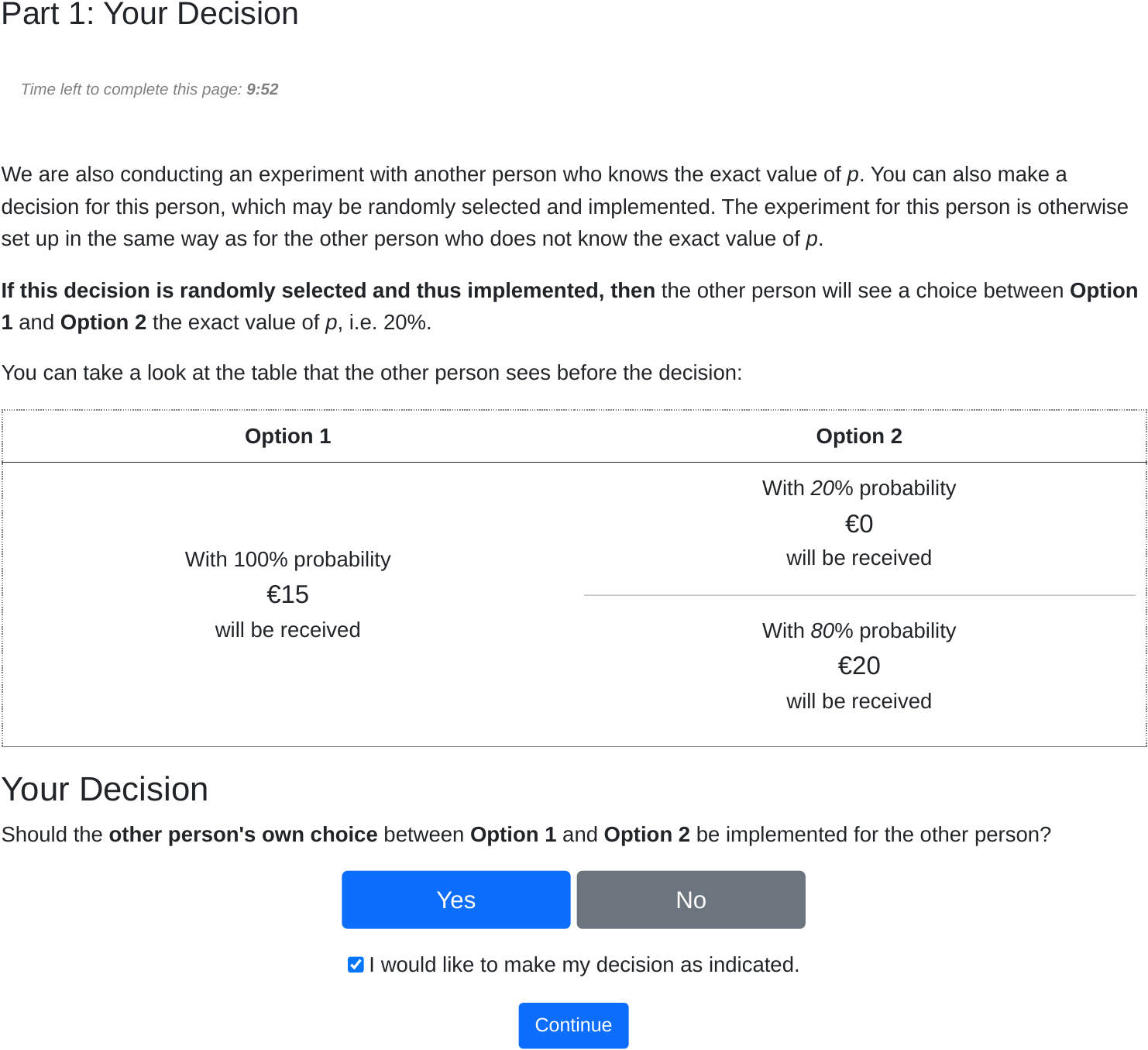}
\end{center}

\subsubsection*{Screen 17}

\begin{center}
    \includegraphics[width=\textwidth]{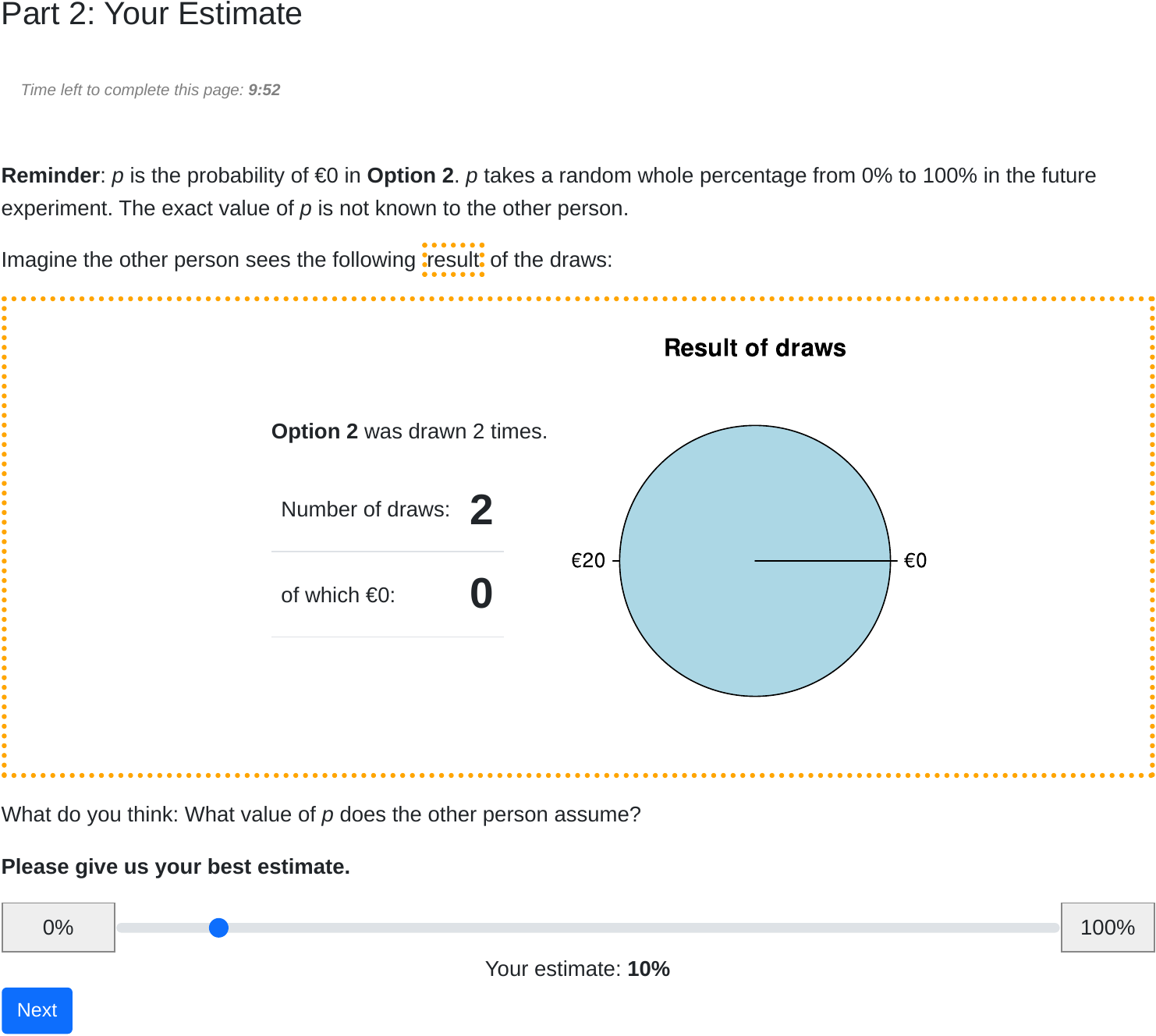}
\end{center}

\subsubsection*{Screen 18}

\begin{center}
    \includegraphics[width=\textwidth]{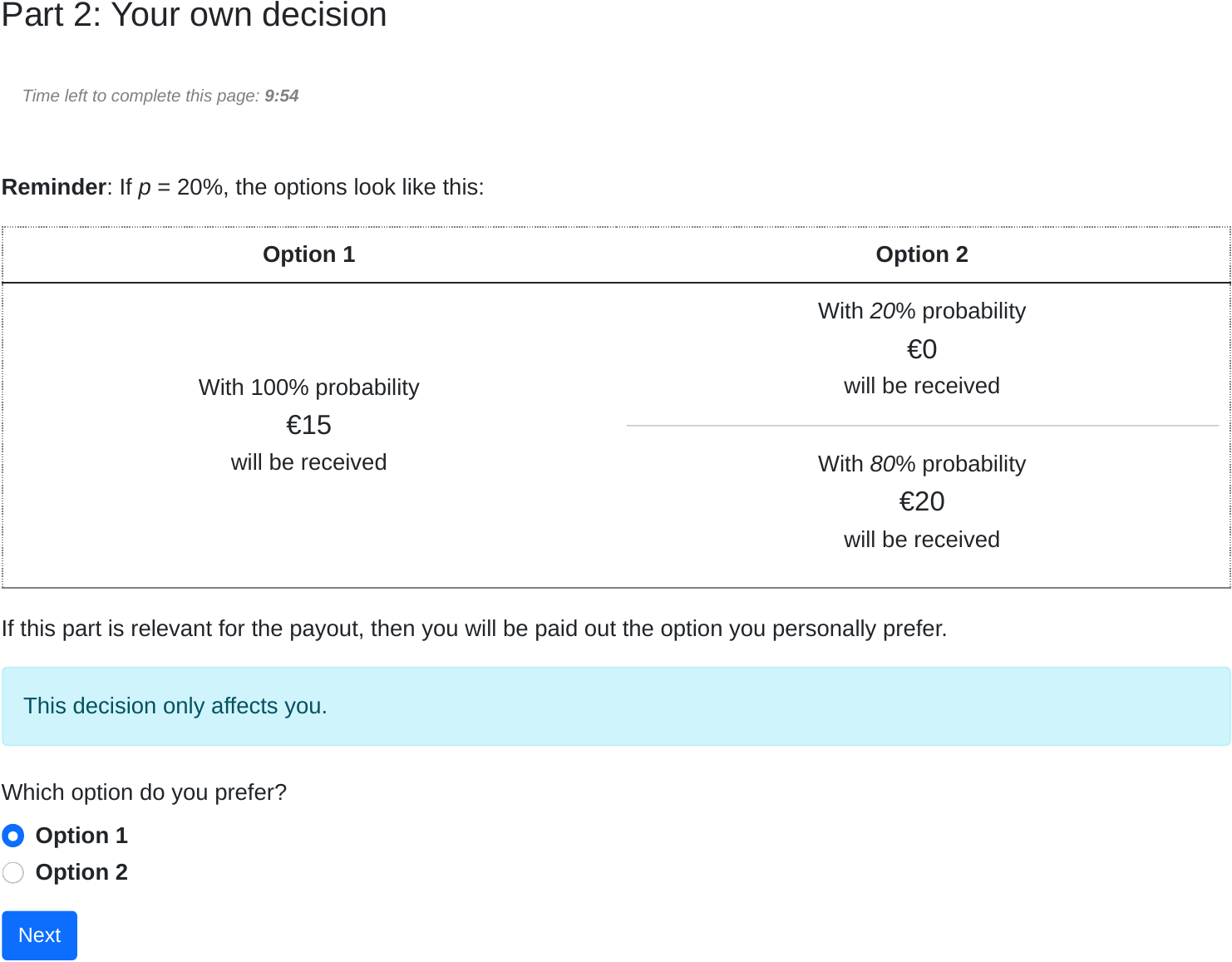}
\end{center}

\subsubsection*{Screen 19}

\begin{center}
    \includegraphics[width=\textwidth]{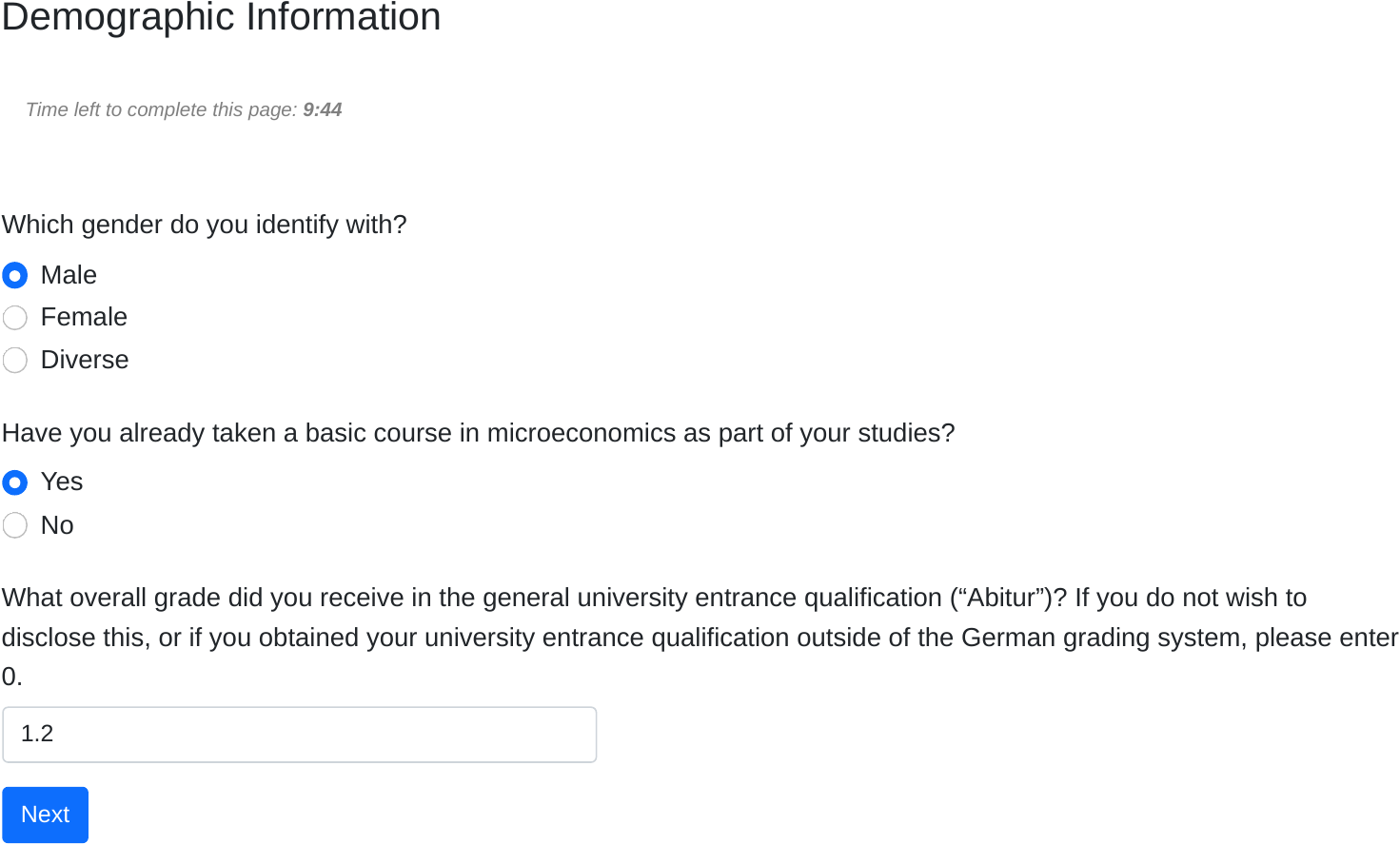}
\end{center}

\subsubsection*{Screen 20}

\begin{center}
    \includegraphics[width=\textwidth]{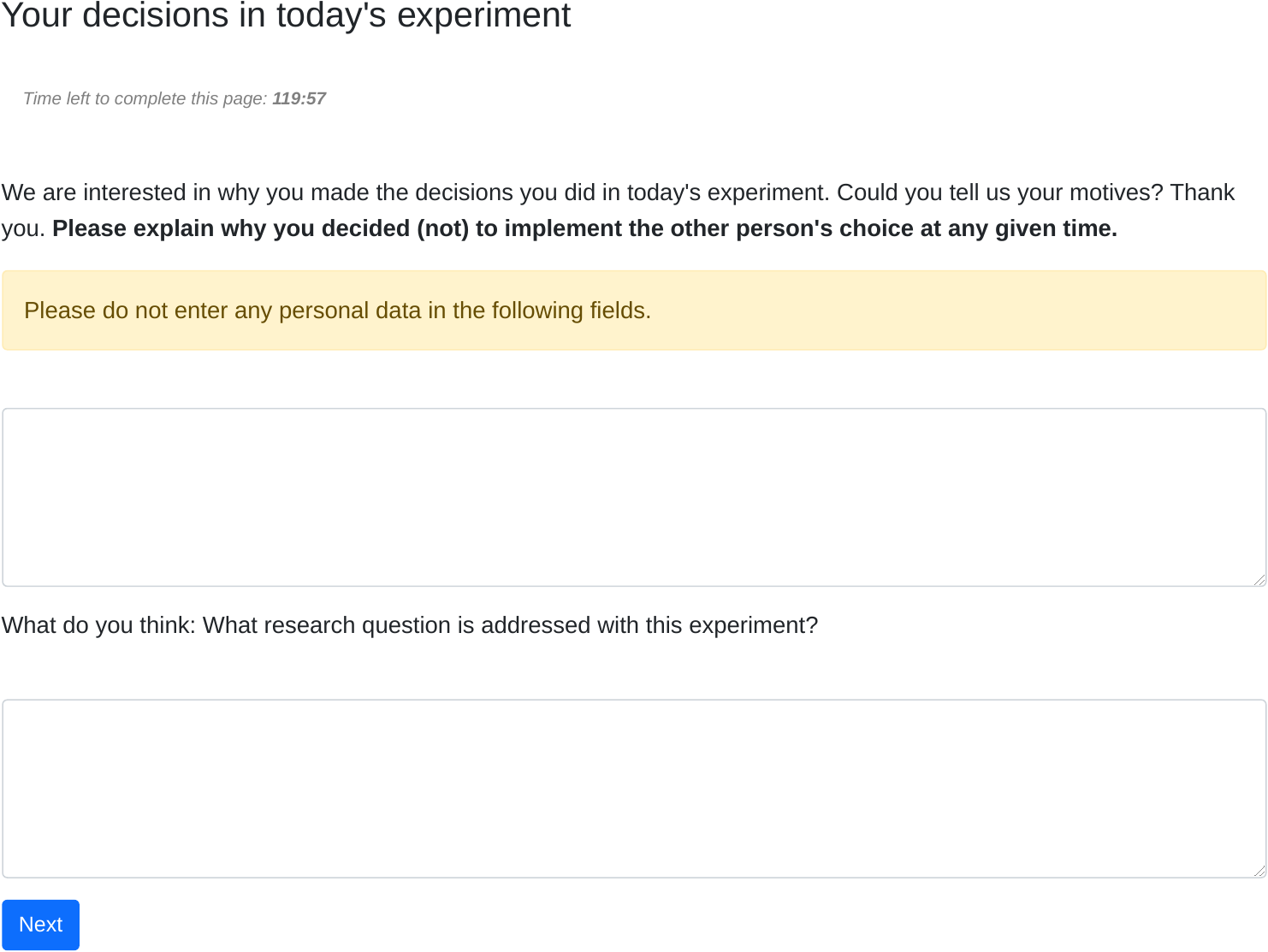}
\end{center}

\subsubsection*{Screen 21}

\begin{center}
    \includegraphics[width=\textwidth]{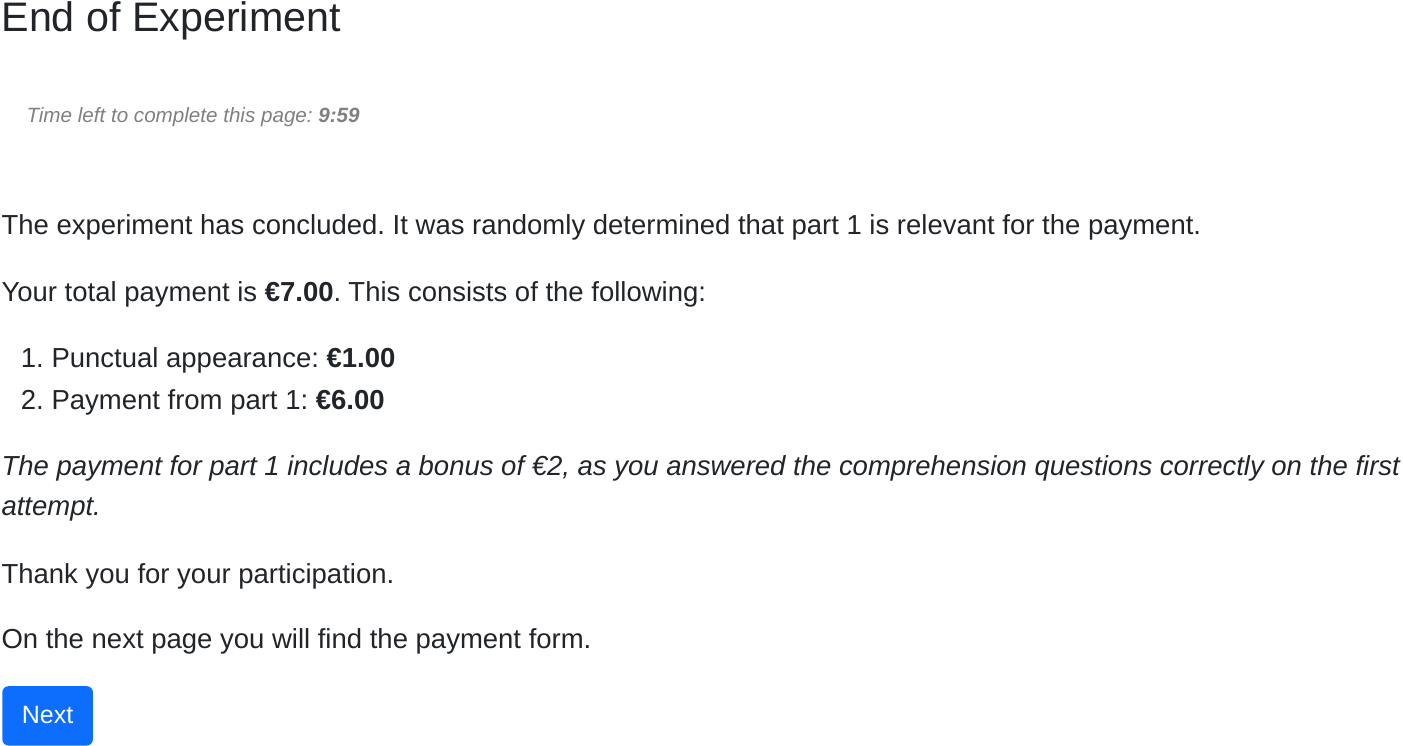}
\end{center}

\subsection{Experiment 2}
\label{app.instr2}

\subsubsection*{Screen 1}

(Consent form.)

\subsubsection*{Screen 2}

\begin{center}
    \includegraphics[width=\textwidth]{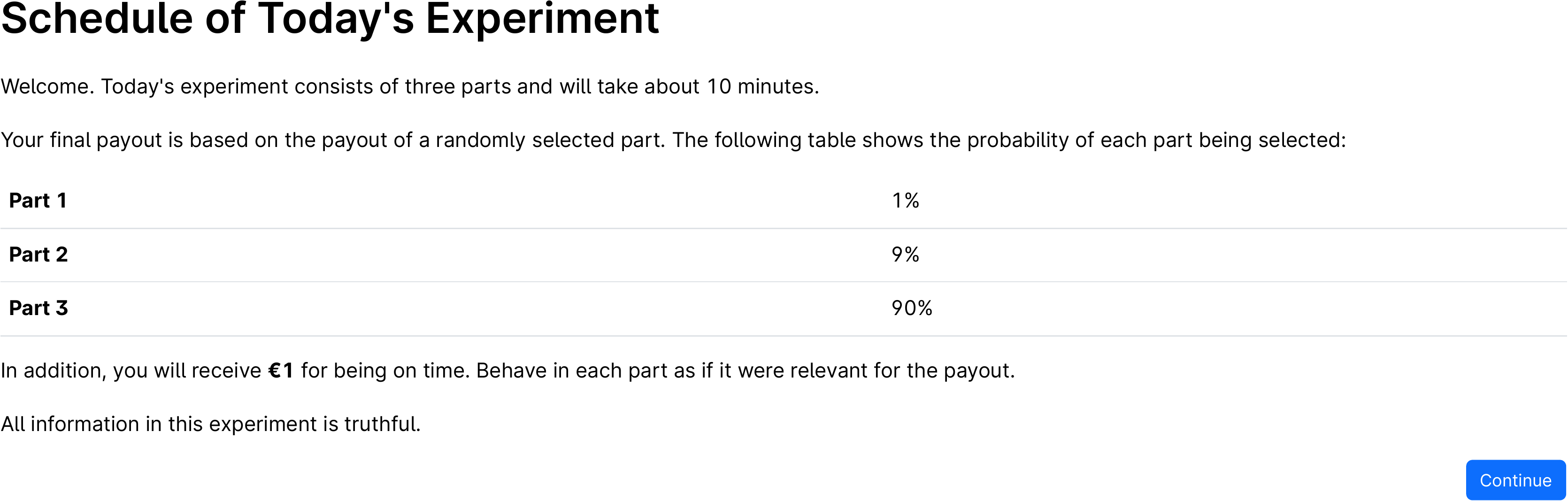}
\end{center}

\subsubsection*{Screen 3}

\begin{center}
    \includegraphics[width=\textwidth]{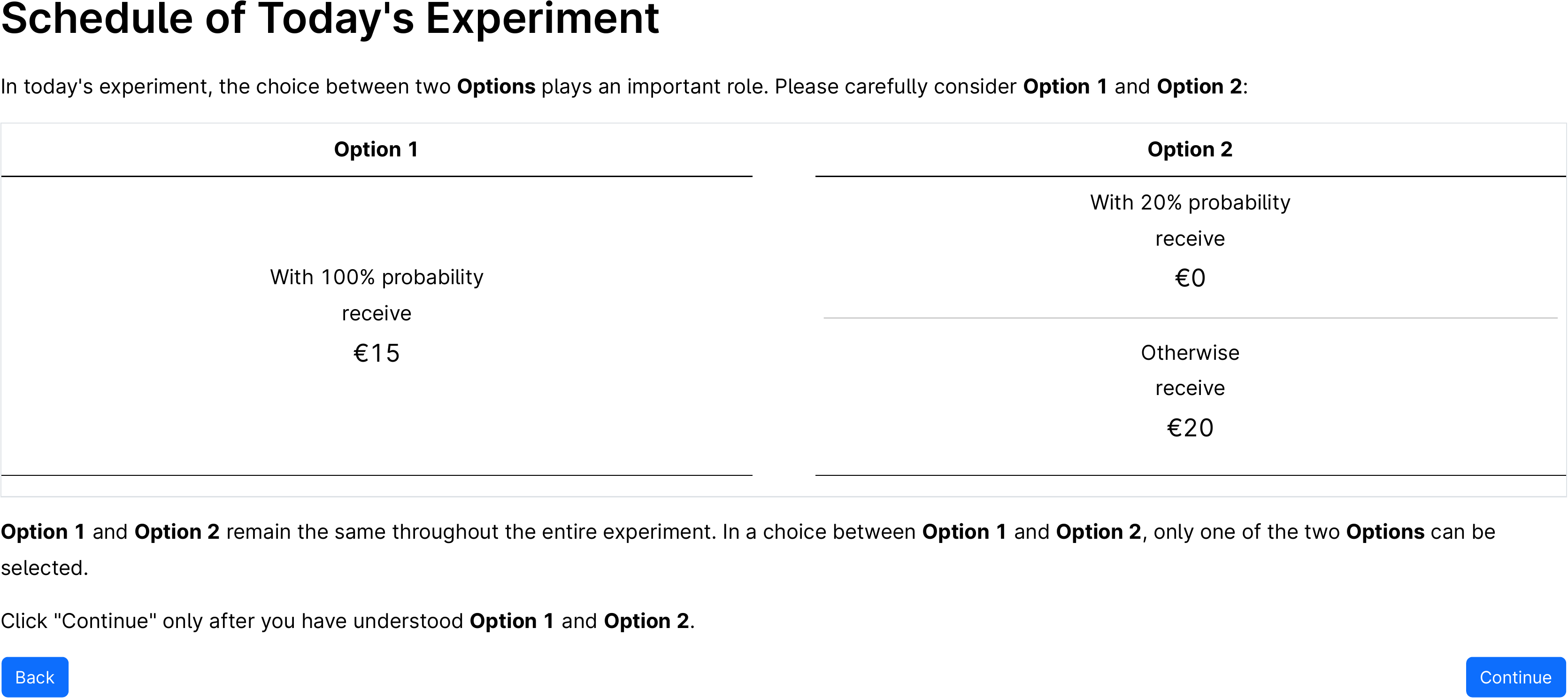}
\end{center}

\subsubsection*{Screen 4}

\begin{center}
    \includegraphics[width=\textwidth]{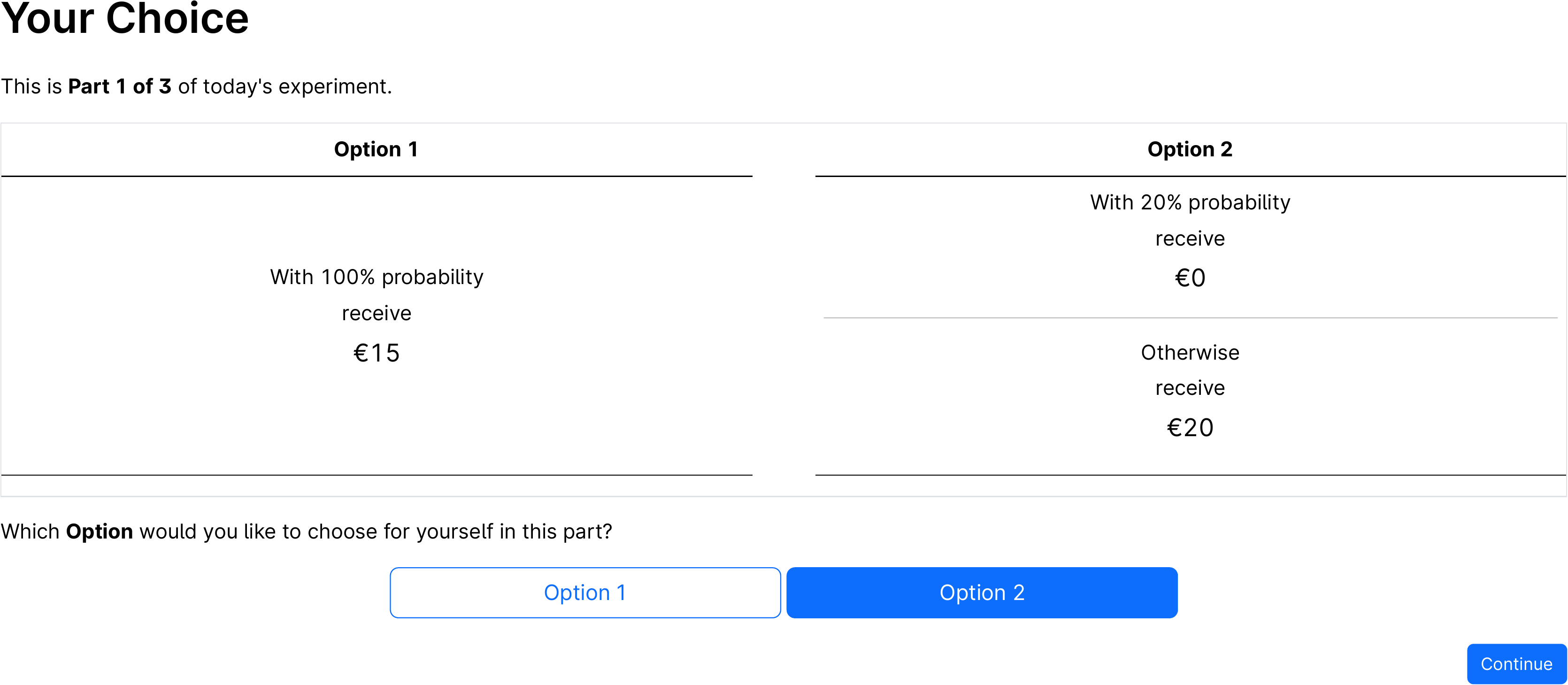}
\end{center}

\subsubsection*{Screen 5}

\begin{center}
    \includegraphics[width=\textwidth]{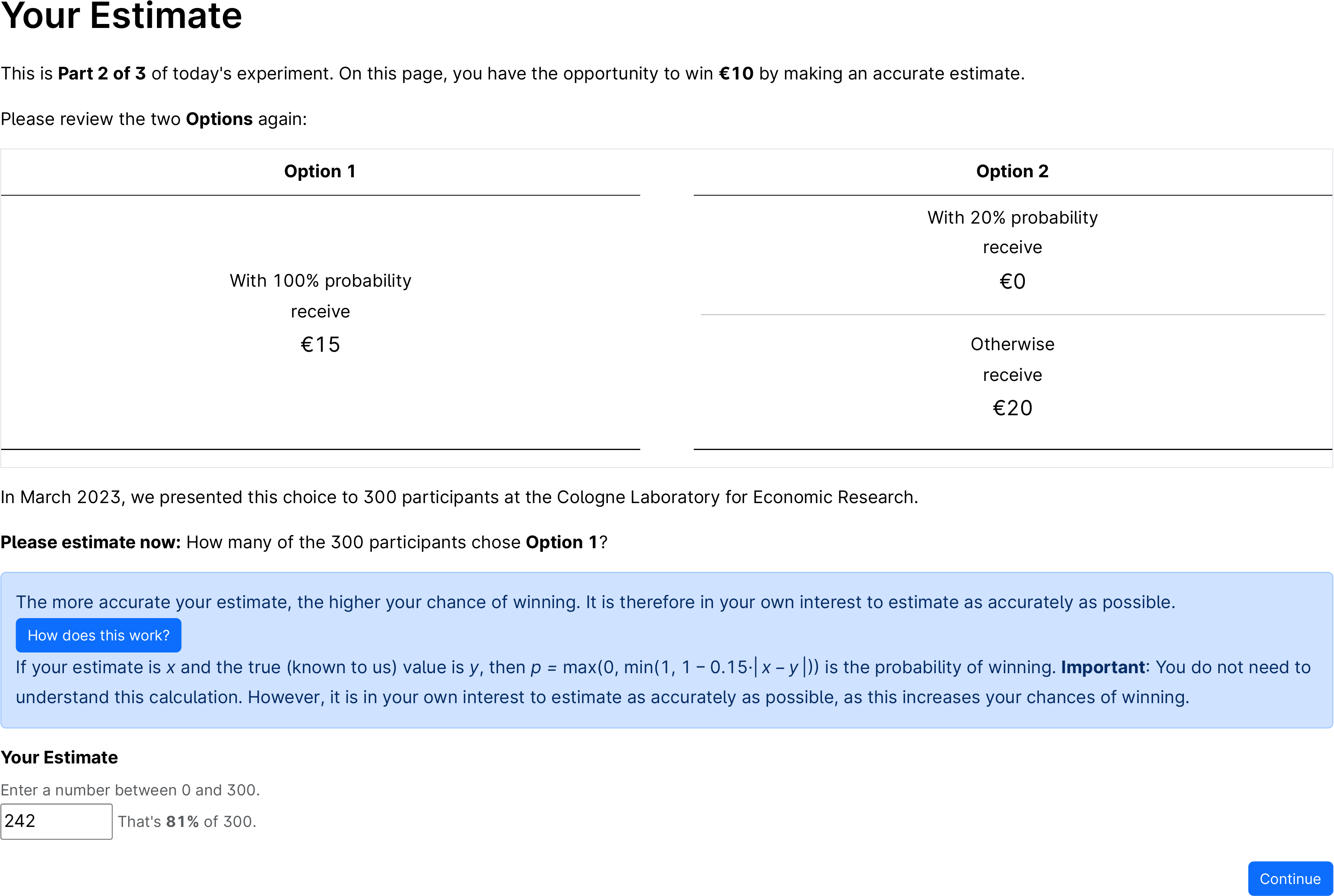}
\end{center}

\subsubsection*{Screen 6}

\begin{center}
    \includegraphics[width=\textwidth]{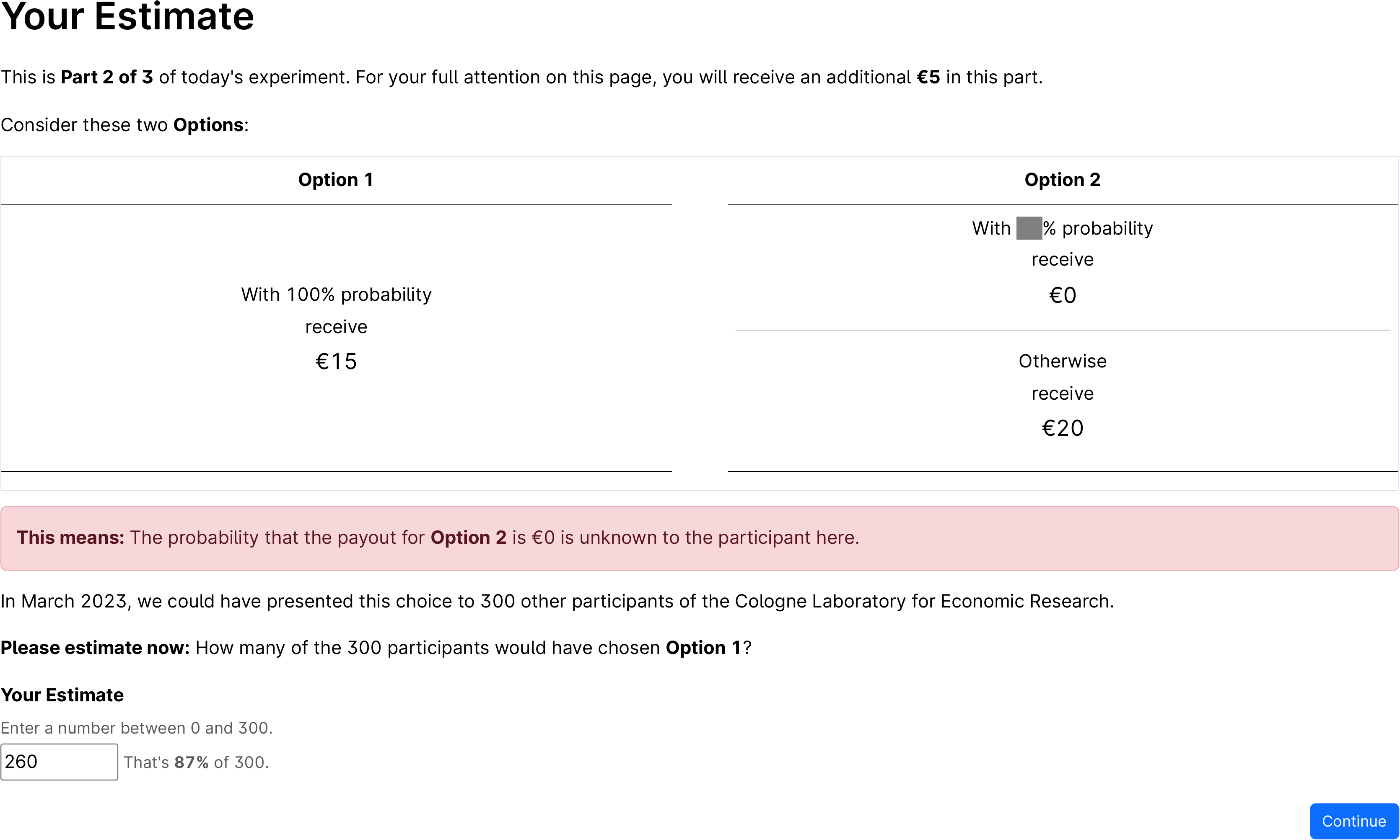}
\end{center}

\subsubsection*{Screen 7}

\begin{center}
    \includegraphics[width=\textwidth]{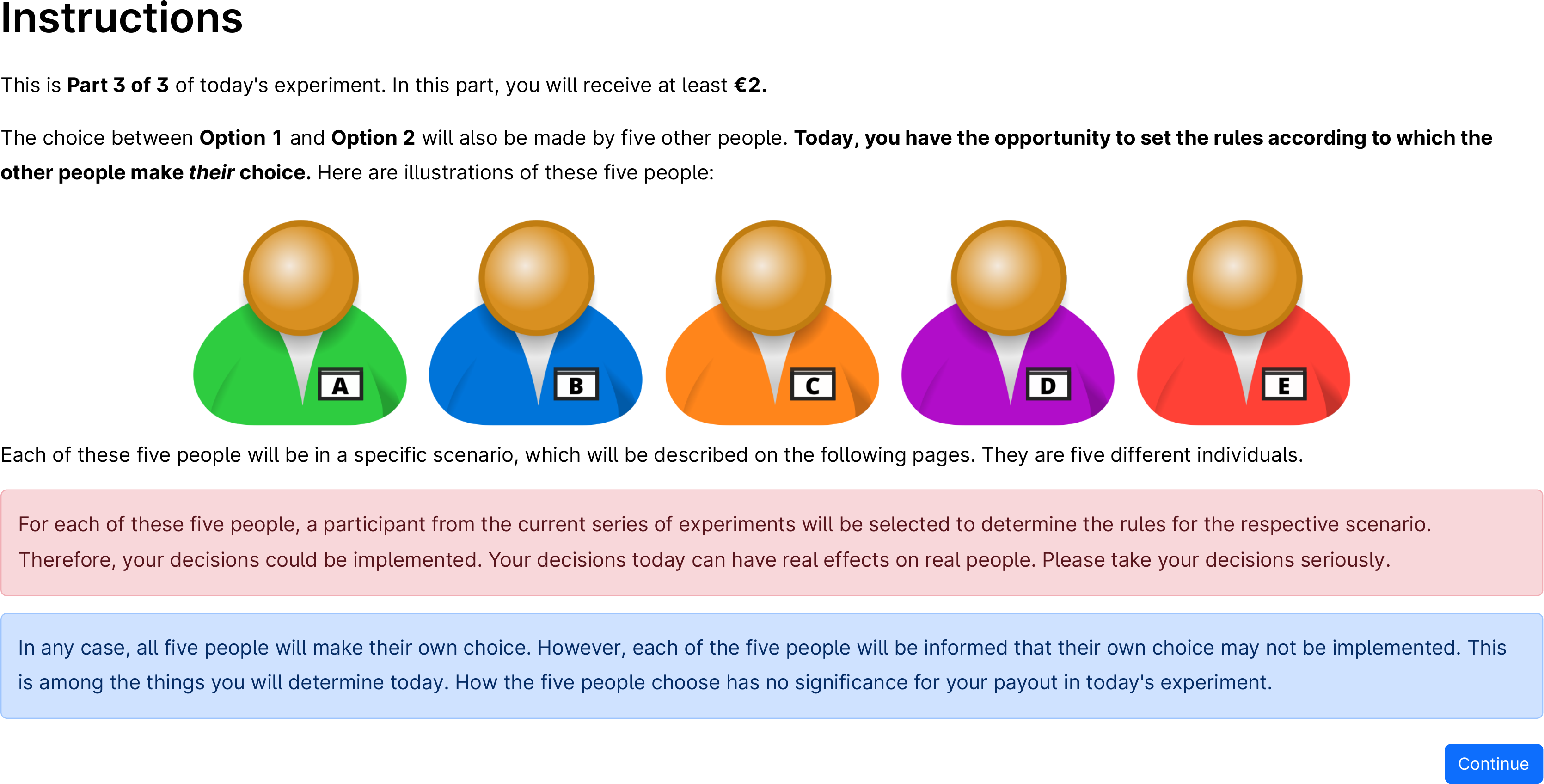}
\end{center}

\subsubsection*{Screen 8}

\begin{center}
    \includegraphics[width=\textwidth]{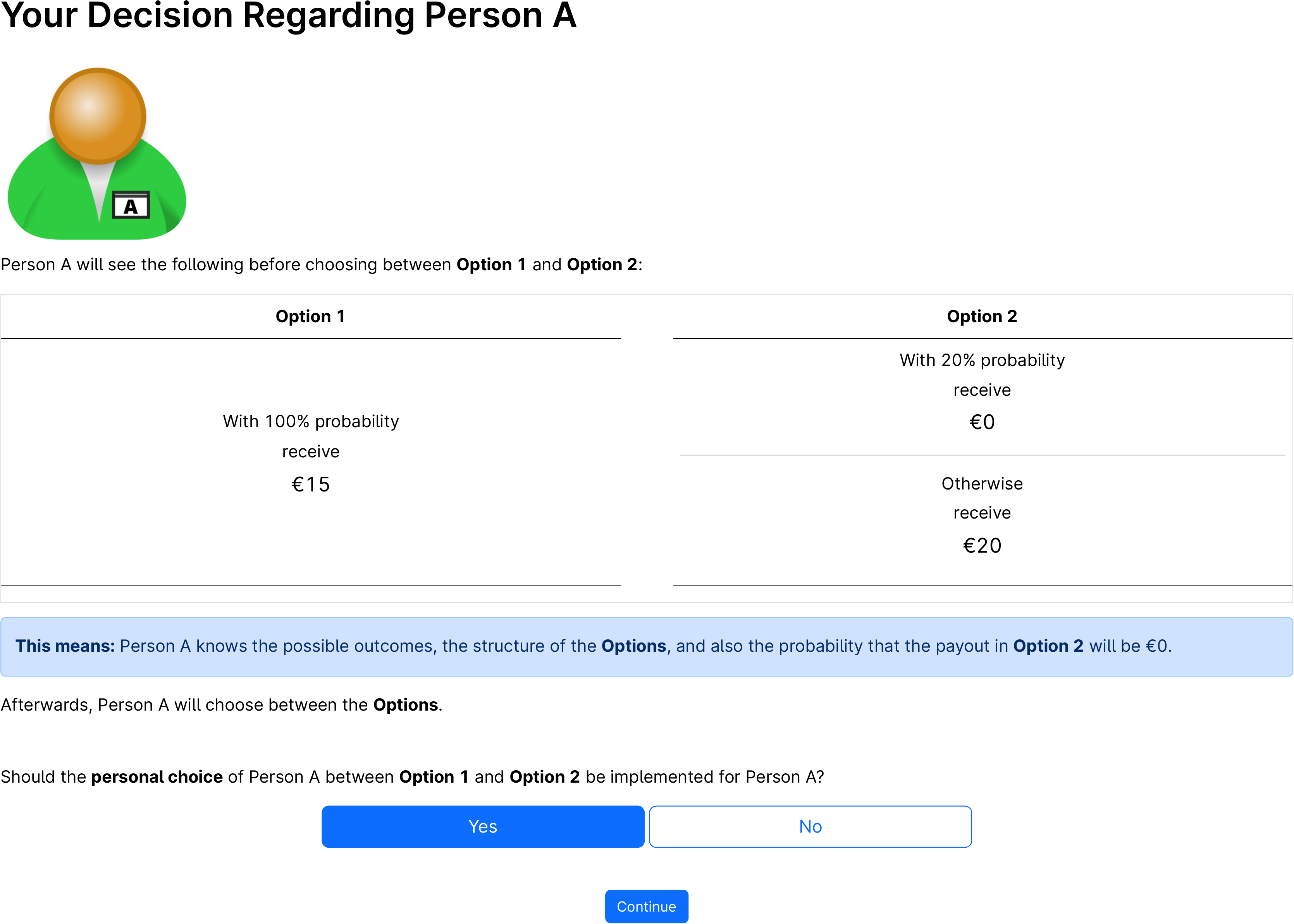}
\end{center}

\subsubsection*{Screen 9}

\begin{center}
    \includegraphics[width=\textwidth]{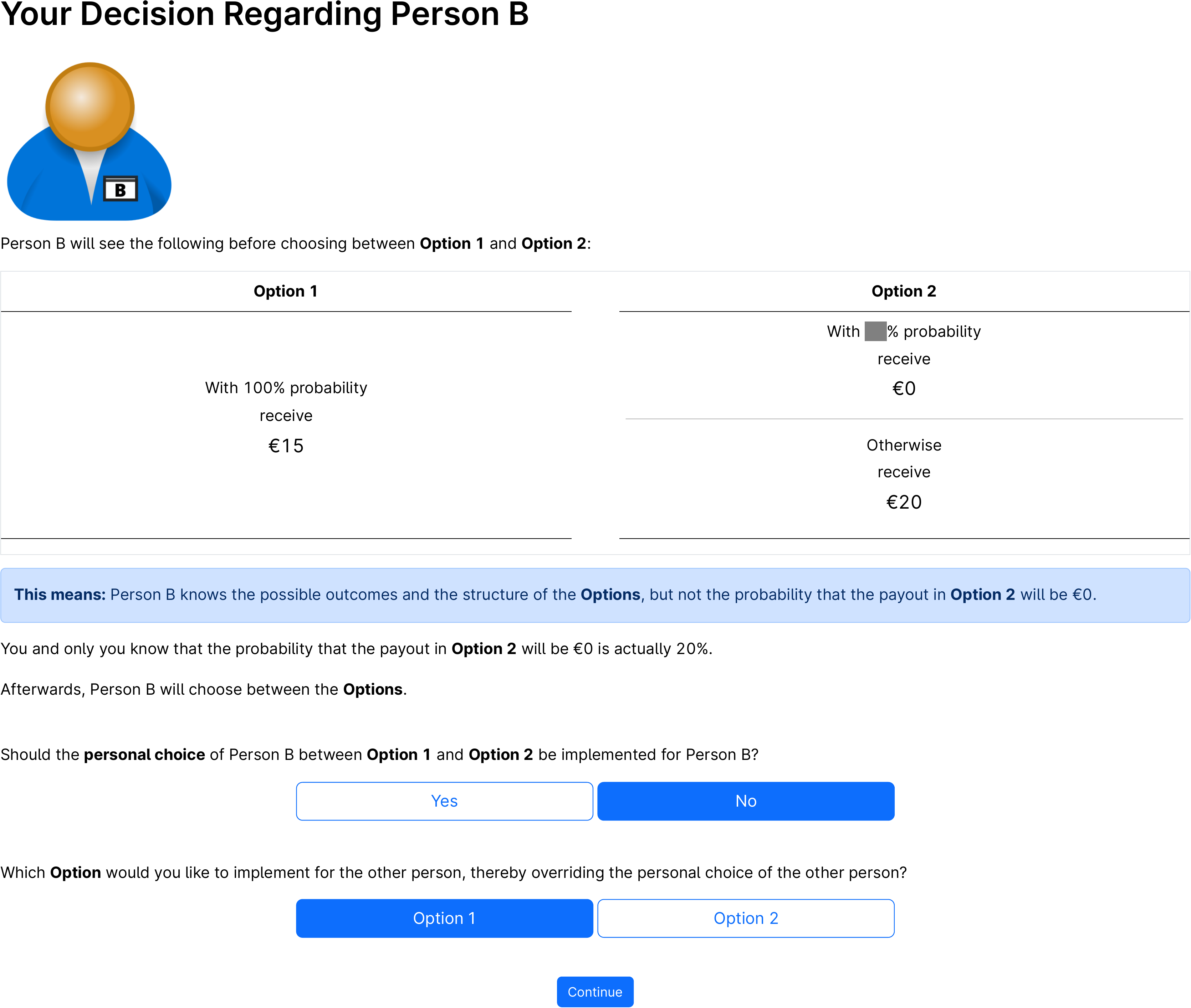}
\end{center}

\subsubsection*{Screen 10}

\begin{center}
    \includegraphics[width=\textwidth]{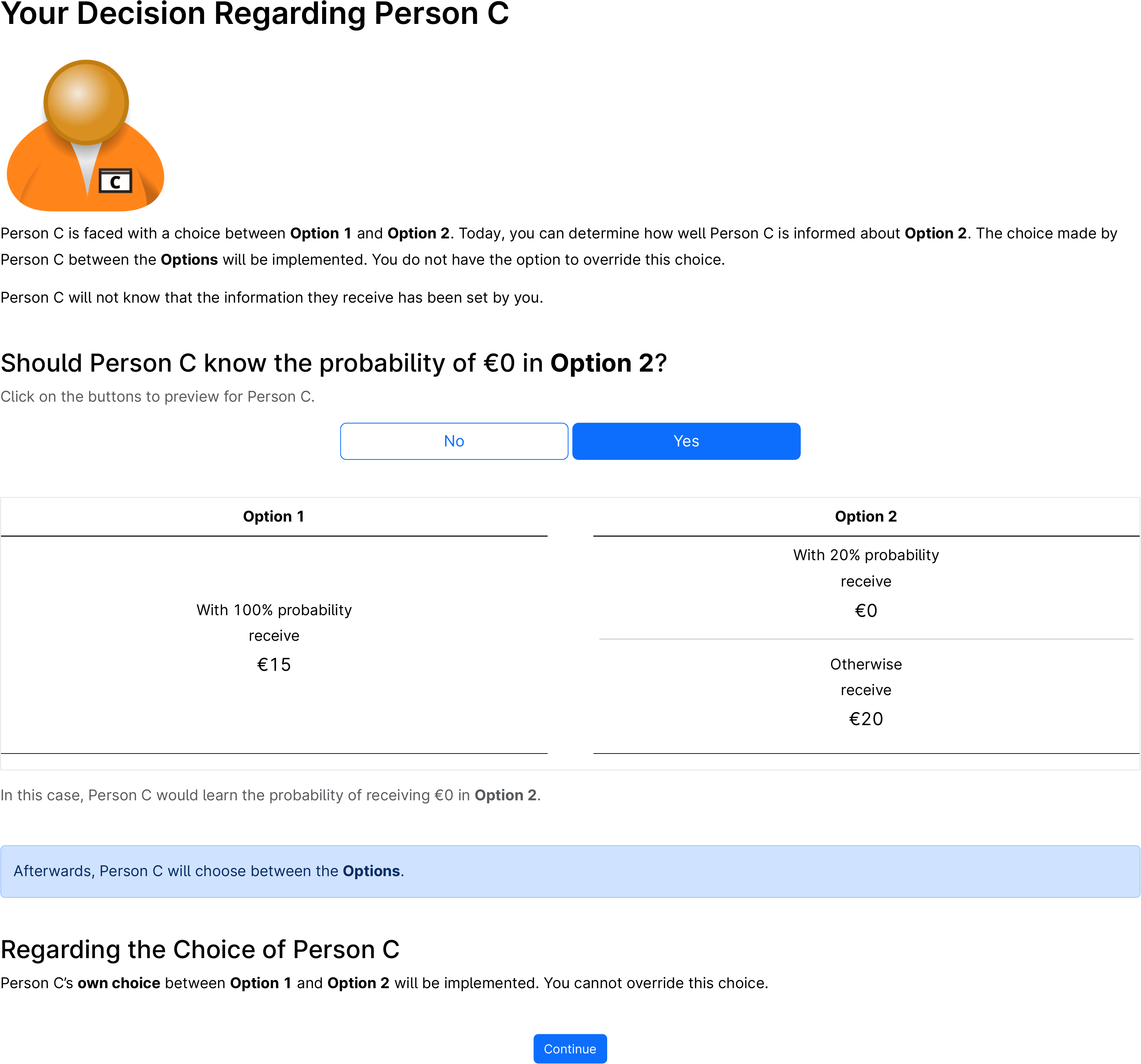}
\end{center}

\subsubsection*{Screen 11}

\begin{center}
    \includegraphics[width=\textwidth]{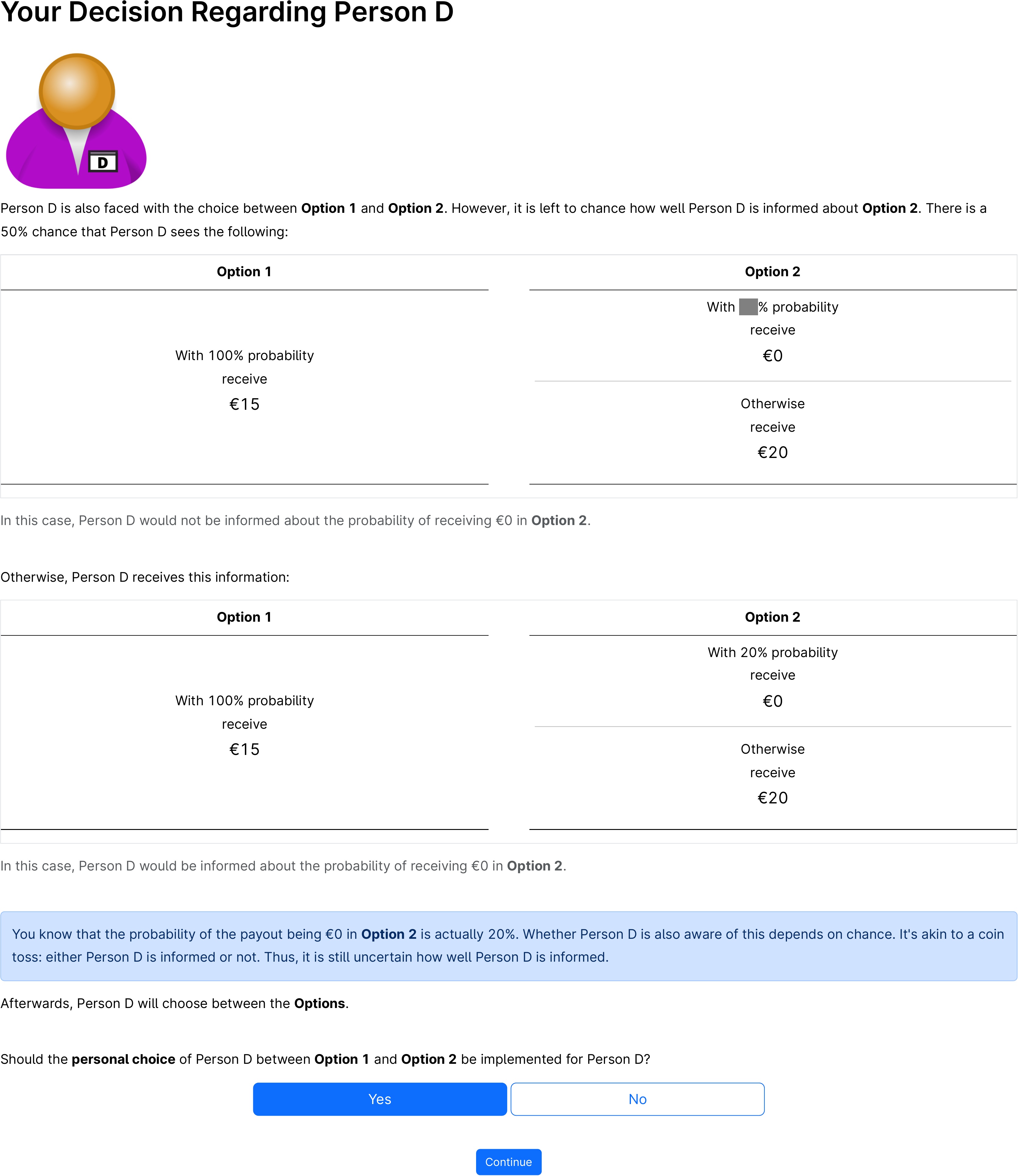}
\end{center}

\subsubsection*{Screen 12}

\begin{center}
    \includegraphics[width=\textwidth]{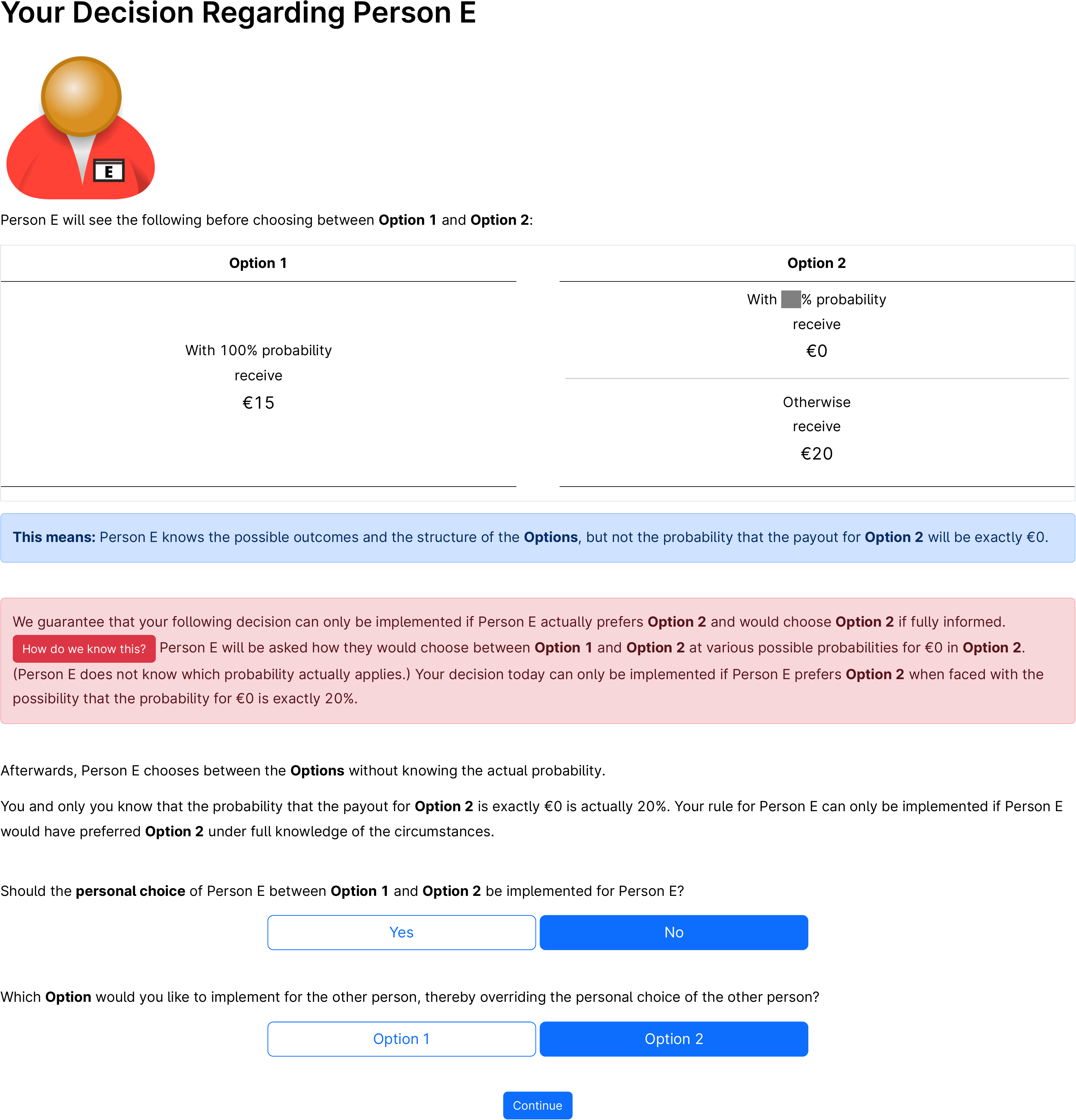}
\end{center}

\subsubsection*{Screen 13}

\begin{center}
    \includegraphics[width=\textwidth]{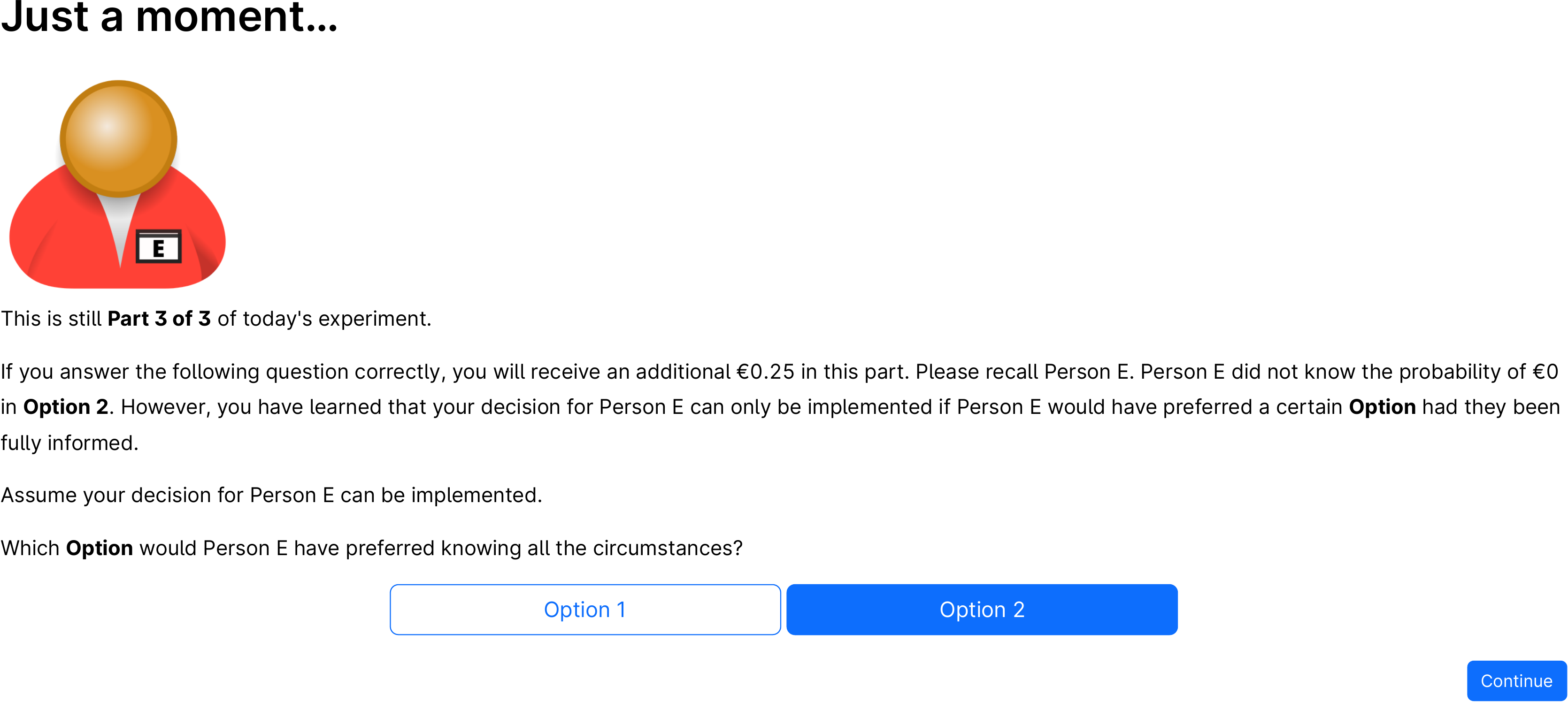}
\end{center}

\subsubsection*{Screen 14}

\begin{center}
    \includegraphics[width=\textwidth]{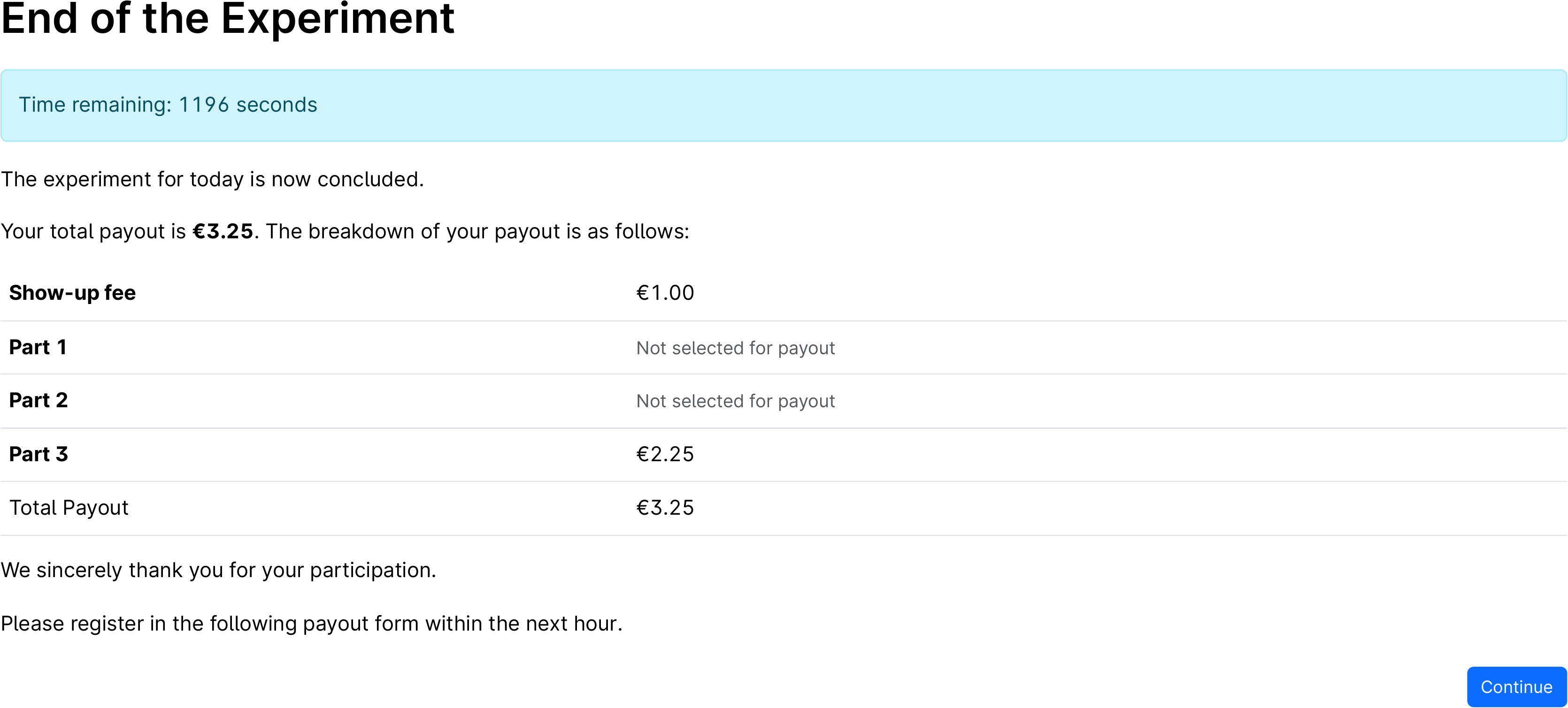}
\end{center}

\end{document}